\documentclass[]{aa520}           
\usepackage{natbib,amssymb,amsmath,curves,epsfig}

\bibpunct{(}{)}{;}{a}{}{,}

\newcommand{\hlf}{\ensuremath{\frac{1}{2}}}
\newcommand{\nux}{{\nu_x}}
\newcommand{\nue}{{\nu_\mathrm{e}}}
\newcommand{\nuae}{{\bar\nu_\mathrm{e}}}
\newcommand{\ye}{{Y_\mathrm{e}}}
\newcommand{\equ}{:=}
\newcommand{\dlin}[1]{\mathrm{d}#1\,}
\newcommand{\dvol}[1]{\mathrm{d}^3\vec{#1}\,}
\newcommand{\Real}{{\rm I\mathchoice{\kern-0.70mm}{\kern-0.70mm}{\kern-0.65mm}{\kern-0.50mm}R}}
\newcommand{\Natural}{{\rm I\mathchoice{\kern-0.55mm}{\kern-0.55mm}{\kern-0.50mm}{\kern-0.40mm}N}}
\newcommand{\Complex}{\rm C\kern-.42em\vrule width.03em height.58em
  depth-.02em\kern.4em}

\newcommand{\gcm}{\ensuremath{\mathrm{g}\,\mathrm{cm}^{-3}}}
\newcommand{\kb}{\ensuremath{k_\mathrm{B}\,}}
\newcommand{\mev}{\ensuremath{\mathrm{MeV}}}
\newcommand{\msol}{\ensuremath{M_{\odot}}}
\newcommand{\ms}{\ensuremath{\mathrm{ms}}}

\newcommand{\reffigN}[2][]{Fig.~\ref{#2}#1}
\newcommand{\reffig}[2][]{(Fig.~\ref{#2}#1)}
\newcommand{\reffigLN}[2][]{Figure~\ref{#2}#1}
\newcommand{\reffigA}[3][]{(#3 Fig.~\ref{#2}#1)}

\begin{document}
   \title{Radiation hydrodynamics with neutrinos}
   \subtitle{Variable Eddington
   factor method for core-collapse supernova simulations}

   \author{M. Rampp \and H.-Th. Janka}

   \offprints{M. Rampp, \email{mjr@mpa-garching.mpg.de}}

   \institute{Max-Planck-Institut f\"ur Astrophysik, 
                 Karl-Schwarzschild-Str.\ 1, D-85741 Garching, Germany}

   \date{Received  / Accepted }

\abstract{Neutrino transport and neutrino interactions 
in dense matter play a crucial role in stellar
core collapse, supernova explosions and neutron star formation.
Here we present a detailed description of a new 
numerical code for treating the time and energy dependent 
neutrino transport in hydrodynamical simulations of such 
events. The code is based on a variable Eddington 
factor method to deal with the integro-differential character 
of the Boltzmann equation. The moments of the neutrino distribution
function and the energy and lepton number exchange with the stellar
medium are determined by iteratively solving the zeroth and
first order moment equations in combination with a model
Boltzmann equation. The latter is discretized on a grid of 
tangent rays. The integration of the transport equations and
the neutrino source terms is performed in a time-implicit way.
In the present version of the program, the transport part is
coupled to an explicit hydrodynamics code which 
follows the evolution of the stellar plasma by a 
finite-volume method with piecewise parabolic interpolation, 
using a Riemann solver for calculating the hydrodynamic states. 
The neutrino source terms are implemented in an operator-split
step. Neutrino transport and hydrodynamics can be calculated
with different spatial grids and different time steppings. 
The structure of the described code is modular and offers 
a high degree of flexibility for an application to relativistic
and multi-dimensional problems at different levels of refinement
and accuracy. We critically evaluate results for a number of test cases,
including neutrino transport in rapidly moving stellar media and
approximate relativistic core collapse, and suggest a path for
generalizing the code to be used in multi-dimensional simulations of
convection in neutron stars and supernovae.

}

\maketitle
%

\section{Introduction}

When the core of a massive star collapses to a neutron
star, a huge amount of gravitational binding energy is 
released mainly in the form of neutrinos which are 
abundantly produced by particle reactions in the hot and
dense plasma. In fact, the emission of neutrinos determines
the sequence of dynamical events which precede the death of
the star in a supernova explosion. Electron neutrinos from electron
captures on protons and nuclei reduce the electron fraction 
and the pressure and thus accelerate the contraction of the 
stellar iron core to a catastrophic implosion. The
position of the formation of the supernova shock at the
moment of core bounce depends on the degree of 
deleptonization during the collapse phase. The outward
propagation of the prompt shock is severely damped by energy 
losses due to the photodisintegration of iron nuclei, and
finally stopped only a few milliseconds later when
additional energy is lost in a luminous outburst of electron
neutrinos. These neutrinos are created in the shock-heated
matter and leave the star as soon as their diffusion is 
faster than the expansion of the shock.
``Prompt'' supernova explosions generated by a direct propagation of
the hydrodynamical shock have been obtained in simulations only when
the stellar iron core is very small \cite[]{barcoo90} and/or the
equation of state of neutron-rich nuclear matter is extraordinarily
soft \cite[]{barcoo85,barbet87}.

At a later stage (roughly a hundred milliseconds after 
the shock formation) the situation has changed. The 
temperature behind the stalled shock has dropped such that
increasingly energetic neutrinos diffusing out from 
deeper layers start to transfer energy to the stellar gas
around the nascent neutron star. If this energy deposition is 
sufficiently strong, the stalled shock can be revived and a
``delayed''
explosion can be triggered (\citealp{wil85,betwil85}; the idea that neutrinos provide the energy of the supernova 
explosion was originally brought up by \citealp{colwhi66}).
A small fraction of less than one per cent of the
energy released in neutrinos can account for the kinetic energy
of a typical type II supernova. This explanation is currently
favored for the explosion of massive stars in the mass range
between about 10$\,M_{\odot}$ and roughly 25$\,M_{\odot}$. 
It is supported by numerical simulations and analytic 
considerations. A finally convincing hydrodynamical simulation,
however, is still missing \cite[a summary of our current 
understanding of the explosion mechanism can be found 
in][]{jan01}. The measurement of a neutrino
signal from a Galactic supernova would offer the most direct
observational test for our theoretical perception of the onset
of the explosion. Due to the central role of neutrinos in
supernovae, the neutrino transport deserves particular 
attention in numerical models.

\subsection{Brief review of the status of supernova models} 

Current hydrodynamic models of neutrino-driven supernova explosions
leave an ambiguous impression and have caused confusion about the 
status of the field outside the small community of supernova
modelers. Simulations by different groups seem to be contradictory
because some models show successful explosions by the neutrino-driven 
mechanism whereas others have found the mechanism to fail. 

Wilson and
collaborators have performed successful simulations for more 
than 15 years now 
\cite[e.g.,][]{wil85,wilmay86,wilmay88,wilmay93,maytav93,totsat98}
Their models were calculated in spherical symmetry, but shock 
revival was obtained by making the assumption that the neutrino
luminosity is boosted by neutron-finger instabilities in the 
nascent neutron star \cite[]{wilmay88,wilmay93}. 
In fact, in their calculations explosion energies comparable to
typically observed values (of the order of $10^{51}\,$erg)
require pions to be abundant in the nuclear matter already 
at moderately high densities \cite[]{maytav93}. The 
corresponding equation of state (EoS) with pions leads to higher
core temperatures and the emission of more energetic neutrinos.
Both the convectively boosted luminosities
and the harder spectra enhance the neutrino heating behind the 
stalled shock. The relevance of neutron-finger instabilities
for efficient energy transport on large scales, however, has not
been demonstrated by direct simulations.
The existence of neutron-finger instabilities 
\cite[in the sense 
of the definition introduced by][]{wilmay93}
is indeed questioned by other investigations \cite[]{brudin96}
and might be a consequence of specific properties of the 
high-density EoS or the treatment of the neutrino transport 
by Wilson and collaborators. Also the implementation
of a pionic component in their EoS is at least controversial
and not in agreement with other, more conventional 
descriptions of nuclear matter (Pethick \& Pandharipande, 
personal communication). 

Rather than finding neutron-finger instabilities, two-dimensional 
hydrodynamic simulations have shown that regions inside the
nascent neutron star can exist where Ledoux or quasi-Ledoux
convection \cite[as defined by][]{wilmay93}
develops on a time scale of tens of milliseconds
after core bounce (\citealp{keijan96,kei97}; also
\citealp{jankif01}). 
The models were constrained to a 
simulation of the neutrino cooling of the nascent neutron
star, but it must be expected that the convective enhancement
of the neutrino luminosity, which becomes appreciable after
about 200 milliseconds \cite[see also][]{jankif01}, can have
important consequences for the revival of the stalled supernova
shock \cite[]{bur87}. Ledoux unstable regions were also
detected in spherical cooling models of newly formed neutron
stars \cite[]{bur87,burlat88,ponred99,mirpon00}.
Nevertheless, their significance is 
controversial, because \cite{brumez95:phyrep} in spherically 
symmetric models and \cite{mezcal98:pnconv} in two-dimensional
simulations observed convection inside the neutrinosphere
only as a transient phenomenon during a relatively short
period after core bounce. 

The question is therefore undecided whether convection plays an important 
role in newly formed neutron stars and
if so, what its implications for the supernova explosion are.
Unfortunately, the various calculations were performed
with different treatments of the neutrino transport, 
one-dimensional vs.\ two-dimensional hydrodynamics, general
relativistic gravitational potential vs.\ Newtonian gravity,
or even an inner boundary condition to replace
the central, dense part of the neutron star core
\cite[]{mezcal98:pnconv}. It is this inner region, however, 
where \cite{keijan96} found convection to develop roughly
100 milliseconds after bounce. Convection at larger radii
and thus closer below the neutrinosphere had indeed 
died out within only 20--30 milliseconds after shock formation
in agreement with the results of \cite{brumez95:phyrep}.
Future studies with a better and more 
consistent handling of the different aspects of the physics 
are definitely needed to clarify the situation. 

A second hydrodynamically unstable region develops
exterior to the nascent neutron star in the neutrino-heated
layer behind the stalled supernova shock. 
Convective overturn in this region is helpful and can lead 
to explosions in cases which otherwise fail 
\cite[]{herben92,herben94,janmue95,janmue96,burhay95}.
In two-dimensional supernova models computed recently by
\cite{fry99}, \cite{fryheg00}, and \cite{fryheg01} these hydrodynamical 
instabilities in the postshock region are crucial for the 
success of the neutrino-driven mechanism, because they help
transporting hot gas from the neutrino-heating
region directly to the shock, while downflows
simultaneously carry cold, accreted matter to the layer
of strongest neutrino heating where a part of this gas readily 
absorbs more energy from the neutrinos.
The existence of this multi-dimensional phenomenon seems to
be generic for the situation which builds up in the stellar core
some time after shock formation. It is therefore no matter of 
dispute. 

All simulations showing explosions as a consequence of
post-shock convection, however, have so far been performed with a 
strongly simplified treatment of the crucial neutrino physics,
e.g., with grey, flux-limited diffusion which is matched to a 
``light bulb'' description at some ``low'' value of the
optical depth \cite[]{herben94,burhay95}.
Alternatively, an inner boundary near the neutrinosphere
had been used where the spectra and luminosities of neutrinos
were prescribed to parameterize our 
ignorance and the potential uncertainties of the exact 
properties of the neutrino emission
from the newly formed neutron star \cite[]{janmue95,janmue96}.
It is therefore not clear whether the instabilities and their
associated effects in fully self-consistent and more accurate
simulations will be strong enough to cause
successful explosions. Doubts in that
respect were raised by \cite{mezcal98:ndconv}, whose 
two-dimensional models showed convective overturn in the
neutrino-heating region but still no explosion. 
\cite{mezcal98:ndconv} combined two-dimensional 
hydrodynamics with neutrino transport results obtained by 
multi-group flux-limited         
diffusion in spherical symmetry. Although not 
self-consistent, their approach is nevertheless an improvement
compared to previous treatments of the neutrino physics by 
other groups.

All computed models bear some pieces of truth. Essentially they 
demonstrate the remarkable sensitivity of the supernova 
dynamics to the different physical aspects of the problem,
in particular
the treatment of neutrino transport and neutrino-matter
interactions, the properties of the nuclear EoS, 
multi-dimensional hydrodynamical processes, and general
relativity. Considering the huge energy reservoir carried
away by neutrinos, the neutrino-driven mechanism appears 
rather inefficient (the often quoted value of 1\% efficiency,
however, misjudges the true situation, because neutrinos can
transfer between 5\% and 10\% of their energy to the stellar 
gas during the critical period of shock revival;
see \citealp{jan01}). Nevertheless, neutrino-driven explosions
are ``marginal'' in the sense that the energy of a 
standard supernova explosion is of the same order as the 
gravitational binding energy of the ejected progenitor
mass. The final success of the supernova shock is the result of
different physical processes which compete against each other. On
the one hand, neutrino heating tries to drive the shock expansion,
on the other hand energy losses, e.g., by neutrinos that are reemitted
from the inward flow of neutrino-heated matter which enters the
neutrino-cooling zone below the heating layer, 
extract energy and thus damp the shock revival.
It is therefore not astonishing that different approximations
or descriptions for one or more physical components 
of the problem can decide between an explosion or failure
of a simulation.

None of the current supernova models includes all relevant 
aspects to a satisfactory level of accuracy, but all of these
models are deficient in one or more respects. Wilson and
collaborators obtain explosions, but their input physics
is unique and cannot be considered as generally accepted.
Two-dimensional \cite[]{herben94,burhay95,fry99,fryheg00,fryheg01}
and three-dimensional (Fryer, personal communication)
models show explosions due to strong postshock convection, 
but the neutrino transport and neutrino-matter interactions
are handled at a level of accuracy which falls back behind 
the most elaborate treatments that have been applied in spherical
symmetry. The sensitivity of the outcome of numerical calculations
to details of the neutrino transport demands improvements.
Self-consistent multi-dimensional simulations with 
a sophisticated and quantitatively reliable treatment of the 
neutrino physics have yet to be performed.

\subsection{Boltzmann solvers for neutrino transport}

Most recently, spherically symmetric Newtonian 
\cite[]{ramjan00,mezlie01}, and general relativistic 
\cite[]{liemez01} hydrodynamical
simulations of stellar core-collapse and post-bounce evolution
including a Boltzmann solver for the neutrino transport
have become possible. Although the models do not yield
explosions, they must be considered as a major achievement
for the modeling of supernovae. Before these calculations only
the collapse phase of the stellar core had been investigated
with solving
the Boltzmann equation \cite[]{mezbru93:coll}. Boltzmann
transport has now superseded multi-group
flux-limited diffusion \cite[]{arn77,bowwil82,bru85,myrblu87,barmyr89,bru89:xp,bru89:eos,bru93,coobar92} 
as the most elaborate treatment of neutrinos in supernova models. 

The differences between both methods in dynamical
calculations have still to be figured out in detail, but the
possibility of solving the Boltzmann equation removes
imponderabilities and inaccuracies associated with the use
of flux-limiters in particular in the region of semi-transparency,
where neutrinos decouple from the stellar background and also
deposit the energy for an explosion 
\cite[]{jan91,jan92,mesmez98,yamjan99}.
For the first time the
transport can now be handled at a level of sophistication where 
the technical treatment is more accurate than our standard
description of neutrino-matter interactions, which includes
various approximations and simplifications (for an overview
of the status of the handling of neutrino-nucleon interactions, 
see \citealp{hor02} and the references therein).

In this paper we describe our new numerical code for solving
the energy and time dependent 
Boltzmann transport equation for neutrinos coupled to the
hydrodynamics of the stellar medium. First results
from supernova calculations with this code were published
before \cite[]{ramjan00}, but a detailed documentation of
the method will be given here. It is based on a variable 
Eddington factor technique where the coupled set of Boltzmann
equation and neutrino energy and momentum equations is 
iterated to convergence. Variable Eddington moments of 
the neutrino phase space distribution are used for closing 
the moment equations, and the integro-differential character
of the Boltzmann equation is tamed by using the zeroth and 
first order angular moments (neutrino density and flux) in the
source terms on the right hand side of the Boltzmann equation.
This numerical approach is fundamentally different from the
so-called S$_{\mathrm N}$ methods 
\cite[]{car67,yuebuc77,mezbru93:code,yamjan99} which employ a
direct discretization of the Boltzmann equation in all variables
including the dependence on the angular direction of the 
radiation propagation.

Some basic characteristics of our code are similar to elements
described by \cite{buryou00}. Different from the latter
paper we shall discuss the details of the numerical implementation 
of the transport scheme and its coupling to a hydrodynamics
code \cite[]{ram00}, which in our case is the 
PROMETHEUS code with the potential of performing 
multi-dimensional simulations.
The variable Eddington factor technique was our method of
choice for the neutrino transport because of its
modularity and flexibility, which offer significant advantages
for a generalization to multi-dimensional problems. We shall
suggest and motivate corresponding approximations which 
we consider as reasonable in the supernova case, at least as a 
first step towards multi-dimensional hydrodynamics with a 
Boltzmann treatment of the neutrino transport.

This paper is arranged as follows: 
In Sect.~\ref{chap:rhdbasics} the equations of radiation
hydrodynamics are introduced.
Sect.~\ref{chap:numrhd} provides a
general overview of the numerical methods used to solve these
equations and contains details
of their practical implementation.
Results for a number of idealized test problems as well as
applications of the new method to the supernova problem are presented in
Sect.~\ref{chap:test}. 
A summary will be given in Sect~\ref{chap:concl}.
In the Appendix the numerical implementation of neutrino
opacities and the equation of state used for core-collapse and
supernova simulations is described.

\section{Radiation-hydrodynamics --- basic equations}\label{chap:rhdbasics}

\subsection{Hydrodynamics and equation of state}\label{chap:rhdbasics.hydro}

For an ideal fluid characterized by the mass density $\rho$, the
Cartesian components of the velocity vector 
$(v_1, v_2, v_3)^\mathrm{T}$, 
the specific energy density $\varepsilon = e + \frac{1}{2} v^2$ and
the gas pressure 
$p$, the Eulerian, nonrelativistic equations of hydrodynamics in
Cartesian coordinates read (sum over $i$ implied): 
\begin{eqnarray}
\lefteqn{
\partial_t\rho + \partial_i(\rho v_i) = 0 \,,}\label{eq:hydro.rho} \\
\lefteqn{
\partial_t(\rho v_k) + \partial_i(\rho v_i v_k+ \delta_{ik}p)  =
- \rho\partial_k\Phi^\mathrm{Newt} \!\!+ {Q_\mathrm{M}}_k \,,}\label{eq:hydro.v} \\
\lefteqn{
\partial_t(\rho \varepsilon) + 
\partial_i(\{\rho \varepsilon +p \}v_i) =
- \rho v_i \partial_i\Phi^\mathrm{Newt} \!\!+ {Q_\mathrm{E}} 
                    \!+ v_i{Q_\mathrm{M}}_i \,,}\label{eq:hydro.etot}
\end{eqnarray}
where $\Phi^\mathrm{Newt}$ denotes the Newtonian gravitational 
potential of the fluid, which can be determined by the Poisson equation
$\partial_i\partial^i \Phi^\mathrm{Newt}=4\pi G\,\rho$ 
($G$ is Newton's constant).
$\vec{Q}_\mathrm{M}$ and $Q_\mathrm{E}$ are the neutrino source 
terms for momentum transfer and
energy exchange, respectively, $\delta_{ik}$ is the Kronecker symbol,
and $\partial_i\equ\partial/\partial x^i$ is an abbreviation for the
partial derivative with respect to the coordinate $x^i$. 
In order to describe the evolution of the chemical composition of the
(electrically neutral) fluid, the hydrodynamical equations are
supplemented by a conservation equation for the electron fraction $\ye$,
\begin{equation}\label{eq:hydro.ye}
\partial_t (\rho \ye) + \partial_i(\rho \ye\, v_i) = 
Q_\mathrm{N}
\,,
\end{equation}
where the source term $Q_\mathrm{N}$ describes the change of the net
electron number density (i.e.~the density of electrons minus that of
positrons) due to 
emission and absorption of electron-flavour neutrinos.
Unless nuclear statistical equilibrium (NSE) holds, an equation like
(\ref{eq:hydro.ye}) has to be solved for the mass fraction $X_k$ of 
each of the $N_\mathrm{nuc}$ individual (nuclear) species. In NSE,
$X_k=X_k(\rho,T,\ye)$ is determined by the Saha equations.

An equation of state is invoked in order to express the pressure as
a function of the independent thermodynamical variables,
i.e., $p=p(\rho,T,\ye)$, if NSE holds, or
$p=p(\rho,T,\ye,{\{X_k\}}_{k=1\dots N_\mathrm{nuc}})$ otherwise (see
Appendix~\ref{appx:eos} for the numerical handling of the equation of 
state). 

\medskip

In the following we will employ spherical coordinates and, unless
otherwise stated, assume spherical symmetry.

\subsection{Equations for the neutrino transport}

\subsubsection{General relativistic transport equation}\label{sect:GR}

\cite{lin66} derived a covariant transfer equation
and specialized it for particles of zero rest mass
interacting in a spherically symmetric medium supplemented with 
the comoving frame metric ($a$ is a Lagrangian coordinate)
$\mathrm{d}s^2=
-\mathrm{e}^{2\Phi(t,a)}c^2\mathrm{d}t^2+
\mathrm{e}^{2\Lambda(t,a)}\mathrm{d}a^2+
R(t,a)^2\mathrm{d}\Omega^2$

The ``Lindquist-equation'', which describes the evolution
of the specific intensity ${\cal I}$ as measured in the comoving frame of
reference, reads: 
\begin{equation}\label{eq:lindquist}
\begin{split}
\frac{1}{c}&{\cal D}_t \,{\cal I} + 
\mu {\cal D}_a\,{\cal I} +
\Gamma\,\frac{1-\mu^2}{R}
   \frac{\partial}{\partial \mu}\,\,{\cal I}  \\
  &+\frac{\partial}{\partial \mu}
 \left[
   (1-\mu^2)\left\{
   \mu\Big(\frac{U}{R}
       -\frac{1}{c}{\cal D}_t\Lambda\Big)
    -{\cal D}_a\Phi \right\}\,{\cal I}
 \right]  \\
  &- \frac{\partial}{\partial \epsilon}
 \left[
  \epsilon\,\Big(
               (1-\mu^2)\frac{U}{R}
               +\mu^2\frac{1}{c}{\cal D}_t\Lambda
               +\mu {\cal D}_a\Phi
           \Big)\,{\cal I} 
  \right]          \\
  &+ \Big(
        (3-\mu^2)\frac{U}{R}+
        (1+\mu^2)\frac{1}{c}{\cal D}_t\Lambda+
        2\mu {\cal D}_a\Phi
     \Big)\,{\cal I}=C        
\,,\hspace{5cm}
\end{split}
\end{equation}
where we use the classical abbreviations 
$\Gamma(t,a)\equ {\cal D}_a R$ and $U(t,a)\equ c^{-1}{\cal D}_t R$ with
${\cal D}_a\equ \mathrm{e}^{-\Lambda} \partial_a$ and
${\cal D}_t\equ \mathrm{e}^{-\Phi}\partial_t$. The latter definition has
been used in Eq.~(\ref{eq:lindquist}) also in order to emphasize that the
time derivative has to be taken at fixed \emph{Lagrangian coordinate}
$a$.   

The functional dependences of the 
metric functions $\Phi(t,a), \Lambda(t,a)$, $R(t,a)$, the specific
intensity ${\cal I}(t,a,\epsilon,\mu)$, and the collision
integral $C(t,a,\epsilon,\mu)$ were suppressed for brevity. 
Momentum space is described by the coordinates $\epsilon$ and $\mu$,
which are the energy and the cosine of the angle of propagation
of the neutrino with respect to the radial direction, both measured in
the locally comoving frame of reference.   
Note that the opacity $\chi$ and the emissivity $\eta$, and thus the 
collision integral $C=\eta - \chi{\cal I}$ in general depend also 
{\em explicitly} on momentum-space 
integrals of ${\cal I}$, which makes the transfer equation an
integro-partial differential equation. 
Examples of the actual computation of the collision integral for a
number of interaction processes of neutrinos with matter can
be found in Appendix~\ref{appx:opa}.

\subsubsection{${\cal O}(v/c)$ transport equations}\label{chap:equations.transp}

In general, the metric functions $\Phi(t,a), \Lambda(t,a)$ and
$R(t,a)$ have to be computed numerically from the Einstein 
field equations.
When working to order ${\cal O}(\beta\!\!\equ\!\! v/c)$ and in a flat
spacetime (usually called the ``Newtonian approximation''), it is 
however possible to express these functions analytically in terms of
only the velocity field and its first time derivative (the fluid
acceleration).
Details of the derivation can be found in \cite{cas72}.
Alternatively one can simply reduce the special relativistic
transfer equation \cite[]{mih80} to order ${\cal O}(v/c)$.

This transfer equation, together with its
angular moment equations of zeroth and first order reads
\cite[e.g.,][ see also \citealp{lowmih01}]{mihmih84}: 
\begin{eqnarray}
(\frac{1}{c}\frac{\partial}{\partial t} &+& 
\beta \frac{\partial}{\partial r})\,{\cal I} + 
\mu\frac{\partial}{\partial r}\,{\cal I} + 
\frac{1-\mu^2}{r}\frac{\partial}{\partial \mu}\,{\cal I} \nonumber  \\ 
  &+&\frac{\partial}{\partial \mu}\left[(1-\mu^2)
\left\{\mu\Big(\frac{\beta}{r}- \frac{\partial \beta}
{\partial r}\Big)-\frac{1}{c}\frac{\partial \beta}{\partial t} \right\}\,
{\cal I}  \right]  \nonumber\\
  &-&\frac{\partial}{\partial \epsilon}\left[\epsilon\,\Big((1-\mu^2)
\frac{\beta}{r}+\mu^2\frac{\partial \beta}{\partial
r}+\mu  \frac{1}{c}\frac{\partial \beta}{\partial t} \Big)
\,{\cal I}  \right] \nonumber\\
  &+&\Big((3-\mu^2)\frac{\beta}{r}+(1+\mu^2)
\frac{\partial \beta}{\partial r} +
\mu  \frac{2}{c}\frac{\partial \beta}{\partial t} \Big)\,{\cal I} = C
\,,\label{eq:BTE_sr}
\end{eqnarray}
\begin{eqnarray}
(\frac{1}{c}\frac{\partial}{{\partial t}} &+& 
\beta\frac{\partial}{{\partial r}}) J +
\frac{1}{r^2}\frac{\partial}{\partial
r}\left(r^2H\right) \nonumber\\ 
  &-&\frac{\partial}{\partial \epsilon}\left[\epsilon\left(\frac{\beta}{r}(J-K)+\frac{\partial \beta}{\partial r}K+\frac{1}{c}\frac{\partial \beta}{\partial t}H
\right)  \right]  \nonumber\\
  &+&\frac{\beta}{r}(3J-K)+\frac{\partial \beta}{\partial
r}(J+K)+\frac{2}{c}\frac{\partial \beta}{\partial t}H=C^{(0)}
\,, \label{eq:J_sr}
\end{eqnarray} 
\begin{eqnarray}
(\frac{1}{c}\frac{\partial}{{\partial t}} &+& 
\beta\frac{\partial}{{\partial r}}) H +
\frac{1}{r^2}\frac{\partial}{\partial
r}\left(r^2K\right) + \frac{K-J}{r} \nonumber\\
  &-&\frac{\partial}{\partial
\epsilon}\left[\epsilon\,\left(\frac{\beta}{r}(H-L)+\frac{\partial \beta}{\partial r}L+\frac{1}{c}\frac{\partial \beta}{\partial t}K
\right)   \right]  \nonumber\\
  &+&2\left(\frac{\partial \beta}{\partial r}+\frac{\beta}{r}\right) H
+\frac{1}{c}\frac{\partial \beta}{\partial t}(J+K)=C^{(1)} 
\,,\label{eq:H_sr}
\end{eqnarray} 
where, in spherical symmetry, the angular moments of the specific
intensity are given by
\begin{equation}
\{J,H,K,L,\dots\}(t,r,\epsilon)\equ
\frac{1}{2}\int\limits_{-1}^{+1}\!\mathrm{d}\mu\,\mu^{\{0,1,2,3,\dots\}}
{\cal I}(t,r,\epsilon,\mu)
\,.
\end{equation}

Note that in Eqs.~(\ref{eq:BTE_sr}--\ref{eq:H_sr}) all physical
quantities, in particular also the collision  
integral and its angular moments $C^{(k)}(t,r,\epsilon)\equ
\frac{1}{2}\int_{-1}^{+1}\mathrm{d}\mu\,\mu^{k}C(t,r,\epsilon,\mu)$ are 
measured in the
comoving \emph{frame}, but the choice of \emph{coordinates} $(r,t)$ is
Eulerian. The simple replacement 
$\partial/\partial t+v\partial/\partial r\to \mathrm{D}/\mathrm{D}t$ yields
the conversion to Lagrangian coordinates.

For reference we also write down the transformations 
(correct to ${\cal O}(\beta)$) which allow one to relate the 
\emph{frequency-integrated} moments in the
comoving (``Lagrangian'') and in the inertial (``Eulerian'') frame of
reference (indicated by the superscript ``Eul'').  
\begin{alignat}{2}\label{eq:momentstrafo}
J^\mathrm{Eul} & = J & + & 2\beta\, H           \,,\nonumber \\
H^\mathrm{Eul} & = H & + &  \beta\, (J + K )    \,,  \\
K^\mathrm{Eul} & = K & + & 2\beta\, H 
\,. \nonumber
\end{alignat}
Eq.~(\ref{eq:momentstrafo}) can easily be deduced
from  a Lorentz-transformation of the radiation stress-energy tensor
\cite[e.g.,][]{mihmih84}. 
In principle also relations for \emph{monochromatic} moments can be
derived by transforming the specific intensity  
${\cal I}$, the angle cosine $\mu$ and the energy $\epsilon$, which,
however, leads to more complicated expressions. 

The system of Eqs.~(\ref{eq:BTE_sr}--\ref{eq:H_sr}) is coupled to the
evolution equations of the
fluid~(Eqs.~\ref{eq:hydro.rho}--\ref{eq:hydro.ye} in spherical
coordinates and symmetry) by virtue of the
definitions of the source terms
\begin{eqnarray}
Q_\mathrm{E}&=&
    -4\pi\int_0^\infty\dlin{\epsilon}C^{(0)}(\epsilon)\,, \label{eq:sourceterm_E}\\
Q_\mathrm{M}&=&
    -\frac{4\pi}{c}\int_0^\infty\dlin{\epsilon}C^{(1)}(\epsilon)\,, \label{eq:sourceterm_M}\\
Q_\mathrm{N}&=&
    -4\pi\,
    m_\mathrm{B}\int_0^\infty\dlin{\epsilon}\mathcal{C}^{(0)}(\epsilon)\,, \label{eq:sourceterm_N}
\end{eqnarray}
where $m_\mathrm{B}$ denotes the baryonic mass, and
$\mathcal{C}^{(0)}(\epsilon)\equ\epsilon^{-1}C^{(0)}(\epsilon)$.

\medskip

Recently, \cite{lowmih01} emphasized the fundamental significance
of a term 
\begin{equation}\label{eq:newvc}
\beta\mu\cdot\frac{1}{c}\frac{\partial}{{\partial t}}\,{\cal I}
\end{equation}
on the left hand side of the special relativistic transport equation.
This term has traditionally been dropped in deriving the 
${\cal O}(v/c)$-approximation (Eq.~\ref{eq:BTE_sr}) by assuming 
the time variation of all dependent variables (like, e.g., ${\cal I}$)
to be on a \emph{fluid time scale} (given by $l/v$, where $l$ is the
characteristic length scale of the system and $v$ a typical fluid
velocity). In this case the term given by Eq.~(\ref{eq:newvc}) is of
order $v^2/c^2$ and can be omitted \cite[]{mihmih84}.  
If, on the other hand, appreciable temporal variations of,
e.g., the specific intensity occur on \emph{radiation time scales} 
(given by $l/c$), Eq.~(\ref{eq:BTE_sr}) is no longer
correct to ${\cal O}(v/c)$ \cite[]{lowmih01}.

In core-collapse supernova simulations carried out so far, the
\emph{dynamics} of the stellar fluid presumably was not affected by
neglecting the term in Eq.~(\ref{eq:BTE_sr}).
However, our tests with Eq.~(\ref{eq:BTE_sr}), including the
additional time derivative of Eq.~(\ref{eq:newvc}) and the corresponding
changes in the moment equations (Eqs.~\ref{eq:J_sr}, \ref{eq:H_sr}),
have shown that the \emph{neutrino signal} computed in a supernova 
simulation is indeed altered compared to the traditional treatment. 
We will therefore take the term of Eq.~(\ref{eq:newvc}) into account 
in future simulations. 

For calculating nonrelativistic problems 
\cite{mihmih84} suggested a form of the radiation momentum
equation~(Eq.~\ref{eq:H_sr}), in which all velocity-dependent terms
except 
for the $\beta\partial/\partial r$-term in the first line of
Eq.~(\ref{eq:H_sr}) are dropped.
When the velocities become sizeable, however, it may be advisable
to solve the momentum equation in its general form 
(Eq.~\ref{eq:H_sr}). 
Doing so, we indeed found that the terms omitted by \cite{mihmih84} 
can have an effect on the solution of the neutrino transport in 
supernovae, in particular on the neutrino energy spectrum.


\section{Numerical implementation}\label{chap:numrhd}

\subsection{Overview of the basic numerical strategy}

For solving the neutrino radiation hydrodynamics problem
we employ the well-known ``operator
splitting''-approach, a popular method to make large problems
computationally tractable by considering them as a combination of
smaller subproblems.
The coupled set of equations of radiation
hydrodynamics (Eqs.~\ref{eq:hydro.rho}--\ref{eq:hydro.ye},
\ref{eq:BTE_sr}--\ref{eq:H_sr}) is split into 
a ``hydrodynamics part'' and a ``neutrino part'', which are solved
independently in subsequent (``fractional'') steps.
In the ``hydrodynamics part'' (Sect.~\ref{chap:prometheus}) a
solution of the equations of hydrodynamics including the effects of
gravity is computed without taking into account neutrino-matter
interactions.  

In what we call the ``neutrino part'' (Sects.~\ref{chap:transp.ME}--\ref{chap:transp.BTE}) the coupled system of equations
describing the neutrino transport and 
the changes of the specific internal energy $e$ and the electron
fraction $\ye$ of the stellar fluid due to the emission and absorption of
neutrinos are solved. The latter changes are given by the equations 
\begin{eqnarray}
\frac{\delta}{\delta t} e&=&
   -\frac{4\pi}{\rho}\int_{0}^{\infty} \mathrm{d}\epsilon\,
\sum_{\nu\in\{\nue,\nuae,\dots\}}
      C^{(0)}_\nu(\epsilon) \,,\label{eq:e_opsplit} \\
\frac{\delta}{\delta t} \ye&=&
   -\frac{4\pi m_\mathrm{B}}{\rho}\int_{0}^{\infty} \mathrm{d}\epsilon\,
\left(\mathcal{C}^{(0)}_\nue(\epsilon)-
      \mathcal{C}^{(0)}_\nuae(\epsilon)\right)
\,.                                        \label{eq:ye_opsplit}
\end{eqnarray}
Here, the notation $\delta/\delta t$ indicates that the full problem
as given, e.g., by Eq.~(\ref{eq:hydro.ye}), is split into the equations 
$\partial_t(\rho \ye) + r^{-2} \partial_r\,(r^2\rho \ye\,v) = 0$ 
and $\delta_t\,\ye=Q_\mathrm{N}/\rho$.

Within the neutrino part, we calculate in a first step the transport
of energy and 
electron-type lepton number by neutrinos and the corresponding
exchange with the stellar fluid 
for electron neutrinos ($\nue$) and antineutrinos
($\nuae$) only, which also means that the sum in
Eq.~(\ref{eq:e_opsplit}) is restricted to $\nu \in \{\nue, \nuae\}$. 
The $\mu$ and $\tau$ neutrinos and their antiparticles, all 
represented by a single species (``$\nux$'') are handled together with
the remaining terms of the sum in Eq.~(\ref{eq:e_opsplit}) in a second
step. 

For the neutrino transport we use the so-called ``variable
Eddington factor'' method, which determines the neutrino phase-space 
distribution by iteratively solving the radiation moment equations
for neutrino energy and momentum coupled to the Boltzmann transport
equation.
Closure of the set of moment equations is achieved by a variable
Eddington factor calculated from a formal solution of the Boltzmann
equation, and the integro-differential character of the latter is
tamed by making use of the integral moments of the neutrino
distribution as obtained from the moment equations.
The method is similar to the one described by \cite{buryou00}.
 
\subsection{The hydrodynamics code}\label{chap:prometheus}

For the integration of the equations of hydrodynamics we employ the
Newtonian finite-volume hydrodynamics code PROMETHEUS developed by
\cite{frymue89}.
PROMETHEUS is a direct Eulerian, time-explicit implementation of the
Piecewise Parabolic Method (PPM) of \cite{colwoo84}, which is a
second-order Godunov scheme employing a Riemann solver.
PROMETHEUS is particularly well suited for following discontinuities
in the  
fluid flow like shocks or boundaries between layers of different
chemical composition.
It is capable of tackling multi-dimensional problems with high
computational efficiency and numerical accuracy.

The version of PROMETHEUS used in our radiation hydrodynamics code
was provided to us by \cite{kei97}. It offers an optional
generalization of the Newtonian gravitational potential to include
general relativistic corrections. The hydrodynamics code was
considerably updated and augmented by K.~Kifonidis who added 
a simplified version of the ``Consistent Multifluid Advection (CMA)'' 
method \cite[]{plemue99} which ensures an accurate advection of
the individual chemical components of the fluid. 
In order to avoid spurious oscillations arising in multidimensional
simulations, when strong shocks are aligned with one of the
coordinate lines (the so-called ``odd-even decoupling'' phenomenon),
a hybrid scheme was implemented \cite[]{kif00,plemue01} which, in the
vicinity of such shocks, selectively switches from the original PPM
method to the more diffusive HLLE solver of \cite{ein88}.

\subsection{Multi-group formulation of the ${\cal O}(v/c)$ moment equations}\label{chap:transp.ME}

In what follows we describe the numerical implementation of the
``Newtonian'' limit of the transport equations including ${\cal
  O}(v/c)$-terms (cf.~Sect.~\ref{chap:equations.transp}). 
The general relativistic version of the code is based on almost exactly
the same considerations. 

\subsubsection{General considerations}\label{chap:transp.ME.general}

In order to construct a conservative numerical scheme for the system of 
moment equations, we integrate the moment equations
(Eqs.~\ref{eq:J_sr}, \ref{eq:H_sr}) over a zone of
the numerical $(r,\epsilon)$-mesh.

\paragraph{Spatial discretization:}

After having performed the integral over the volumes $\Delta V$
 of the radial mesh zones, the moment equations can be rewritten as
evolution equations for the volume-integrated moments (e.g.,~$\int_{\Delta
  V}J\,\mathrm{d}V$). 
In case of a moving radial grid, where the
volume $\Delta V$ of a grid cell is time dependent, one has to 
apply the ``moving grid transport theorem'' \cite[see e.g.,][]{mue91}
in order to interchange the operators $\mathrm{D}_t$ and 
$\int_{\Delta V}\mathrm{d}V$. For the special case of a Lagrangian grid,
where the grid moves with the velocity of the stellar fluid, the
latter reads:
\begin{equation}\label{eq:movtransp.lag}
\int_{\Delta V}\!\!\dlin{V}(\mathrm{D}_t\,\Xi)=
\mathrm{D}_t\int_{\Delta V}\!\! \dlin{V} \Xi -
\int_{\Delta V}\!\!\dlin{V}(\Xi\, \mathrm{div} \vec{v})
\,.
\end{equation}
Here and in the following the symbol $\Xi$ represents one of the moments
$J$ or $H$ (and also ${\cal J}\equ \epsilon^{-1}J$ or 
${\cal H}\equ\epsilon^{-1}H$).

\medskip

The computational domain $r_\mathrm{min} \le r \le r_\mathrm{max}$ is
covered by $N_r$ radial zones.
As the zone center $r_{i+\hlf}$ we define the volume-weighted mean
of the interface coordinates $r_i$ and $r_{i+1}$:
\begin{equation}
r_{i+\hlf}\equ \left(\frac{1}{2}\,(r_i^3+r_{i+1}^3)\right)^{1/3},\quad
i=0,\dots,N_r-1
\,.
\end{equation}
Scalar quantities like $J_{i+\hlf}$, $K_{i+\hlf}$ are defined on zone 
centers, whereas vectors like $H_{i}$ and $L_{i}$
live on the interfaces of radial zones.
``Miscentered'' values like, e.g,.~$J_{i}$ or $H_{i+\hlf}$, are
in general computed by linear interpolation, using the values of
the nearest neighbouring coordinates ($J_{i-\hlf}$ and $J_{i+\hlf}$, or
$H_{i}$ and $H_{i+1}$, respectively).

\paragraph{Spectral discretization:}

The spectral range $0\le \epsilon \le \epsilon_\mathrm{max}$ is covered
by a discrete energy grid consisting of $N_\epsilon$ energy ``bins'',
where the centers of these zones are given in terms of the interface values as 
\begin{equation}
\epsilon_{j+\hlf}\equ \frac{1}{2}\,(\epsilon_{j}+\epsilon_{j+1}),\quad
j=0,\dots,N_\epsilon-1
\,,
\end{equation}
and the width of an energy bin is denoted by 
${\Delta\epsilon}_{j+\hlf} \equ \epsilon_{j+1}-\epsilon_{j}$. 
The radiation moments are interpreted as average values within energy
bins: 
\begin{equation}\label{eq:freqdis1}
\Xi_{i,j+\hlf}\equ
\frac{1}{{\Delta \epsilon}_{j+\hlf}}\cdot
\int_{\epsilon_j}^{\epsilon_{j+1}}\mathrm{d}\epsilon\,\Xi_i(\epsilon)
\,.
\end{equation}

\paragraph{Time differencing:}

We employ a time-implicit finite differencing scheme for the moment
equations~(Eqs.~\ref{eq:J_sr}, \ref{eq:H_sr}) in combination with the
equations which describe the exchange of 
internal energy and electron fraction with the stellar medium
(Eqs.~\ref{eq:e_opsplit}, \ref{eq:ye_opsplit}).
Doing so we avoid the restrictive Courant Friedrichs Lewy (CFL)
condition which explicit numerical schemes have to obey for stability
reasons: For neutrino transport in the optically thin regime the CFL
condition would require $c\Delta t < \Delta r$, where
$\Delta r$ is the size of the smallest zone and $\Delta t$ the
numerical time step.
Even more importantly, the time-implicit discretization allows one to
efficiently cope with the different time scales of the problem:  
The ``stiff'' source terms
introduce a characteristic time scale proportional to the mean free
path $\lambda$ of the neutrinos, $c\Delta t = 
\lambda=\Delta r/\Delta\tau$, which can be orders of
magnitude smaller than the CFL time step in the
opaque interior of a protoneutron star, 
where the optical depth $\Delta\tau$ of individual grid cells becomes
very large.

\subsubsection{Finite differencing of the moment equations}\label{sect:ME_FD}

Applying the procedures described in Sect.~\ref{chap:transp.ME.general} to the moment
equations~(Eqs.~\ref{eq:J_sr}, \ref{eq:H_sr}) we obtain for each
neutrino 
species $\nu\in\{\nue,\nuae,\nu_{\mu},\dots\}$ (the corresponding
index is suppressed in the notation below) the finite
differenced moment equations. On an Eulerian grid they read:
\begin{eqnarray}\label{eq:MEj_fd}
\lefteqn{
\frac{J_{i+\hlf,j+\hlf}^{n+1}-J_{i+\hlf,j+\hlf}^{n}}
     {ct^{n+1}-ct^{n}}}\nonumber \\
&+&\frac{4\pi}{\Delta V_{i+\hlf}} 
  (r_{i+1}^2H_{i+1,j+\hlf}^{n+1}-r_{i}^2H_{i,j+\hlf}^{n+1})  \nonumber \\
&+&\frac{4\pi}{\Delta V_{i+\hlf}}
             (r_{i+1}^2\beta_{i+1}\;J_{\iota(i+1),j+\hlf}^{n+\hlf}-
         r_{i}^2\beta_{i}J_{\iota(i),j+\hlf}^{ n+\hlf}) \nonumber \\
&-&\frac{1}{{\Delta \epsilon}_{j+\hlf}}
    \left\{ \epsilon_{j+1} {[u\;J^{n+1}]}_{i+\hlf,j+1} 
    -\epsilon_j {[u\;J^{n+1}]}_{ i+\hlf,j}
    \right\}                                          \nonumber \\
&+&\left[\frac{\beta}{r}\right]_{i+\hlf}{[(1-f_K)\; J^{n+1}]}_{i+\hlf,j+\hlf}          \nonumber \\
&+&\left[\frac{\partial \beta}{\partial r}\right]_{
i+\hlf}{f_K}_{ i+\hlf,j+\hlf}\; J_{ i+\hlf,j+\hlf}^{n+1} \nonumber \\
&+&\frac{2}{c}\left[\frac{\partial \beta}{\partial t}\right]_{i+\hlf}
     {[{f_H}\; J^{n+1}]}_{ i+\hlf,j+\hlf}        
= {C^{(0)}}^{n+1}_{i+\hlf,j+\hlf}
\,,
\end{eqnarray}
\begin{eqnarray}\label{eq:MEh_fd}
\lefteqn{
\frac{H_{i,j+\hlf}^{n+1}-H_{i,j+\hlf}^{n}}
     {ct^{n+1}-ct^{n}}}\nonumber \\
&+&\frac{4\pi}{\Delta V_{i}}
 (r_{i+\hlf}^2{[f_K J^{n+1}]}_{i+\hlf,j+\hlf}-
      r_{i-\hlf}^2{[f_K J^{n+1}]}_{i-\hlf,j+\hlf}) \nonumber \\
&+&\frac{2\pi}{\Delta V_{i}}(r_{i+\hlf}^2-r_{i-\hlf}^2 )\,
        {[(f_K-1)\,J^{n+1}]}_{i,j+\hlf}
                     \nonumber \\
&+&\frac{4\pi}{\Delta V_{i}}(
      r_{i+\hlf}^2\beta_{i+\hlf}\;H_{\iota(i+\hlf),j+\hlf}^{n+\hlf}-
         r_{i-\hlf}^2\beta_{i-\hlf}H_{\iota(i-\hlf),j+\hlf}^{
     n+\hlf})\nonumber \\ 
&-&\frac{1}{{\Delta \epsilon}_{j+\hlf}}
    \left\{ \epsilon_{j+1} {[w\;J^{n+1}]}_{i,j+1} 
    -\epsilon_j {[w\;J^{n+1}]}_{ i,j}
    \right\}                                          \nonumber \\
&+&\left[\frac{\partial \beta}{\partial r}\right]_{i}\;H_{ i,j+\hlf}\nonumber \\
&+&\frac{1}{c}\left[\frac{\partial \beta}{\partial t}\right]_{i}
     {[(1+{f_K})\; J^{n+1}]}_{ i,j+\hlf}
={C^{(1)}}^{n+1}_{i,j+\hlf}
\,.
\end{eqnarray}
Note that by virtue of introducing the 
``Eddington factors'' (see the following section for their
actual computation) ${f_H}\equ{H}/{J}, {f_K}\equ{K}/{J}$ and
${f_L}\equ{L}/{J}$, the system of moment
equations has been closed. 
The ``flux'' in energy space across the boundaries between energy
zones 
appearing in the fourth and fifth line of Eq.~(\ref{eq:MEj_fd}) and
Eq.~(\ref{eq:MEh_fd}), respectively, is 
given in terms of the interface values $J_{i+\hlf,j}$, which are defined
as the \emph{geometrical} mean of $J_{i+\hlf,j-\hlf}$ and
$J_{i+\hlf,j+\hlf}$, and the advection velocities 
\begin{eqnarray}
u_{i+\hlf,j}&\equ&
  \left[\frac{\beta}{r}\right]_{i+\hlf}{[1-f_K]}_{i+\hlf,j} 
+ \left[\frac{\partial \beta}{\partial r}\right]_{i+\hlf}
                                         {[f_K]}_{i+\hlf,j}\nonumber\\  
&+& \frac{1}{c}\left[\frac{\partial \beta}{\partial t}\right]_{i+\hlf}
                                         {[f_H]}_{i+\hlf,j}\,,\label{eq:adv_e_j}
\end{eqnarray}
\begin{eqnarray}
w_{i,j}&\equ&
  \left[\frac{\beta}{r}\right]_{i}{[f_H-f_L]}_{i,j} 
+ \left[\frac{\partial \beta}{\partial r}\right]_{i}
                                         {[f_L]}_{i,j}\nonumber\\ 
&+& \frac{1}{c}\left[\frac{\partial \beta}{\partial t}\right]_{i}
                                         {[f_K]}_{i,j}
\,.\label{eq:adv_e_h}
\end{eqnarray}
The  nonlinear interpolation of $J$ in energy space turned out to be
crucial for obtaining a numerically stable algorithm. 
The Eddington factors $f_H$, $f_K$ and $f_L$ are interpolated linearly in energy
space in order to avoid problems in case the odd moments ($H$ or $L$)
change their signs.

Since the radiation moments are defined on a staggered mesh,
the finite-difference equations are second-order accurate in space.  
First-order accuracy in time is achieved by employing fully implicit
time-differences (``backward Euler''). 
Only the advection terms in the third and forth line of 
Eq.~(\ref{eq:MEj_fd}) and Eq.~(\ref{eq:MEh_fd}), respectively, 
are treated differently (see Sect.~\ref{chap:test.sn2.newt} for a motivation):
\begin{equation}\label{eq:time_interpolation}
\Xi^{n+\hlf}\equ
(1-\zeta)\,\Xi^{n}+\zeta\,\Xi^{n+1}
\,.
\end{equation}
For stability reasons, the interpolation parameter $\zeta=0.51$ 
was chosen slightly larger than 0.5, which makes the treatment
of the advection terms ``almost'' second-order accurate in time
\cite[see, e.g.,][]{dor98}. 
The interface values $J_{\iota}$ and 
$H_{\iota}$ are taken as the corresponding \emph{upwind}
quantities with 
\begin{equation}
\iota(i)\equ
\begin{cases}
i-\hlf &\text{for\quad} \beta_i > 0 \,, \\
i+\hlf &\text{else}\,. \\
\end{cases} 
\end{equation}

In order to damp spurious oscillations behind sharp
neutrino pulses travelling through nearly transparent zones,  
an additional diffusive term is added to the left-hand side of the first
moment equation~(Eq.~\ref{eq:MEh_fd}). Its finite-differenced form is
\cite[]{blieas98}: 
\begin{align}\label{eq:rvisc}
{\cal D}_\mathrm{visc}=\frac{{\cal A}^n_{i,j+\hlf}}{2\,r_i^2}\cdot
\Big(
  &\frac{H^{n+1}_{i+1,j+\hlf}r_{i+1}^2-H^{n+1}_{i,j+\hlf}r_{i}^2}{r_{i+1}-r_{i}}\nonumber\\
 -&\frac{H^{n+1}_{i,j+\hlf}r_{i}^2-H^{n+1}_{i-1,j+\hlf}r_{i-1}^2}{r_{i}-r_{i-1}}
\Big)
\,.
\end{align}
In effect this term resembles the diffusivity of the donor cell
scheme and hence allows one to selectively switch from the
second order radial discretization to a first order scheme.
Following \citet{blieas98} the (energy dependent) ``artificial
radiative viscosity'' ${\cal A}$ is set
to unity in transparent zones and goes to zero for optical depth
$\Delta \tau \gtrsim 1$.
By construction the diffusive term vanishes when the luminosity
$H r^2$ does not depend on radius and hence is only active during
transient phases when sharp neutino pulses propagate across the grid.

Using a similar diffusive term in the zeroth
moment equation~(Eq.~\ref{eq:MEj_fd}) allowed us also to overcome
stability problems with the numerical handling of the velocity-dependent
terms in the general form of the radiation momentum equation,
Eq.~(\ref{eq:H_sr}).
    
\subsubsection{Boundary conditions}\label{sect:ME.boundary}

For the solution of the moment equations~(Eqs.~\ref{eq:J_sr},
\ref{eq:H_sr}), boundary conditions have to be specified at
$r=r_\mathrm{min}$, $r=r_\mathrm{max}$, $\epsilon=0$ and 
$\epsilon=\epsilon_\mathrm{max}$. 
In the radial domain the values of quantities at the boundaries 
are iteratively obtained from the solution of the Boltzmann 
equation (see Sect.~\ref{sect:BTE.boundary} for the boundary conditions
employed there), which has the advantage that in the moment 
equations no assumptions have to be made about the angular
distribution of the specific intensity at the
boundaries. 
Specifically, at the inner boundary $H_{0,j+\hlf}$ is given by
$H(t,r=r_\mathrm{min},\epsilon)$ as computed from the Boltzmann 
equation. 
To handle the outer boundary, we apply Eq.~(\ref{eq:MEh_fd})
on the ``half shell'' between $r_{N_r-\hlf}$ and
$r_{N_r}$ \cite[]{mihmih84}. Where necessary, $J_{N_r+\hlf,j+\hlf}$ is
replaced by $H_{N_r,j+\hlf}/[{f_H}]_{N_r,j+\hlf}$ in 
Eqs.~(\ref{eq:MEj_fd}, \ref{eq:MEh_fd}). 

At  $\epsilon=0$ the flux in energy space vanishes and hence we simply
set $\epsilon_0 u_{i+\hlf,0}J_{ i+\hlf,0}^{n+1}=0$ 
in Eq.~(\ref{eq:MEj_fd}). At the upper boundary of the spectrum the flux
in energy space, 
$\epsilon_{N_\epsilon} 
u_{i+\hlf,N_\epsilon}J_{i+\hlf,N_\epsilon}^{n+1}$ is computed by a
(geometrical) 
extrapolation of the moments $J_{i+\hlf,N_\epsilon-\frac{3}{2}}$ and
$J_{i+\hlf,N_\epsilon-\frac{1}{2}}$ and by a linear extrapolation of the
advection velocity using $u_{i+\hlf,N_\epsilon-\frac{3}{2}}$ and
$u_{i+\hlf,N_\epsilon-\frac{1}{2}}$. 
Analogous expressions are used for Eq.~(\ref{eq:MEh_fd}).

\subsubsection{Finite differencing of the source term}

The finite differenced versions of the operator-splitted source terms
in the energy and lepton number equations (Eqs.~\ref{eq:e_opsplit},
\ref{eq:ye_opsplit}) of the stellar fluid read: 
\begin{alignat}{2}
\frac{e_{i+\hlf}^{n+1}-e_{i+\hlf}^{n}}
     {t^{n+1}-t^{n}}&=& \nonumber \\ 
- \frac{4\pi}{\rho_{i+\hlf}}
&\sum_{j=0}^{N_\epsilon-1}\,
\sum_{\nu\in\{\nue,\nuae,\dots\}}\,
{C^{(0)}}^{n+1}_{i+\hlf,j+\hlf,\nu} \,,\label{eq:STe_fd} \\
\frac{{\ye}_{i+\hlf}^{n+1}-{\ye}_{i+\hlf}^{n}}
     {t^{n+1}-t^{n}}&=&  \nonumber \\
- \frac{4\pi m_\mathrm{B}}{\rho_{i+\hlf}}
&\sum_{j=0}^{N_\epsilon-1}\,\left(
  {\mathcal{C}^{(0)}}^{n+1}_{i+\hlf,j+\hlf,\nue}
- {\mathcal{C}^{(0)}}^{n+1}_{i+\hlf,j+\hlf,\nuae}
                        \right)    \label{eq:STy_fd} 
\,.
\end{alignat}

\subsubsection{Neutrino number transport}\label{chap:numconsv}

In the finite-differenced version of the (monochromatic) neutrino energy
equation~(Eq.~\ref{eq:MEj_fd}) the derivative with respect to energy,
$\partial/\partial\epsilon$, has been written in a conservative form.  
When a summation over all energy bins is performed in
Eq.~(\ref{eq:MEj_fd}), the terms containing 
fluxes across the boundaries of the energy zones telescope and an
appropriate finite differenced version of the total 
(i.e.~spectrally integrated) neutrino energy equation
is recovered. 
This essential property does, however, not hold automatically also for
the neutrino number density 
${\cal N}\equ 
4\pi/c \, \int_0^\infty\dlin{\epsilon} {\cal J}(\epsilon)$,
when the naive definition 
${\cal J}_{i+\hlf,j+\hlf,\nu}\equ J_{i+\hlf,j+\hlf,\nu}/\epsilon_{j+\hlf}$
is adopted. By inserting the latter expression into Eq.~(\ref{eq:MEj_fd})
and summing over all energy bins, it can easily be verified that the
corresponding 
fluxes across the boundaries of the energy bins do not cancel anymore due
to the finite cell size $\Delta\epsilon_{j+\hlf}$.

In order to avoid this problem, we instead derive the moment
equations for ${\cal J}$ and  ${\cal H}$ (by multiplying
Eqs.~\ref{eq:J_sr}, \ref{eq:H_sr} with $\epsilon^{-1}$) and
recast them into a conservative form: 
\begin{eqnarray}
(\frac{1}{c}\frac{\partial}{{\partial t}} &+& 
 \beta\frac{\partial}{{\partial r}})\, {\cal J} +
\frac{1}{r^2}\frac{\partial}{\partial
r}\left(r^2{\cal H}\right) \nonumber \\ 
  &-&\frac{\partial}{\partial \epsilon}\left[
        \epsilon\,\left(
          \frac{\beta}{r}({\cal J}-{\cal K})+
          \frac{\partial \beta}{\partial r}{\cal K}+
          \frac{1}{c}\frac{\partial \beta}{\partial t}{\cal H}
                \right)  
                                        \right]           \nonumber   \\
  &+&(2\frac{\beta}{r}+\frac{\partial \beta}{\partial r}){\cal J}
     +\frac{1}{c}\frac{\partial \beta}{\partial t}{\cal H}
 =\mathcal{C}^{(0)}
\,, \label{eq:JN_sr}
\end{eqnarray} 
\begin{eqnarray}
(\frac{1}{c}\frac{\partial}{{\partial t}} &+& 
 \beta\frac{\partial}{{\partial r}})\, {\cal H} +
 \frac{1}{r^2}\frac{\partial}{\partial r}\left(r^2{\cal K}\right) + 
 \frac{{\cal K}-{\cal J}}{r}                       \nonumber     \\
  &-&\frac{\partial}{\partial\epsilon}\left[
        \epsilon\,\left(
          \frac{\beta}{r}({\cal H}-{\cal L})+
          \frac{\partial \beta}{\partial r}{\cal L}+
          \frac{1}{c}\frac{\partial \beta}{\partial t}{\cal K}
                \right)   
                                       \right]     \nonumber      \\
  &+&\frac{\beta}{r}\left({\cal H}\!+\!{\cal L}\right)
     +\frac{\partial \beta}{\partial r}\left(2{\cal H}\!-\!{\cal L}\right)
     +\frac{1}{c}\frac{\partial \beta}{\partial t}{\cal J}
=\mathcal{C}^{(1)} 
.\label{eq:HN_sr}
\end{eqnarray}
Equations~(\ref{eq:JN_sr}, \ref{eq:HN_sr}) are then discretized as a
separate set of equations. 
Consequently, the finite-difference representation of the monochromatic 
number density ${\cal J}_{i+\hlf,j+\hlf}$ and
the monochromatic number flux  ${\cal H}_{i,j+\hlf}$ are 
now considered as new variables, which are determined by solving
an extra set of two moment equations.
Note that treating ${\cal J}_{i+\hlf,j+\hlf}$ and 
${\cal H}_{i,j+\hlf}$ as additional variables is not necessarily in
conflict 
with the relations ${\cal J}(\epsilon)\equiv J(\epsilon)/\epsilon$ and 
${\cal H}(\epsilon)\equiv H(\epsilon)/\epsilon$ (${\cal K}$ and ${\cal L}$
are defined likewise), which constitute
dependences between the quantities. Since $J_{i+\hlf,j+\hlf}$ and 
${\cal J}_{i+\hlf,j+\hlf}$, $H_{i,j+\hlf}$ and 
${\cal H}_{i,j+\hlf}$ are defined as \emph{average values} within  
the energy interval $[\epsilon_j,\epsilon_{j+1}]$ of 
\emph{different energy moments}  of the same function (rather than
values of a function at 
point  $\epsilon_{j+\hlf}$; cf.~Eq.~\ref{eq:freqdis1}) they can 
be considered as variables that are independent of each other,
provided that the inequalities $\epsilon_j \le \langle\epsilon\rangle_{i+\hlf,j+\hlf} \le
\epsilon_{j+1}$ with 
$\langle\epsilon\rangle_{i+\hlf,j+\hlf} 
  = J_{i+\hlf,j+\hlf}/{\cal J}_{i+\hlf,j+\hlf}$ or  
$\langle\epsilon\rangle_{i,j+\hlf} 
  = |H_{i,j+\hlf}|/|{\cal H}_{i,j+\hlf}|$
are fulfilled (we thank our referee, A.~Mezzacappa, for stressing the
importance of this point).
In practice, we find that this is always well fulfilled at high
optical depths and low velocities. The constraint is violated only in
the steeply falling, high-energy tail of the neutrino
spectrum. Systematically inspecting results from simulations we
find that a part of the spectrum is affected where
$J_{i+\hlf,j+\hlf}\Delta\epsilon_{i+\hlf,j+\hlf}$ is less than one
percent of the 
maximum value, which means that the energy integral over this spectral
tail accounts for a negligible fraction of the total neutrino (energy)
density.

\subsubsection{Numerical solution}

With the Eddington factors ${f_H}$, ${f_K}$, ${f_L}$ and flow field
$\beta$ being given, 
the nonlinear system of equations (Eqs.~\ref{eq:MEj_fd},
\ref{eq:MEh_fd}, \ref{eq:JN_sr}, \ref{eq:HN_sr}, \ref{eq:STe_fd},
\ref{eq:STy_fd}) is solved  
for the variables $\{J_{i+\hlf,j+\hlf}, H_{i,j+\hlf}, 
{\cal J}_{i+\hlf,j+\hlf}, {\cal H}_{i+\hlf,j+\hlf}, e_{i+\hlf},
{\ye}_{i+\hlf}\}$ by a Newton-Raphson iteration
\cite[e.g][]{preteu92}, using the aforementioned boundary conditions.

This requires the inversion of a block-pentadiagonal  
system with $2 N_r+1$ rows of blocks. The dimension of an individual
block of the 
Jacobian is $(4N_\epsilon +2)$ in case of the transport of
electron neutrinos and antineutrinos (the number of energy bins
$N_\epsilon$ is multiplied by a factor of two because $\nue$ and
$\nuae$ are treated as separate particles, the other factor of two
comes from the additional transport of neutrino number). In contrast,
muon and tau neutrinos and their antiparticles are considered here as
one neutrino species because their interactions with supernova matter
are roughly the same. In the absence of the corresponding charged
leptons, muon and tau neutrinos are only produced (or annihilated) as
$\nu\bar\nu$ pairs and hence do not transport lepton number. 
Therefore the blocks have a dimension of $(N_\epsilon +1)$ for these
flavours. 

For setting up the Jacobian all partial derivatives with respect to
the radiation moments can be calculated analytically, whereas the
derivatives with respect to electron fraction and specific internal
energy are approximated by finite differences.

\subsection{Formal solution of the Boltzmann equation}\label{chap:transp.BTE}

In order to provide the closure relations (the ``variable Eddington
factors'') for the truncated system of moment equations, we have to 
solve the Boltzmann equation. Since the emissivity $\eta$ 
and the opacity $\chi$ depend in general also on angular moments 
of ${\cal I}$, this is actually a nonlinear, integro-differential
problem.  
However, $\eta$ and $\chi$ become known functions of only the
coordinates $t$, $r$, $\epsilon$ and $\mu$, when 
$J$ and $H$ are used as given solutions of the moment equations. 
This allows one to calculate a formal solution of the Boltzmann
equation. 

\subsubsection{Model equations}

Making the change of variables
\cite[cf.][]{yor80,mihmih84,koe92,basefi97} 
\begin{equation}\label{eq:tanrcoords}
(r,\mu)\mapsto(s\equ\mu r,p\equ r\sqrt{1-\mu^2}),\qquad\mu\ge 0
\,,
\end{equation}
and introducing the symmetric and antisymmetric averages of the
specific intensity 
\begin{align}
j(t,s,p)&\equ
\frac{1}{2}\big({\cal I}(\mu)+{\cal I}(-\mu)\big)
\,,\label{eq:jdef}\\
h(t,s,p)&\equ
\frac{1}{2}\big({\cal I}(\mu)-{\cal I}(-\mu)\big)
\,,   \label{eq:hdef} 
\end{align}
Eq.~(\ref{eq:BTE_sr}) can be split into two equations for $j$ and $h$,
\begin{eqnarray}\label{eq:j_modl}
\frac{1}{c}\frac{\mathrm{D}}{\mathrm{D} t}j&+&
\frac{\partial}{\partial s} h \\
&-&\frac{\partial}{\partial \epsilon}
  \left\{\epsilon
  \Big[(1-\mu^2)\frac{\beta}{r}\,j
       +\mu^2\frac{\partial \beta}{\partial r}\,j
       +\mu \frac{1}{c}\frac{\partial \beta}{\partial t}\,h
  \Big]\right\} \nonumber \\
&+&(3-\mu^2)\frac{\beta}{r}\,j
   +(1+\mu^2)\frac{\partial \beta}{\partial r}\,j
   +\mu \frac{2}{c}\frac{\partial \beta}{\partial t}\, h 
   = s_\mathrm{E} \nonumber \,,
\end{eqnarray}
\begin{eqnarray}\label{eq:h_modl}
\frac{1}{c}\frac{\mathrm{D}}{\mathrm{D}t}h&+&
\frac{\partial}{\partial s} j \\
&-&\frac{\partial}{\partial \epsilon}
  \left\{\epsilon
  \Big[(1-\mu^2)\frac{\beta}{r}\,h
       +\mu^2\frac{\partial \beta}{\partial r}\,h
       +\mu \frac{1}{c}\frac{\partial \beta}{\partial t}\,j
  \Big]\right\} \nonumber \\
&+&2\left(\frac{\partial \beta}{\partial r}+\frac{\beta}{r}\right) h
+\mu \frac{1}{c}\frac{\partial \beta}{\partial t}\, j = u_\mathrm{E}
\nonumber \,.
\end{eqnarray}
The symmetric ($s_\mathrm{E}$) and antisymmetric ($u_\mathrm{E}$)
averages of the source term $C$ are defined in analogy to
Eqs.~(\ref{eq:jdef}, \ref{eq:hdef}).
Following \citet[][ with our Eq.~\ref{eq:h_modl_replacement} correcting a misprint in their Eq.~24]{buryou00} we
have made the replacements  
\begin{eqnarray}
(\mu-\mu^3)\left(\frac{\partial \beta}{\partial
    r}-\frac{\beta}{r}\right)\,\frac{\partial j}{\partial \mu}&\to&
(3\mu^2-1)\left(\frac{\partial \beta}{\partial
    r}-\frac{\beta}{r}\right)\,j  \,,\label{eq:j_modl_replacement}\\
(\mu-\mu^3)\left(\frac{\partial \beta}{\partial
    r}-\frac{\beta}{r}\right)\,\frac{\partial h}{\partial \mu}&\to&
(4\mu^2-2)\left(\frac{\partial \beta}{\partial
    r}-\frac{\beta}{r}\right)\,h  \,,\label{eq:h_modl_replacement}
\end{eqnarray}
in deriving Eq.~(\ref{eq:j_modl}) and Eq.~(\ref{eq:h_modl}), 
respectively, from Eq.~(\ref{eq:BTE_sr}).
This modification can be motivated as follows:
As a consequence of
Eqs.~(\ref{eq:j_modl_replacement}, \ref{eq:h_modl_replacement})   
no more partial derivatives of the intensity
(and hence $j$ and $h$) with respect to $\mu$ appear in the model
equations (Eqs.~\ref{eq:j_modl}, \ref{eq:h_modl}) of the original
Boltzmann equation (Eq.~\ref{eq:BTE_sr}). 
This quality allows us to construct an efficient numerical solution 
scheme to be described in the next section.
At the same time it is ensured that the monochromatic ${\cal O}(\beta)$
moment equations 
(Eqs.~\ref{eq:J_sr}, \ref{eq:H_sr}) can be recovered when 
Eq.~(\ref{eq:j_modl}) and $\mu$ times Eq.~(\ref{eq:h_modl}) are
integrated over angles.  
Thus, Eqs.~(\ref{eq:j_modl}, \ref{eq:h_modl}) are
consistent with the monochromatic moment equations without
including the exact aberration terms \cite[see also][]{buryou00}.
Moreover, the effects of angular aberration on the solution
of the neutrino transport are
found to be small (Liebend\"orfer, personal communication).
Therefore we do not expect significant changes, in particular because
only \emph{normalized} moments $H/J$, $K/J$, $L/J$ are
computed from 
the solution of the Boltzmann equation to be used as closure relations
in the moment equations.

\subsubsection{Finite-difference equations}\label{sect:BTE_FD}

For the moment we shall ignore terms which contain frequency derivatives
$\partial /\partial\epsilon$ (second line of
Eqs.~\ref{eq:j_modl}, \ref{eq:h_modl}).  
These Doppler terms will be included in an operator-split manner.

Equations~(\ref{eq:j_modl}, \ref{eq:h_modl}) are then
discretized on a so-called ``tangent ray grid''
(for an illustration, see Fig.~\ref{fig:tanray_interp}), the
geometry of which being an immediate consequence of the transformation of
variables given by Eq.~(\ref{eq:tanrcoords}). 
Applying this transformation, partial derivatives of only one
momentum-space coordinate $s$ remain, whereas the second coordinate
$p$ appears only in a parametric way \cite[cf.][]{basefi97}. 
This greatly facilitates the numerical solution of the system.  

A ``tangent ray'' $k$ is defined by its ``impact parameter'' $p_k=r_k$ at
$\mu=0$. 
The coordinate $s$ serves to measure 
the path length along the ray. 
On each tangent ray $k$, a staggered numerical mesh is introduced for
the coordinate $s$.  
The zone boundaries (centers) of this mesh are given by
the ray's intersections with the zone boundaries (centers) of the
radial grid (cf.~Fig.~\ref{fig:tanray_interp}). 
With the ``flux-like'' variable $h$ being defined at the zone boundaries
$s_{ik}$ 
and the ``density-like'' variable $j$ being defined at the zone centers
$s_{i+\hlf,k}$, 
the finite-differenced versions of 
Eqs.~(\ref{eq:j_modl}, \ref{eq:h_modl}) finally can be written down
\cite[cf.][ \S V.2]{mihkle82}:
\begin{multline}\label{eq:BTEj_fd}
\frac{{j^*}^{n+1}_{i_k+\hlf,k}-\widetilde{j}^n_{i_k+\hlf,k}}{ct^{n+1}-ct^{n}}+
\frac{h^{n+1}_{i_k+1,k}-h^{n+1}_{i_k,k}}
     {s^{n+1}_{i_k+1,k}-s^{n+1}_{i_k,k}}  \\
  +A_{i_k+\hlf,k}^{n+1}\,j^{n+1}_{i_k+\hlf,k}
  +B_{i_k+\hlf,k}^{n+1}\,h^{n+1}_{i_k+\hlf,k}=0   \,, 
\end{multline}
\begin{multline}\label{eq:BTEh_fd} 
\frac{{h^*}^{n+1}_{i_k,k}-\widetilde{h}^n_{i_k,k}}{ct^{n+1}-ct^{n}}+
\frac{j^{n+1}_{i_k+\hlf,k}-j^{n+1}_{i_k-\hlf,k}}
     {s^{n+1}_{i_k+\hlf,k}-s^{n+1}_{i_k-\hlf,k}}     \\
   +A_{i_k,k}^{n+1}\,j^{n+1}_{i_k,k}
   +B_{i_k,k}^{n+1}\,h^{n+1}_{i_k,k}=0\,,  
\end{multline}
with the indices $k=K_0,\dots,N_r-1$ and $i_k=k,\dots,N_r-1$.
If an inner boundary condition at some finite radius $r_\mathrm{min}>0$ is
used, a number of ``core rays'' $k=K_0,K_0+1,\dots,-1$ 
($-K_0\in \Natural$), which penetrate the innermost radial shell ($p_k
\le r_\mathrm{min}$), 
can be defined in order to describe the angular distribution of the
radiation at the inner boundary. For details, see
e.g.,\cite{yor80}, \cite{koe92}, \cite{dor98}. 
The coefficients $A$ and $B$ are combinations of the velocity and
angle-dependent terms and the source terms of Eqs.~(\ref{eq:j_modl},
\ref{eq:h_modl}). 
The definition of $\widetilde{j}$ and $\widetilde{h}$ will be
given later, for the moment one may ignore the tilde symbol.
We add a diffusive term to Eq.~(\ref{eq:BTEh_fd}) analogous to the
discussion in Sect.~\ref{sect:ME_FD}, replacing $H\,r^2$ in
Eq.~(\ref{eq:rvisc}) by $h$. 

\bigskip
The frequency derivatives of Eqs.~(\ref{eq:j_modl}, \ref{eq:h_modl}),
which were ignored in the finite-difference versions,
Eqs.~(\ref{eq:BTEj_fd}, \ref{eq:BTEh_fd}), are
taken into account in a separate step by operator-splitting (the
partially updated values for $j$ and $h$ were marked by asterisks in
Eqs.~\ref{eq:BTEj_fd}, \ref{eq:BTEh_fd}).
The discretization of the corresponding terms is explicit in time and
--- provided the acceleration terms in the second line of
Eqs.~(\ref{eq:j_modl}, \ref{eq:h_modl}) is neglected ---
can be performed straightforwardly using upwind differences in energy
space ($l=0,\dots,N_\epsilon-1$):
\begin{multline}\label{eq:BTEj_eshift}
j^{n+1}_{\_,\_,l+\hlf}={j^*}^{n+1}_{\_,\_,l+\hlf}+\frac{ct^{n+1}-ct^{n}}{\Delta\epsilon_{l+\hlf}}\cdot
\\
\left(
 \epsilon_{l+1}\upsilon_{\_,\_}\,{j^*}^{n+1}_{\_,\_,\iota(l+1)}
-\epsilon_{l}  \upsilon_{\_,\_}  \,{j^*}^{n+1}_{\_,\_,\iota(l)}
\right)
\,,
\end{multline}
\begin{multline}\label{eq:BTEh_eshift}
h^{n+1}_{\_,\_,l+\hlf}={h^*}^{n+1}_{\_,\_,l+\hlf}+\frac{ct^{n+1}-ct^{n}}{\Delta\epsilon_{l+\hlf}}\cdot
\\
\left(
 \epsilon_{l+1}\upsilon_{\_,\_}\,{h^*}^{n+1}_{\_,\_,\iota(l+1)}
-\epsilon_{l}  \upsilon_{\_,\_}  \,{h^*}^{n+1}_{\_,\_,\iota(l)}
\right) \,,
\end{multline}
with
\begin{equation}\label{eq:BTE_seshift}
\iota(l)\equ
\begin{cases}
l-\hlf &\text{for\quad} \upsilon_{\_,\_} > 0 \,, \\
l+\hlf &\text{else}\,, \\
\end{cases} 
\end{equation}
and
\begin{equation}
\upsilon_{i_k,k}\equ
(1-\mu_{i_k,k}^2)\left[\frac{\beta}{r}\right]_{i_k}
  +\mu_{i_k,k}^2\left[\frac{\partial\beta}{\partial r}\right]_{i_k}
\,,
\end{equation}
where, for better readability, dashes replace the appropriate index
pairs corresponding to radius and impact parameter, 
$(i_k+1/2,k)$ in Eq.~(\ref{eq:BTEj_eshift}) and $(i_k,k)$ in
Eq.~(\ref{eq:BTEh_eshift}), respectively.

It should be possible to proceed along similar lines for including the
exact aberration effects in the Boltzmann solution which, for the time
being, includes aberration only in an approximate way on the basis of 
the replacements given by Eqs.~(\ref{eq:j_modl_replacement},
\ref{eq:h_modl_replacement}).   
In practice, however, the adopted tangent-ray geometry does not allow
for a conservative 
discretization of the angular derivatives in a way as simple as in the
case of the frequency derivatives. We plan to spend extra work on the
omitted angular derivatives of the intensity in order to include them
in future version of our code.

\subsubsection{Boundary conditions}\label{sect:BTE.boundary}

On the tangent ray grid boundary conditions must be specified for
each ray $k$ at $s_{k,k}$ and at $s_{N_r,k}$. 
At the inner core radius ($i=0;\, k=K_0,\dots, 0$ and
$s_{-\hlf,k} \equ s_{0,k}$) we set the boundary condition
$j_{-\hlf,k}\equ {\cal I}^+_{0,k}-h_{0,k}$,  
with ${\cal I}^+_{0,k}\equ {\cal I}(t,r_\mathrm{min},\mu_{0,k})$. 
For the remaining rays ($k=1,\dots, N_r$), symmetry and
Eq.~(\ref{eq:hdef}) imply $h_{k,k}=0$, since $\mu_{k,k}=0$.

At the outer radius
($i=N_r;\, k=K_0,\dots,N_r$ and 
$s^{n+1}_{N_r+\hlf,k}\equ s^{n+1}_{N_r,k}$) 
we consider Eq.~(\ref{eq:BTEh_fd}) on the ``half shell'' between 
$r_{N_r-\hlf}$ and $r_{N_r}$ (cf.~Sect~\ref{sect:ME.boundary}) and
make the replacement $j_{N_r+\hlf,k}\equ {\cal I}^-_{N_r,k}+h_{N_r,k} $,
with ${\cal I}^-_{N_r,k}\equ {\cal I}(t,r=r_\mathrm{max},-\mu_{N_r,k})$.

The physical boundary conditions are described by the functions 
${\cal I}(t,r_\mathrm{min},\mu)$ and 
${\cal I}(t,r_\mathrm{max},-\mu)$ with $0\le \mu \le 1$, which specify
the specific intensity entering the computational volume at the
inner and outer surfaces, respectively. 

At the boundaries of the energy grid we use the same type of boundary
conditions as described for the moment equations
(cf.~Sect.~\ref{sect:ME.boundary}). 

\subsubsection{Numerical solution}

By virtue of the approximations used to derive the model 
equations (Eqs.~\ref{eq:j_modl}, \ref{eq:h_modl}) and upon
introducing the tangent ray grid,
the system of Eqs.~(\ref{eq:BTEj_fd}, \ref{eq:BTEh_fd}) with suitably
chosen boundary conditions can be
solved \emph{independently} for
each impact parameter $p_k$, each type of neutrino (note that for 
simplicity we have dropped the index $\nu$ 
in our notation), and --- because Doppler shift terms are split off --- each
neutrino energy bin $\epsilon_{j+\hlf}$ (index also suppressed).

For the same reasons as detailed in Sect.~\ref{chap:transp.ME},
we have employed fully implicit (``backward Euler'')
time differencing. Solving Eqs.~(\ref{eq:BTEj_fd},
\ref{eq:BTEh_fd}) therefore requires the  
\emph{separate} solution of $N_k \times N_\epsilon \times
N_\nu$ ($N_k\equ N_r-K_0+1$ is the number of tangent rays) 
pentadiagonal linear systems of dimension $\le N_r$.
On vector computers, this can be done very
efficiently by employing a vectorization over the index $k$. 
Once the numerical solution for $j$ and $h$ has been obtained, 
the monochromatic angular moments and thus the Eddington factors
$f_H=H/J$, $f_K=K/J$ and $f_L=L/J$ can be computed using the 
numerical quadrature formulae 
\begin{alignat}{3}
&J(r_i)&=&\int_0^1\mathrm{d}\mu\,j(r_i,\mu)&\approx
           \sum_{k=K_0}^{i} j_{ik} a_{ik}& \,,\label{eq:numq.moments0} \\
&H(r_i)&=&\int_0^1\mathrm{d}\mu\,\mu h(r_i,\mu)&\approx
           \sum_{k=K_0}^{i} h_{ik} b_{ik}& \,,\label{eq:numq.moments1} \\
&K(r_i)&=&\int_0^1\mathrm{d}\mu\,\mu^2j(r_i,\mu)&\approx
           \sum_{k=K_0}^{i} j_{ik} c_{ik}& \,,\label{eq:numq.moments2} \\
&L(r_i)&=&\int_0^1\mathrm{d}\mu\,\mu^3h(r_i,\mu)&\approx
           \sum_{k=K_0}^{i} h_{ik} d_{ik}&
\,.\label{eq:numq.moments3}
\end{alignat}
Explicit expressions for the quadrature weights 
$a_{ik}, b_{ik}$ and $c_{ik}$ can be found
in \cite{yor80}, $d_{ik}$ is calculated in analogy.

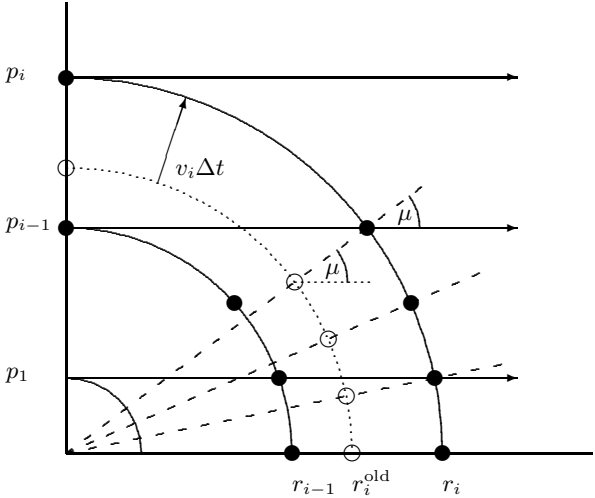
\begin{figure}[t]
\setlength{\unitlength}{1cm}
\hspace{1cm}
\begin{picture}(8.25,7)

\put(0,1){\vector(1,0){6}}
\put(0,0){\arc(0.0,1.0){-90}}
\put(-0.8,1){$p_{1}$}

\put(0,0){\line(0,1){6}}
\put(0,0){\line(1,0){7}}
\put(0,0){\arc(0.0,3.0){-90}}
\put(-0.8,3){$p_{i-1}$}
\put(0,3){\vector(1,0){6}}
\curvedashes[0.1mm]{0,1,8}
\put(0,0){\arc(0.0,3.8){-90}}
\curvedashes{}
\put(0,0){\arc(0.0,5.0){-90}}
\put(-0.8,5){$p_{i}$}
\put(0,5){\vector(1,0){6}}

\put(3,-0.5){$r_{i-1}$}
\put(5,-0.5){$r_{i}$}
\put(3.8,-0.5){$r_{i}^\mathrm{old}$}
\put(1.212,3.602){\vector(1,3){0.379}}
\put(1.45,3.7){$v_{i}\Delta t$}

\put(0.0,5.0){\circle*{0.2}}
\put(4.0,3.0){\circle*{0.2}}
\put(4.0,3.0){\arc(0.7,0.0){36.87}}\put(4.4,3.1){$\mu$}
\put(4.583,2.0){\circle*{0.2}}
\put(4.899,1.0){\circle*{0.2}}
\put(5.0,0.0){\circle*{0.2}}

\put(0.0,3.0){\circle*{0.2}}
\put(2.236,2.0){\circle*{0.2}}
\put(2.828,1.0){\circle*{0.2}}
\put(3.0,0.0){\circle*{0.2}}

\put(0.0,3.8){\circle{0.2}}
\put(3.04,2.28){\circle{0.2}}
\put(3.04,2.28){\arc(0.7,0.0){36.87}}\put(3.44,2.38){$\mu$}
\curvedashes[0.1mm]{0,1,8}
\put(3.04,2.28){\curve[10](0,0,1,0)}
\curvedashes{}

\put(3.483,1.52){\circle{0.2}}
\put(3.723,0.76){\circle{0.2}}
\put(3.8,0.0){\circle{0.2}}

\curvedashes[1mm]{0,1,2}
\put(0,0){\curve[50](0,0,5.8788,1.2)}
\put(0,0){\curve[50](0,0,5.4996,2.4)}
\put(0,0){\curve[50](0,0,4.8000,3.6)}
\curvedashes[1mm]{}

\end{picture}\\
[5mm]
\caption[]{Eulerian radial and tangent ray grid of the computation
  (solid lines) and virtual Lagrangian shell at time level
 $t^{n}=t^{n+1}-\Delta t$ (dotted). 
 For simplicity we have assumed a radial motion of only one
 shell with index $i$. The filled circles mark the positions where $j$
 and $h$ are known as 
 the solutions of the previous time step on the Eulerian grid. 
 For $v\not\equiv 0$ these positions do not coincide with 
 the locations $(r_{i}^\mathrm{old},\mu_{ik}^{n+1})$, 
 where initial values of the radiation field are required for computing
 the current time step (open circles). 
 The latter positions are subject to the requirement that the 
 $\mathrm{D}/\mathrm{D}t$ operator has to be evaluated at 
 fixed \emph{radial Lagrangian} \emph{and angular} coordinates
 (The radial dashed lines connect points with constant angle cosines
 $\mu$ of tangent rays.).}
\label{fig:tanray_interp}
\end{figure}

\bigskip
Unless the velocity field vanishes identically (in this case
$\widetilde{j}\equiv j$ and $\widetilde{h}\equiv h$ in Eqs.~\ref{eq:BTEj_fd} and
\ref{eq:BTEh_fd}, respectively),
an additional complication arises, since the partial derivative 
$\mathrm{D}/\mathrm{D}t$ has to be
evaluated not only at fixed Lagrangian radius 
(the index $i$ in our case)
but also for a fixed value of the angle cosine $\mu$.
The angular grid $\{\mu_{ik}^n\}_{k=K_0\dots i}$, which is associated
with 
the radius $r_i^n$, however, changes as $r_i$ moves with time.
As a consequence, e.g., the values of $h$ at the old
time level used in Eq.~(\ref{eq:BTEh_fd}) are
$\widetilde{h}^n_{ik}\equ h(t^n,r_{i}^n,\mu_{ik}^{n+1})$ and not 
$h^n_{ik}\equ h(t^n,r_{i}^n,\mu_{ik}^n)$, the solution 
of the previous time step \cite[cf.][ \S 98]{mihmih84}.
At the beginning of the time step we therefore have to interpolate
the radiation field for each fixed radial index $i$ from 
the $\{\mu_{ik}^n\}_{k=K_0\dots i}$-grid onto the 
$\{\mu_{ik}^{n+1}\}_{k=K_0\dots i}$-grid.
If an Eulerian grid is used for the computation, 
an additional interpolation in the radial direction is
performed to map from the fixed $r_{i}$-grid to the coordinates given
by $r_{i}^\mathrm{old}\equ r_{i}-v_i\Delta t$ (see
Fig.~\ref{fig:tanray_interp}). 
This seemingly complicated procedure has the advantage of
being computationally much more efficient than directly
discretizing Eqs.~(\ref{eq:j_modl}, \ref{eq:h_modl}) in Eulerian
coordinates. In this case the $v\partial_r$-term, which 
originates from making the replacement 
$\mathrm{D}_t=\partial_t+v\partial_r$, would
couple grid points of different impact parameters and would therefore
defy an independent treatment of the tangent rays.

%

\subsection{Iteration procedure}

A transport time step starts by adopting guess values (e.g., given by the
solution of the previous time step) for $J$, $H$, $\ye$, and $e$.
The quantities $\ye$, $e$ and the density $\rho$ determine the
thermodynamic state and thus the temperature 
and chemical potentials. These, together with $J$ and $H$ are used
to evaluate the rhs.~of the Boltzmann equation.
This allows one to compute its formal
solution and the Eddington factors as detailed in
Sect.~\ref{chap:transp.BTE}.  
The Eddington factors are then fed into the system of moment and
source-term equations, the solution of which (Sect.~\ref{chap:transp.ME})
yields improved values for $J$, $H$, $\ye$, and $e$. At this point, 
where required (e.g.,~for evaluating $\mathcal{C}^{(0)}$), 
the neutrino number density ${\cal J}$ and 
number flux density ${\cal H}$ are conventionally replaced by
$\epsilon^{-1}J$ and $\epsilon^{-1}H$. This procedure
is iterated until numerical convergence is achieved. 
With the Eddington factors being known from the described
iteration procedure, the complete system of source-term and moment
equations for neutrino energy \emph{and} number is solved (once) in
order to accomplish lepton number conservation. In this step 
${\cal J}$ and ${\cal H}$ are treated as additional variables 
(cf.~Sect.~\ref{chap:numconsv}).

\subsection{Details of coupling neutrino transport and hydrodynamics}
\label{chap:transp.rhd}

Our experience with the operator-splitting technique has shown
that considerable care is necessary in how precisely the
equations of hydrodynamics are coupled with the neutrino transport
and how the fractional time steps are scheduled.
In the following we therefore describe the used procedures in some
detail.

Since the hydrodynamics code and the transport solver in general
have different requirements for numerical resolution, accuracy and
stability, the discretization in both space and time of the two code
components is preferably chosen to be independent of each other. 
In supernova simulations, for example, the time step limit (from the
CFL condition) for the
hydrodynamic evolution computed by a time-explicit method is typically
more restrictive than in the time-implicit treatment of the transport
part. Since the transport is computationally much more expensive, one
wants to use a larger time step than for the hydrodynamics part.
This means that a transport time step is divided into a suitable
number of hydrodynamical substeps.

\subsubsection{Schedule of updates}\label{sect:transp.rhd.schedule}

Starting from the hydrodynamic state $(\rho^n,e^n,\ye^n,v^n)$ at the
old time level $t^n$, the time-explicit PPM algorithm first computes a
solution of the hydrodynamical conservation laws without the
effects of gravity and neutrinos.
From the updated density, the gravitational potential can be
calculated by virtue of Poisson's equation and finally the
gravitational source 
terms are applied to the total energy equation (Eq.~\ref{eq:hydro.etot})
and the momentum equation (Eq.~\ref{eq:hydro.v}). 
This completes the ``hydrodynamics'' time step with size
$\Delta t_\mathrm{Hyd}$, which is limited by the CFL-condition \cite[see
e.g.,][]{lve92}. 
By performing a total of $N_\mathrm{Hyd}^n$ substeps
the hydrodynamic state is evolved from the time level $t^n$ to the new
time level $t^{n+1}$ given by
\begin{equation} 
t^{n+1}=
t^n+\Delta t^n=
t^n+\sum_{i=1}^{N_\mathrm{Hyd}^n}\Delta t^{i}_\mathrm{Hyd}
\,,
\end{equation}
where $N_\mathrm{Hyd}^n\in\Natural$ is determined by the condition
$\Delta t^n \le \Delta t^\mathrm{max}_\mathrm{Tr}$. 
Here $\Delta t^\mathrm{max}_\mathrm{Tr}$ is the maximum 
size for the transport time step, which is compatible with
the constraints that the relative changes are at most 10\% for the
monochromatic neutrino energy and number density, and the relative
changes of the 
density,  temperature, internal energy density and electron fraction
must not exceed values of a few percent during the time
interval $\Delta t^n$.  
This leads to a partially updated hydrodynamic state
$(\rho^{n+1},e^*,\ye^*,v^{*})$ which includes the effects due to
hydrodynamic fluid motions and the acceleration by gravitational
forces and neutrino pressure (see below).
The corresponding state variables are then mapped onto 
the transport grid. 
During the transport time step of size $\Delta t^n$,
the transport equations in combination with the evolution
equations for the electron fraction and
internal energy of the stellar medium in response to neutrino
absorption and emission (Eqs.~\ref{eq:STe_fd}, \ref{eq:STy_fd}) are
solved. This yields the source terms for energy and lepton number as
$Q_\mathrm{E}=\rho^{n+1}\cdot (e^{n+1}-e^{*})/\Delta t^n$ and
$Q_\mathrm{N}=\rho^{n+1}\cdot (\ye^{n+1}-\ye^{*})/\Delta t^n$. 
Since the transport grid in general does not necessarily coincide with
the  hydrodynamics grid, the new specific internal energy
$e^{n+1}$ and electron fraction $\ye^{n+1}$ as computed on the
transport grid do not directly describe the new hydrodynamic
state. Instead of interpolating the quantities 
$e^{n+1}$ and $\ye^{n+1}$ back onto the hydrodynamics grid, we
map the source terms by a conservative procedure and then
update the electron number density and the 
total energy density on the hydrodynamics grid according to 
\begin{alignat}{2}
[\rho \varepsilon]^{n+1}=&
\rho^{n+1}\varepsilon^{*}&+&
(Q_\mathrm{E}^{n+1}+v^{n+1}Q_\mathrm{M}^{n+1})\cdot\Delta t^n \,,\label{eq:opspl_e}\\
[\rho \ye]^{n+1}=&
\rho^{n+1} \ye^*&+&
Q^{n+1}_\mathrm{N}\cdot\Delta t^n\label{eq:opspl_ye}
\,.
\end{alignat}
Equations~(\ref{eq:opspl_e}) and (\ref{eq:opspl_ye}) express the
effective influence of the source terms $Q_\mathrm{E}+vQ_\mathrm{M}$ and
$Q_\mathrm{N}$ over the time of a transport step, but in fact we apply
the corresponding sources (as given at the old time level) during each
substep of the hydrodynamics. The quantities $\varepsilon^{*}$ and 
$\ye^*$ as needed for the transport part of the code, are recovered a
posteriori by subtracting the accumulated effects of the neutrino
sources again.
The momentum equation of the stellar fluid is treated in a similar way.
Accounting for the momentum transfer (acceleration) by
neutrinos after each individual hydrodynamical substep 
(by using the old momentum source term $Q^n_\mathrm{M}$), however,
is of crucial importance here. When a sizeable contribution of the
neutrino pressure is ignored during a larger number of hydrodynamical
substeps a potentially hydrostatic
configuration can be severely driven out of mechanical equilibrium.
Correspondingly, the fluid velocity $v^{*}$ used in the transport step
includes the effects due to the acceleration by neutrinos.

After the transport time step has been completed, the new neutrino
stress $Q_\mathrm{M}^{n+1}$ is used for correcting $v^{*}$ to give the
new velocity $v^{n+1}$ at time level $t^{n+1}$:
\begin{equation}
[\rho v]^{n+1}=\rho^{n+1} v^{*}+
(Q_\mathrm{M}^{n+1}- Q_\mathrm{M}^{n}) 
\cdot \Delta t^n
\,.
\end{equation}

\subsubsection{Arrangement of spatial grids}\label{sect:transp.rhd.grids}

Radial discretization of the transport and hydrodynamic equations is done
on independent grids. Therefore there is freedom to choose the number
of zones and their coordinate values separately in both parts of the
code. 
 
Only quantities which obey conservation laws have to be communicated
from the hydro to the transport grid. By invoking the equation of state
all other thermodynamical quantities can be derived
from the density, the total energy density, momentum
density and the number densities of electrons and nuclear species. 
In the other direction it is the neutrino source terms which have to
be mapped back onto the hydro grid.

For the mapping procedure we assume the conserved quantities to be 
piecewise linear functions of the radius inside the grid cells, with 
parameters determined by the cell average values and so-called
``monotonized slopes'' \cite[for details see, e.g.,][ and references
cited therein]{ruf92}.  
In order to achieve global conservation of integral values we then
simply average this function within each cell of the target grid of 
the mapping \cite[cf.][]{daiwoo96}.  

\medskip

Using this procedure in dynamical supernova simulations, we discovered
spurious radial and temporal fluctuations of the
temperature, electron fraction, entropy and related quantities inside
the opaque protoneutron star unless the hydro and transport grids
coincide exactly in this region.  
The scale of these fluctuations is sensitive to the
radial cell size and the size of a time step,
the relative amplitude was found to be typically of the order of 
a few percent, with the exact number varying between different
quantities \cite[see][]{ramjan00}. 
These fluctuations are understood from the fact that the mapping of
the source terms on the one hand and the internal
energy density (resp.~temperature) on the other hand implies deviations from local
thermodynamical equilibrium between neutrinos and stellar medium.  
The attempt of the neutrino transport to restore this equilibrium
within a given grid cell leads to large net production or absorption
rates, driving the temperature in the opposite direction and causing
it to perform oscillations in time around a mean value. 
Despite these oscillations and fluctuations, the use of different
numerical grids inside the protoneutron star did not cause noticeable
problems for the accuracy of the ``global'' evolution of our models,
because the conservative mapping describes the exchange of energy and
lepton number between neutrinos and the stellar fluid correctly on
average. 

Being cautious, we take advantage of the option of using different hydro
and transport grids only during the early phases of
the collapse and in regions of the star, where neutrinos are far from
reaching equilibrium with the stellar fluid. In this case no 
visible fluctuations occur.

\subsubsection{Conservation of total energy and lepton number}

The numerical treatment of the radiation hydrodynamics problem 
should guarantee that the conservation laws for energy and lepton number
are fulfilled. 

Provided the acceleration terms
($\propto \partial\beta/\partial t$), which are of order $(v^2/c^2)$
and thus usually very small compared to the other terms, are ignored,
the constancy of the total lepton number is ensured by, (a) a
conservative 
discretization of the neutrino number equation (Eq.~\ref{eq:JN_sr}),
(b) a conservative handling of the electron number equation
(Eq.~\ref{eq:hydro.ye}), and (c) the exact numerical balance of the
source terms (cf.~Eq.~\ref{eq:sourceterm_N}) $-4\pi\,
m_\mathrm{B}\int\dlin{V}\int_0^\infty\dlin{\epsilon}\mathcal{C}^{(0)}(\epsilon)$
(defined on the transport grid) and
$\int\dlin{V}Q_\mathrm{N}$ (defined on the hydro grid).
Point (a) requires that in Eq.~(\ref{eq:JN_sr}) the flux divergence is
discretized in analogy to the second line in Eq.~(\ref{eq:MEj_fd}) and
that the $\beta\,\partial{\cal J}/\partial r$ and
$(2\beta/r+\partial\beta/\partial r)\,{\cal J}$ terms are combined to
$\mathrm{div} (\beta\,{\cal J})$ to be discretized in analogy to the
third line in Eq.~(\ref{eq:MEj_fd}).
The energy derivative in Eq.~(\ref{eq:JN_sr}) is treated in a
conservative way as described in Sect.~(\ref{chap:numconsv}).
Point (b) is achieved by the use of a conservative numerical
integration of the electron number equation
(Eq.~\ref{eq:hydro.ye}) in the spirit of the PROMETHEUS code, and
requirement (c) is fulfilled by employing a conservative 
procedure for mapping the electron
number source term from the transport grid to the hydro grid (see
Sects.~\ref{sect:transp.rhd.schedule}, \ref{sect:transp.rhd.grids}).
Doing so, the total lepton number remains constant in principle at 
the level of machine accuracy.

Different from the number transport, where the zeroth order
moment equation for neutrinos by itself defines a conservation law, 
the derivation of a conservation
law for the total energy implies a combination of the radiation
energy and momentum equations. The use of a staggered radial mesh for
discretizing the latter equations defies a suitable contraction of
terms in analogy to the analytic case. Therefore our numerical
description does not conserve neutrino energy with the same accuracy
as neutrino number and the quality of total energy conservation has to
be verified empirically for a given problem and numerical resolution. 

For our supernova simulations, tests showed that neutrino number is
conserved to an accuracy of 
better than $10^{-11}$ per time step, while for neutrino energy a
value below $10^{-7}$ is achieved. With a typical
number of about $50\,000$ transport time steps for a supernova
simulation we thus find an empirical upper limit for the violation of
energy conservation of $0.5\%$ of the neutrino energy.
This translates to $0.05\%$ of the internal energy of the
collapsed stellar core, i.e.~a few times $10^{49}~\text{erg}$ in
absolute number. 
Errors of the same magnitude are introduced by the
non-conservative treatment of the gravitational potential as a source
term in the fluid-energy equation (Eq.~\ref{eq:hydro.etot}).
Note that the use of different grids for the hydrodynamics and the
transport does not affect the energy budget because we employ a
conservative mapping of the neutrino source term between the grids 
(see
Sects.~\ref{sect:transp.rhd.schedule}, \ref{sect:transp.rhd.grids}).

\subsection{Approximate general relativistic treatment}\label{chap:transp.gr_rhd}

We have not yet coupled our general relativistic version of the
neutrino transport to a general relativistic hydrodynamics code. 
For the time being we work with a basically Newtonian code, which was
extended to include post-Newtonian corrections of the gravitational
potential. We hope that the deeper gravitational potential can account
for the main effects of general relativity on stellar core collapse
and the formation of neutron stars which do not approach gravitational
instability to become black holes \cite[cf.][]{brunis01}.
Because the general relativistic changes of the space-time metric are
ignored, a consistent description of the neutrino transport requires
that the fully relativistic equations are simplified such that only
the effects of gravitational redshift and time dilation are retained.

\subsubsection{Modified gravitational potential}

By comparing the Tolman-Oppenheimer-Volkoff equation for hydrostatic
equilibrium in general relativity \cite[see,
e.g.,][ Sect.~2.6]{kipwei90} with its Newtonian counterpart
(cf.~Eq.~\ref{eq:hydro.v}) one can define a modified ``gravitational
potential'' which includes correction terms due to pressure and energy
of the stellar medium and the neutrinos:
\begin{multline}\label{eq:grpotential}
\Phi^\mathrm{GR}(r)= \\
G \int\limits_\infty^r\!\dlin{r'}
\frac{1}{{r'}^2}
\left(\!m+\frac{4\pi{r'}^3(p+P)}{c^2}\right)
\frac{1}{\Gamma^2}
\left(\frac{\rho_\mathrm{tot}c^2+p}{\rho c^2}\right)
\,,
\end{multline}
where $\rho_\mathrm{tot}c^2\!\!\equ\!\!\rho (c^2 + e)$ is the total
(``relativistic'') energy density and 
$P=4\pi/c\int_0^\infty \mathrm{d}\epsilon\, K$ the neutrino pressure.
The calculation of the gravitational mass
$m(r)\!\equ\!\int_0^r\dlin{r'}4\pi
{r'}^2(\rho_\mathrm{tot}+c^{-2}E+c^{-3}U F/\Gamma)$ takes into account 
contributions of neutrino energy density
$E=4\pi/c\int_0^\infty \mathrm{d}\epsilon\, J$ and flux
$F=4\pi\int_0^\infty \mathrm{d}\epsilon\, H$. 
The metric function $\Gamma$ is calculated as
$\Gamma(r)=\sqrt{1+U(r)^2-2Gm(r)/rc^2}$ with the term $U^2$ accounting
for the effects of fluid motion.

Equation~(\ref{eq:grpotential}) can be used in the Newtonian hydrodynamic
equations~(Eqs.~\ref{eq:hydro.v}, \ref{eq:hydro.etot}) in
order to approximately take into account general relativistic effects
\cite[cf.][]{kei97}.
The quality of this approach has to be ascertained empirically by
comparison with fully general relativistic calculations.
In our case such a comparison yields quite satisfactory results (see
Sect.~\ref{chap:test.sn2}).

\subsubsection{Approximate GR transport}

The general relativistic moment equations describing transport of
neutrino energy, momentum and neutrino number can be derived from the
Lindquist-equation (cf.~Eq.~\ref{eq:lindquist}, Sect.~\ref{sect:GR}). 
They are:
\begin{multline}\label{eq:J_gr}
\frac{1}{c}\frac{\mathrm{D}}{\mathrm{D}t}\, J 
+ \frac{\Gamma}{R^2}
                    \frac{\partial}{\partial R}(R^2H\mathrm{e}^{\Phi}) 
+  \Gamma \partial_R \mathrm{e}^{\Phi}\, H  \\
 - \frac{\partial}{\partial \epsilon}\left[
     \epsilon\,\Big(
                  \mathrm{e}^{\Phi}\frac{U}{R}(J-K)
                 +c^{-1} \mathrm{D}_t \Lambda\, K
                 +\Gamma 
                     \partial_R \mathrm{e}^{\Phi}\, H  
               \Big)                     \right]  \\
 +\mathrm{e}^{\Phi}\frac{U}{R}\,(3J-K)
    +c^{-1}\mathrm{D}_t \Lambda\,(J + K)
=\mathrm{e}^{\Phi}\,C^{(0)} 
\,,
\end{multline}
\begin{multline}\label{eq:H_gr}
\frac{1}{c}\frac{\mathrm{D}}{\mathrm{D}t}\,  H + 
  \frac{\Gamma}{R^2}\frac{\partial}{\partial R}(R^2K\mathrm{e}^{\Phi})
+ \Gamma\partial_R \mathrm{e}^{\Phi}\, J 
+\mathrm{e}^{\Phi}\Gamma\,\frac{K-J}{R} \\
 - \frac{\partial}{\partial \epsilon}\left[
     \epsilon\,\Big(
                 \mathrm{e}^{\Phi}\frac{U}{R}(H-L)
                +c^{-1} \mathrm{D}_t \Lambda\, L
                +\Gamma 
                     \partial_R \mathrm{e}^{\Phi}\, K   \Big) 
        \right]  \\
    + 2(\mathrm{e}^{\Phi}\frac{U}{R}
    +c^{-1}\mathrm{D}_t \Lambda)\,H
    = \mathrm{e}^{\Phi}\,C^{(1)}                         
\,,
\end{multline}
for the energy transport, and
\begin{multline}\label{eq:JN_gr}
\frac{1}{c}\frac{\mathrm{D}}{\mathrm{D}t}\, {\cal J} 
+ \frac{\Gamma}{R^2}
              \frac{\partial}{\partial R}(R^2{\cal H}\mathrm{e}^{\Phi}) 
  \\
 - \frac{\partial}{\partial \epsilon}\left[
     \epsilon\,\Big(
                  \mathrm{e}^{\Phi}\frac{U}{R}({\cal J}-{\cal K})
                 +c^{-1} \mathrm{D}_t \Lambda\, {\cal K}
                 +\Gamma\partial_R \mathrm{e}^{\Phi}\, {\cal H}  
               \Big)                     \right]  \\
 +(2\,\mathrm{e}^{\Phi}\frac{U}{R}+c^{-1}\mathrm{D}_t \Lambda)\,{\cal J}
=\mathrm{e}^{\Phi}\,\mathcal{C}^{(0)} 
\,,
\end{multline}
\begin{multline}\label{eq:HN_gr}
\frac{1}{c}\frac{\mathrm{D}}{\mathrm{D}t}\,  {\cal H} + 
  \frac{\Gamma}{R^2}\frac{\partial}{\partial R}(R^2{\cal
  K}\mathrm{e}^{\Phi})    
  +\mathrm{e}^{\Phi}\Gamma\,\frac{{\cal K}-{\cal J}}{R} 
  +\Gamma \partial_R \mathrm{e}^{\Phi}\,({\cal J}-{\cal K}) \\
 - \frac{\partial}{\partial \epsilon}\left[
     \epsilon\,\Big(
                 \mathrm{e}^{\Phi}\frac{U}{R}({\cal H}-{\cal L})
                +c^{-1} \mathrm{D}_t \Lambda\, {\cal L}
                +\Gamma 
                     \partial_R \mathrm{e}^{\Phi}\, {\cal K}   \Big) 
        \right]  \\
    + \mathrm{e}^{\Phi}\frac{U}{R}\,({\cal H}+{\cal L})
    + c^{-1} \mathrm{D}_t \Lambda\,(2{\cal H}-{\cal L})
    = \mathrm{e}^{\Phi}\,\mathcal{C}^{(1)}                         
\,,
\end{multline}
for the number transport.

In the approximate treatment we neglect the distinction between
coordinate radius and proper radius ($\partial_R \to \partial_r$,
$\Gamma=1$) in Eqs.~(\ref{eq:J_gr}--\ref{eq:HN_gr}), and 
identify corresponding quantities with their Newtonian counterparts 
($R\to r$, $\mathrm{e}^{\Phi}\,U\to\beta$, 
$c^{-1}\mathrm{D}_t\Lambda \to \partial \beta/\partial r$). 
The same approximations are made in the ``parent'' Boltzmann equation
from which the moment equations of the relativistic approximation can
be consistently derived.
 
Accordingly, the approximations to 
Eqs.~(\ref{eq:J_gr}--\ref{eq:HN_gr}) contain only general 
relativistic redshift and time dilation effects for neutrinos. 
Coupling the transport with the Newtonian equations of hydrodynamics,
these restrictions to a fully relativistic treatment
are necessary in order to verify conservation
laws for energy and lepton number of the coupled system. 

Finite-difference versions of the moment equations and the
corresponding parent Boltzmann equation 
for the approximate GR transport are obtained by applying the techniques
described in Sects.~\ref{chap:transp.ME} and \ref{chap:transp.BTE}.
The lapse function is calculated by integrating the general
relativistic Euler equation $\partial \ln\, \mathrm{e}^{\Phi}/\partial
r=-(\rho_\mathrm{tot}c^2+p)^{-1}\,(\partial p/\partial
r-Q_\mathrm{M}/\Gamma)$ inward from the surface, where the boundary
condition $\mathrm{e}^{\Phi}=\Gamma$ is applied \cite[]{vri79}.


\subsection{Approximate multi-dimensional neutrino transport}\label{chap:transp.2d}

Multi-dimensional frequency dependent neutrino transport in moving
media and relativistic environments is a challenging problem for future
work. Since convective phenomena were recognized to be highly
important in supernovae \cite[][ and
refs.~therein]{herben94,burhay95,janmue96,keijan96} one would of
course like to perform simulations with Boltzmann neutrino transport also
in two and three dimensions. Here we suggest an approximate approach
based on a straightforward generalization of our variable Eddington
factor method, which offers some advantages concerning computational
efficiency. The approximation should be considered as an
intermediate step between spherically symmetric and fully
multi-dimensional models. Of course, the quality of the approximation
for the neutrino transport which we will describe below, will finally
have to be checked by a comparison with fully multi-dimensional transport
calculations.

\subsubsection{Basic considerations}

Our approximate treatment may be a reasonably accurate and physically
justifyable approach for describing neutrino transport in situations
where the star does not show a global deformation (e.g., due to
rotation) in layers which are opaque to neutrinos, but where
inhomogeneities and anisotropies are present only on smaller scales
(e.g., due to convection).
Multi-dimensional hydrodynamical simulations suggest that
convective processes occur in two distinct regions of the supernova
core:
\begin{itemize}
\item[(a)] 
  Convection inside the opaque protoneutron star causes deviations of the
  structural and thermodynamical quantities (like density or
  temperature) from spherical 
  symmetry typically of the order of a few percent
  \cite[]{keijan96,kei97}. Moreover, the time scale for changes of
  such local fluctuations is
  short compared to the neutrino diffusion time scale. Hence, on the
  neutrino diffusion time scale no persistent local gradients in lateral
  and azimuthal directions are present in the dense stellar interior.
  This allows one to 
  disregard the neutrino transport in these directions in a first
  approximation, which is at least correct in the temporal average. 
\item[(b)] 
  Convective overturn motions occur between the neutrinospheres and the
  stalled hydrodynamic shock. Deviations from
  spherical symmetry are present there on larger angular
  scales
  \cite[]{herben92,herben94,burhay95,janmue96,muejan97,
          mezcal98:ndconv,kifple00}, 
  but the neutrinos are only loosely coupled to the stellar medium in
  these regions. Therefore neutrino transport in the lateral
  directions does not play an important role and cannot have a
  significant influence on the local
  emission and absorption of neutrinos.
\end{itemize}

\subsubsection{``Ray-by-ray'' transport}

Under these circumstances the specific intensity  
${\cal I}(t,r,\vartheta,\varphi, \epsilon,\vec{n})$
can be assumed to depend mainly on $r$ but only weakly on longitude
$\varphi$ and latitude $\vartheta$ of the background medium. 
Hence, like in the spherically
symmetric case, the dependence of the specific intensity on the direction
of propagation $\vec{n}$ can be described by only one angle
$\mu\equ\vec{n}\cdot\vec{r}/|\vec{r}|$. 
The flux is thus approximated as 
$\vec{F}=4\pi\vec{r}/|\vec{r}|\cdot H(t,r,\vartheta,\varphi,
\epsilon)$ and the scalar $P=4\pi/c\cdot K(t,r,\vartheta,\varphi,
\epsilon)$ is sufficient to define the radiation stress tensor.

In the moment equations,
gradients in the $\vartheta$- and the $\varphi$-direction, which describe
the transport of energy and neutrino number into these lateral and
azimuthal directions, are neglected, yet the parametric dependence of
the (scalar) moments on the 
coordinates $\vartheta$ and $\varphi$ is retained and the radial
transport is computed using the moment equations independently in each
angular zone of the stellar model.
In order to close the set of moment equations, variable Eddington
factors are computed by iterating the Boltzmann equation and
the corresponding moment equations on a 
spherically symmetric ``image'' of the stellar
background. The latter is defined as the angular average of physical
quantities $\xi\in\{\rho,T,\ye,\beta,\dots\}$ according to
$\xi(t,r)\equ (4\pi)^{-1}\int_{-1}^{+1}\dlin{\cos
  \vartheta}\int_{0}^{2\pi}\dlin{\varphi}\xi(t,r,\vartheta,\varphi)$.
Note that the variable Eddington factors are normalized moments of the
neutrino phase space distribution function and thus should not show
significant 
variation with the angular coordinates of the star. This justifies
replacing them by quantities that depend only on the radial position
and time.

Since for each latitude $\vartheta$ and longitude $\varphi$ the moment
equations~(Eqs.~\ref{eq:J_sr}, \ref{eq:H_sr}) in our approach are solved 
together with the evolution equations of electron fraction and
internal energy due to neutrino sources (Eqs.~\ref{eq:e_opsplit},
\ref{eq:ye_opsplit}), local radiative equilibrium can be attained
properly and conservation of energy can be fulfilled. 
It is not obvious to us how these fundamental requirements could be
met in a more simple approximation where one uses a
one-dimensional transport scheme to compute the transport on a
spherically symmetric ``mean star'', which is obtained at each time
step by averaging the multi-dimensional hydrodynamical stellar model
over angles.

\subsubsection{Computational efficiency}

Besides having significant advantages for easy and efficient
implementation on vector and parallel computer architectures, the
suggested approach also saves computer time compared to a multiple
application of a one-dimensional Boltzmann solver.
Let $\Delta t^\mathrm{CPU}_{1\mathrm{D}}$ be the computation time required for
calculating a time step of the full one-dimensional transport problem
and let $n^\mathrm{It}\in \Natural$ be the number of steps that are
necessary for the iteration between the moment equations and the
Boltzmann equation to achieve convergence. 
If $n_{\vartheta,\varphi}\equ n_{\vartheta}\cdot n_{\varphi}$ is the total
number of angular zones that discretize the background star, the
computation time for calculating one time step of the 
multi-dimensional problem with the described approximate method is
\begin{equation}
\Delta t^\mathrm{CPU}=
  \left(1+\frac{n_{\vartheta,\varphi}}{n^\mathrm{It}}\right)\cdot \Delta t^\mathrm{CPU}_{1\mathrm{D}}
\,.
\end{equation}
Note that the iteration procedure ($n^\mathrm{It}> 1$) has to be
performed only once, namely for calculating the Eddington factors on
the angularly averaged ``image'' of the background medium. 
With typical values of 
$n^\mathrm{It}=3\dots 10$ and large values of $n_{\vartheta,\varphi}$
($\simeq 100$) the  computational effort for solving the
multi-dimensional problem becomes about 
\begin{equation}
\Delta t^\mathrm{CPU}\approx (0.1\dots 0.3)\,
n_{\vartheta,\varphi}\cdot \Delta t^\mathrm{CPU}_{1\mathrm{D}}
\,,
\end{equation}
to be compared with 
$\Delta t^\mathrm{CPU}=
 n_{\vartheta,\varphi}\cdot \Delta t^\mathrm{CPU}_{1\mathrm{D}}$,
which would be necessary for a straightforward multiple application of
the one-dimensional method to a number of $n_{\vartheta,\varphi}$ angular
zones.

\section{Test cases}\label{chap:test}

Stationary solutions of the
Boltzmann equation of radiative transfer can be calculated
analytically, semi-analytically or numerically to high
accuracy in a few cases, where a particularly simple form of the
emissivity and the opacity is assumed.
These problems provide test cases for our new code for solving
the equations of neutrino transport. 
To this end we impose suitably chosen initial and boundary conditions
and solve the time-dependent neutrino transport equations numerically, 
until a stationary state is reached.
The numerical solutions can then be compared to the reference solutions. 
This first class of problems, presented in
Sects.~\ref{chap:test.static} and \ref{chap:test.stationary}, does
neither refer to specific properties of neutrinos nor will the 
equations of neutrino radiation hydrodynamics, i.e.~the coupling of the
neutrino transport equation to the equations of hydrodynamics, be
tested. Time-dependent hydrodynamical tests will be described in
Sect.~\ref{chap:test.sn2}. 

\subsection{Radiative transfer on a static background}\label{chap:test.static}

We first consider
the simplest class of radiative transfer problems, namely those where
the ``background medium'' is assumed to be time-independent and
static, i.e.~the radial 
profiles of the emissivity and opacity do not change with time and 
the velocity and acceleration vanish everywhere and at all times.
As usual, the background medium is assumed to be spherically symmetric.

\paragraph{Propagation of a light pulse:}

Here, our method for computing the variable Eddington
factor (see Sect.~\ref{chap:transp.BTE}) by solving the Boltzmann
equation with a finite difference scheme \cite[]{mihkle82} 
is compared with results obtained from a 
general quadrature solution along characteristics \cite[]{yor80,koe92,basefi97}.
For convenience, we set $c\equiv 1$ in this section.

\begin{figure*}[t]
 \begin{tabular}{lr}
\put(0.9,0.3){{\Large\bf a}}
\epsfxsize=8.8cm \epsfclipon \epsffile{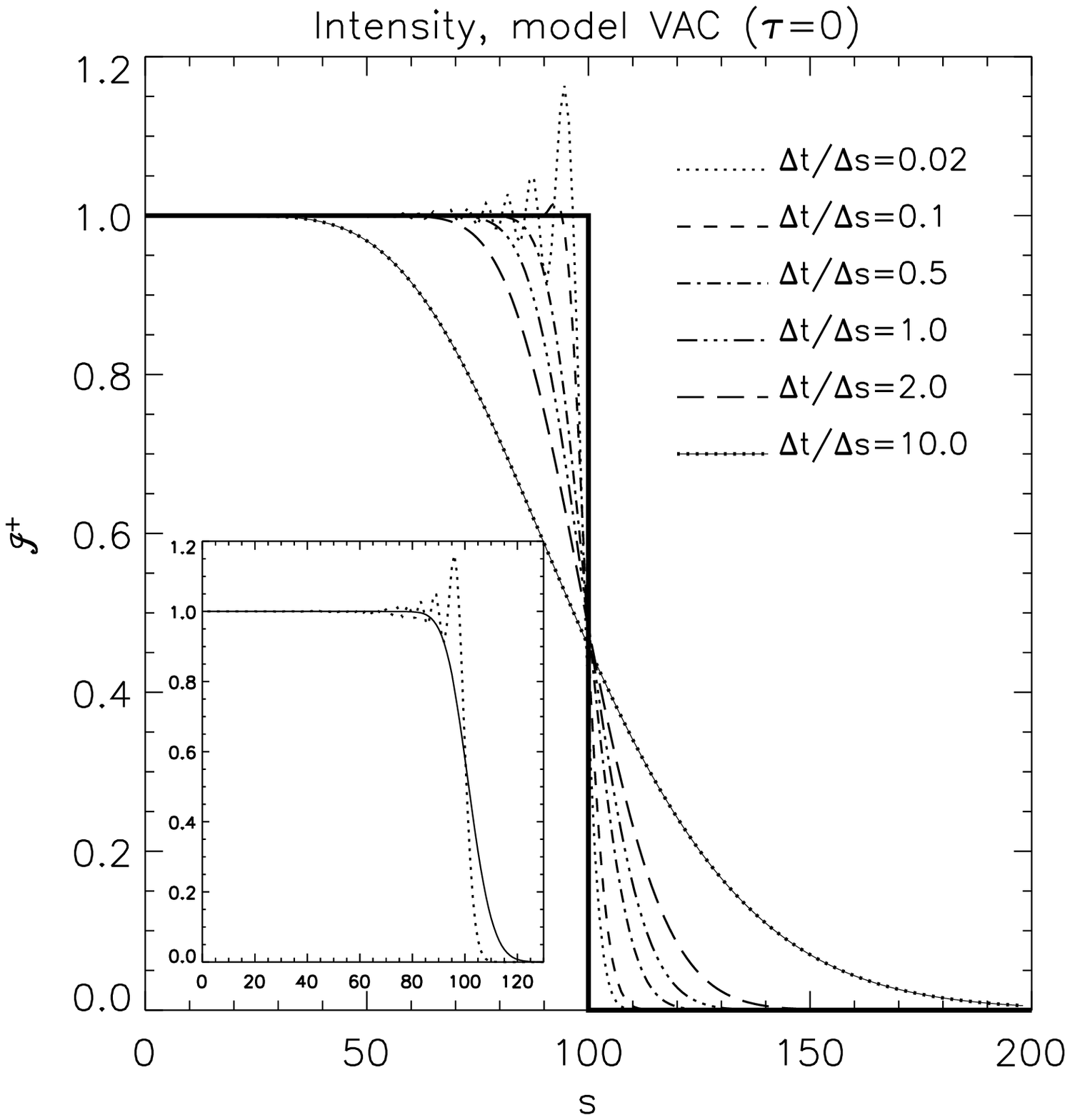} &
\put(0.9,0.3){{\Large\bf b}}
\epsfxsize=8.8cm \epsfclipon \epsffile{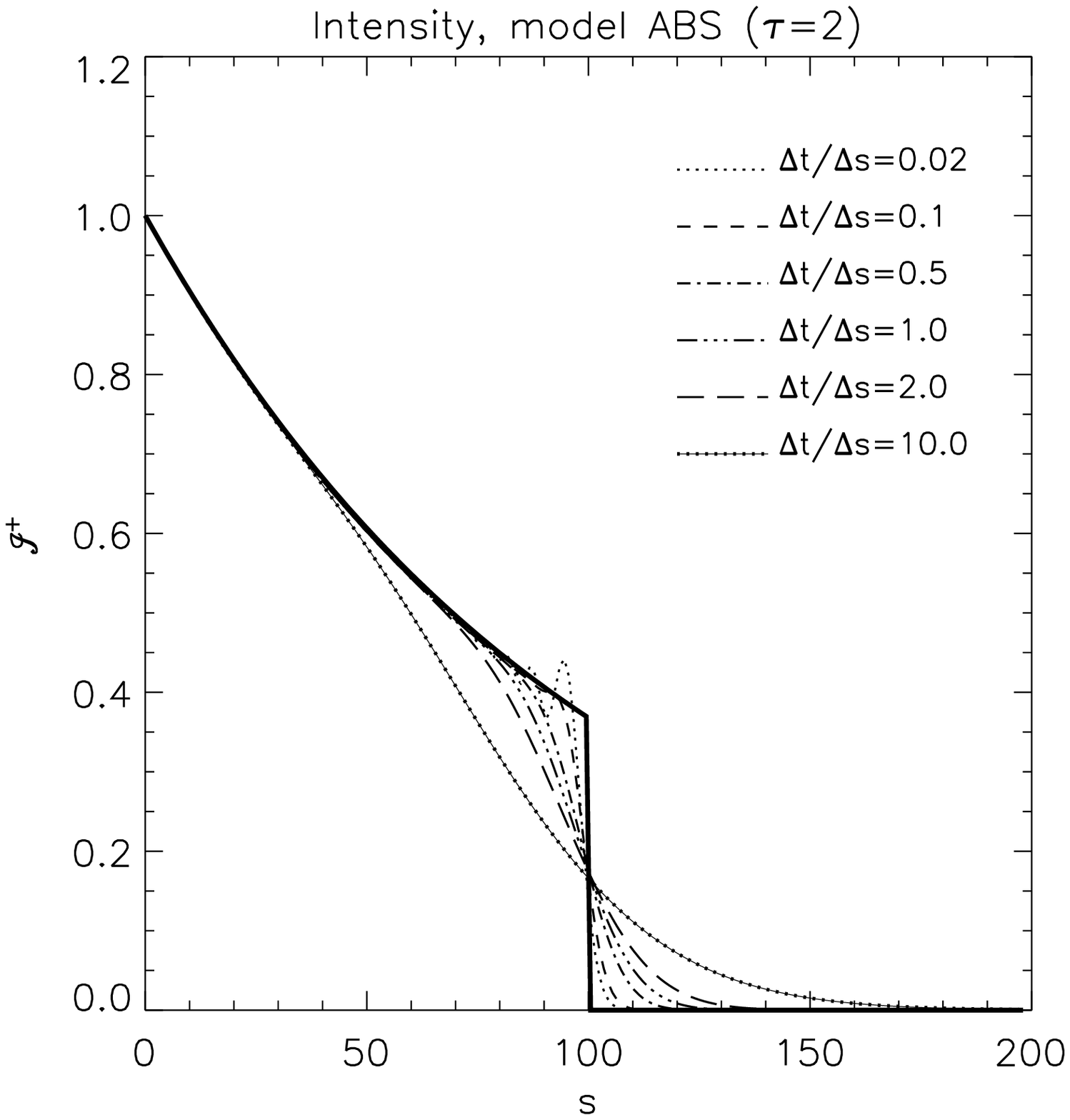}
\end{tabular}
\caption[]{Outwardly directed  intensity ${\cal I}^+$
  (normalized to the true postfront value ${\cal I}_0$) as a function
  of the  position $s$, calculated with our Boltzmann solver at time
  $t=100$ for different values of the computational time step.
  The bold line represents the analytical solution. 
  Panel~{\bf a} shows Model ``VAC'' with no absorption, in Panel~{\bf b}
  Model ``ABS'' with the absorptive shell is displayed. 
  Employing an additional diffusive term in the transport equations
  helps damping the spurious post-front oscillations, which are most
  prominent for $\Delta t/\Delta s=0.02$ (cf.~the solid line compared
  to the dotted line in the plot inserted in 
  Panel~{\bf a}).}\label{fig:pulse} 
\end{figure*}

As a test problem we consider a central point source which emits
radiation with an intensity ${\cal I}_0$ into a spherical, static and homogeneous
shell of matter bounded by two spheres of radius $r_0$ and $R > r_0$.
The only interaction of the radiation with the medium is by
absorption ($C=-\kappa_\mathrm{a}\,{\cal I}$).
The central source is active during the time interval $-r_0 \le t < \infty$.  
Initially ($t=0$), there is no radiation inside the shell.

By symmetry, the intensity vanishes for all but the radial
direction of propagation. In the absence of scattering,
it is therefore sufficient to follow the propagation of the light pulse
along a single radial (characteristic) ray. The dependence of
the  intensity on the angle can thus conveniently be
suppressed in the notation, writing
\begin{eqnarray}
{\cal I}^+(t,s)&\equ &
                \delta(\mu-1)\cdot{\cal I}(t,s,\mu)\,, \nonumber \\
{\cal I}^-(t,s)&\equ &
                \delta(\mu+1)\cdot{\cal I}(t,s,\mu)
\,,
\end{eqnarray}
with $s\equ r-r_0$ and $\delta(x)$ being the Dirac $\delta$-function.
The equations to be solved are (cf.~Eq.~\ref{eq:BTE_sr})
\begin{eqnarray}
\partial_t{\cal I}^+(t,s)+\partial_s{\cal I}^+(t,s)&=&
-\kappa_\mathrm{a}{\cal I}^+(t,s)  \,,\label{eq:pulse0}\\
\partial_t{\cal I}^-(t,s)-\partial_s{\cal I}^-(t,s)&=&
-\kappa_\mathrm{a}{\cal I}^-(t,s)\,,  \label{eq:pulse1}
\end{eqnarray}
subject to the boundary conditions
${\cal I}^+(t,0)={\cal I}_0$ (for $t \ge 0$) and 
${\cal I}^-(t,s_\mathrm{max})=0$, and the
initial condition ${\cal I}^{\pm}(0,s)=0$ 
(for $0 < s < s_\mathrm{max}\equ R-r_0)$.
The analytical (weak) solution can easily be verified to read:
\begin{align}\label{eq:lp_an}
  {\cal I}^+(t,s)&=
  \begin{cases}
     {\cal I}_0\cdot \exp(-\kappa_\mathrm{a}\,s) 
                &\quad \text{for $s \le t$}, \\
     0          
                &\quad \text{else,\hspace{10cm}}             
  \end{cases} \nonumber
  \intertext{and}            
{\cal I}^-(t,s)&\equiv 0 \,.
\end{align}
With our implementation of the finite-difference method 
adopted from \cite{mihkle82},
we computed two models with a shell spanning the range
$0 < s \le 200$. We use an equidistant radial grid with a resolution of
 $\Delta s=1$.
Model ``VAC'' assumes a vacuum shell ($\kappa_\mathrm{a}\equiv 0$),
whereas the absorbing shell of model ``ABS'' is characterized by
$\kappa_\mathrm{a}\equiv 0.01$, resulting in a total optical depth of
$\tau=2$.

\reffigLN{fig:pulse} shows our results for ${\cal I}^+$ at the time
$t=100$, together with the analytical solution~(Eq.~\ref{eq:lp_an}).  
We define the position of the numerically broadened light
front as the radius, where the average of the true 
pre and postfront value of the intensity is reached. 
In all cases the mean {\em propagation speed} is well reproduced, with 
some minor loss of accuracy  for very large time steps.
The {\em shape} of the light front, however, deviates from the true
solution. One observes an artificial precursor together with
a reduced intensity behind the front, 
both effects resulting in a spatial broadening of the front.
A clear trend towards larger diffusive broadening with increasing
time steps can be seen from \reffigN{fig:pulse} and 
Table~\ref{tab:pulse}. This phenomenon was also observed by
\cite{mihkle82}, who reported very similar results for their tests.

\begin{table}[!h]
\centerline{
\begin{tabular}{|rr|rr|rr|}
\hline 
 & &\multicolumn{2}{|c|}{model VAC} &\multicolumn{2}{|c|}{model ABS}  \\ 
 $\Delta t/\Delta s$ & \# &  
$\delta s$ (FD)       & $\delta s$ (CH)      &
$\delta s$ (FD)       & $\delta s$ (CH)    \\ 
\hline\hline
0.02       &  5000     & 6       &  25 & 6       &  26           \\
0.1        &  1000     & 8       &  -- & 8       &  --           \\
0.5        &   200     & 18      &  16 & 16      &  20           \\
1.0        &   100     & 26      &  1  & 22      &  1            \\
2.0        &    50     & 37      &  26 & 29      &  26           \\
10.0       &    10     & 80      &  -- & 55      &  --           \\ 
\hline
\end{tabular}
}
\caption[]{Width of the light front $\delta s$ (defined as the number
  of grid points for which ${\cal I}^+$ lies between 0.1 and 0.9 times
  its true postfront value) as a function of the
  time step $\Delta t$ (or the number of time steps \# required for the
  light front to reach $s=100$). Computations with the finite difference
  method \cite[``FD'';][]{mihkle82} and  with the method of
  characteristics \cite[``CH'';][]{yor80} are compared. Numbers obtained
  with the latter method are taken directly from \citet[ Tables
  1,2]{yor80}. They were, in addition, confirmed by an implementation
  of this numerical scheme by \cite{koe92}.}
\label{tab:pulse}
\end{table}
Table~\ref{tab:pulse} summarizes our results for the width of the front 
and compares them to
simulations done with an implementation of the method of characteristics
\cite[]{yor80,koe92}.
The latter method has the advantage to exactly reproduce the
analytical solution without any smearing of the front, if the
light front is advanced by exactly one grid cell per time step. 
Though the ability to reproduce the exact solution appears to be a 
very appealing property of the method of characteristics, 
the necessary condition  $\Delta t=\Delta s$ can hardly ever be 
accomplished in realistic simulations. This is simply due to the fact
that in typical astrophysical simulations the radial resolution can be
varying over several orders of magnitude,  whereas in most
applications the global value of the time step is determined in
some region of the grid.
When doing radiation hydrodynamical simulations, for example, one is usually
interested in resolving time scales different from those
given by the speed of propagating light fronts.  
Accordingly, the time steps can be very different from the optimum for
describing the light front propagation.
Our finite difference scheme therefore seems to be preferable to the
method of characteristics as far as the resolution of travelling
discontinuities is concerned (cf.~Table~\ref{tab:pulse}).

For time steps $\Delta t$ significantly smaller than the
radial zone width, a case not discussed by \cite{mihkle82}, we
observe some spurious postfront oscillations 
\cite[see also][ \S 5]{mihwea82}.   
Their maximum amplitude, however, does not
grow with time, which is in agreement with a local von Neumann
stability analysis for the fully implicit implementation of the equations
\cite[]{mihkle82}. 
If present in practical applications, the oscillations can be damped
by employing an additional diffusive term in Eq.~(\ref{eq:pulse1}),
which is active in nearly transparent regions 
($\tau\lesssim 1$) of the computational grid 
(cf.~Sects.~\ref{sect:ME_FD}, \ref{sect:BTE_FD}; \citealp{blieas98}).
Due to the correspondingly increased diffusivity of the
finite-difference scheme 
the light front is smeared over a larger number of zones ($\delta
s=18$ for $\Delta t/\Delta s=0.02$; see also the inserted panel in
Fig.~\ref{fig:pulse}a) as compared to the original method 
of \cite{mihkle82}. Note, however, that the representation of the
front is still better than the results obtained with the method of
characteristics.

\paragraph{Homogeneous sphere:}\label{chap:test:hs}

\begin{figure*}[!ht]
 \begin{tabular}{lr}
\put(0.9,0.3){{\Large\bf a}}
\epsfxsize=8.8cm \epsfclipon \epsffile{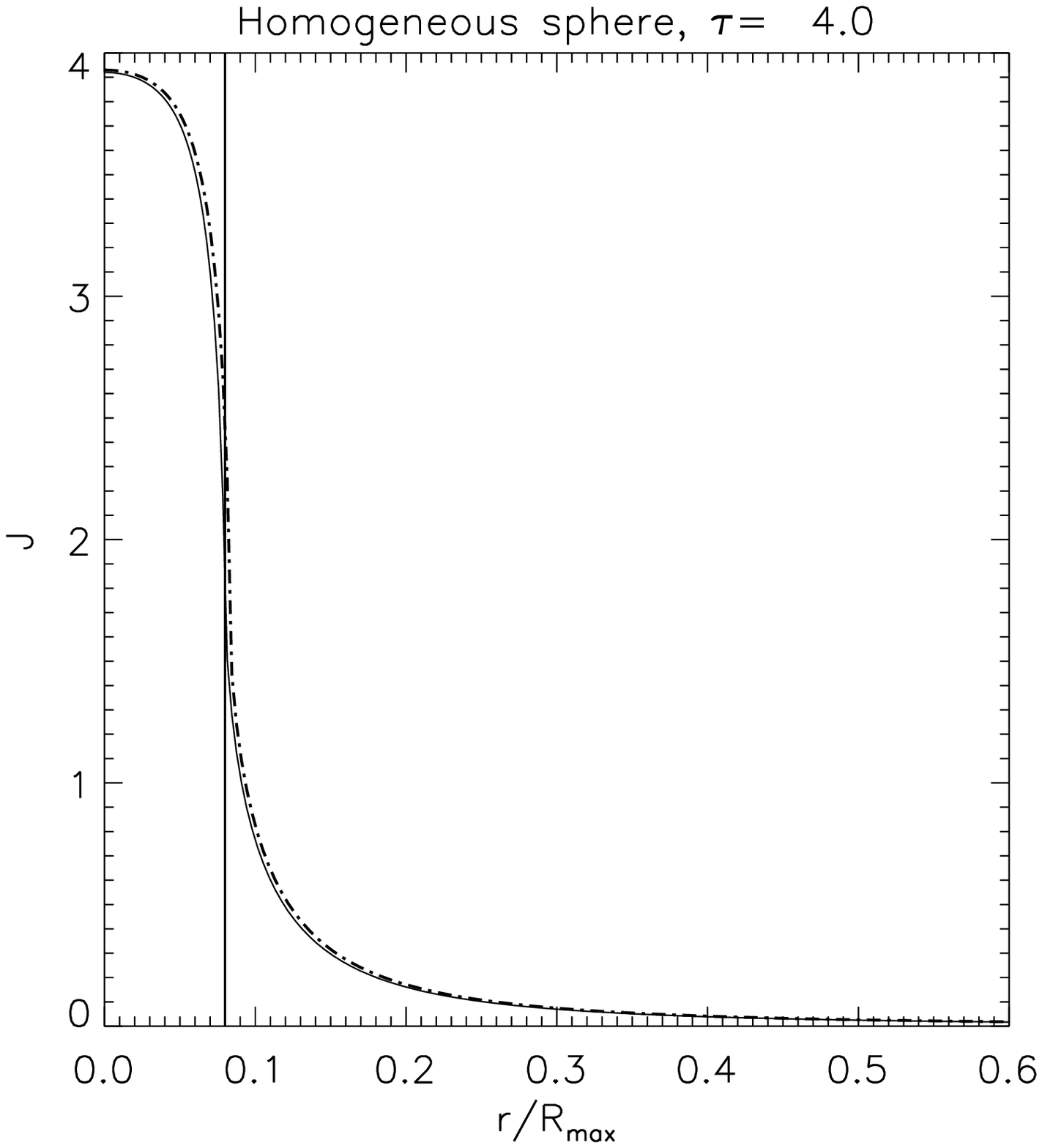} &
\put(0.9,0.3){{\Large\bf b}}
\epsfxsize=8.8cm \epsfclipon \epsffile{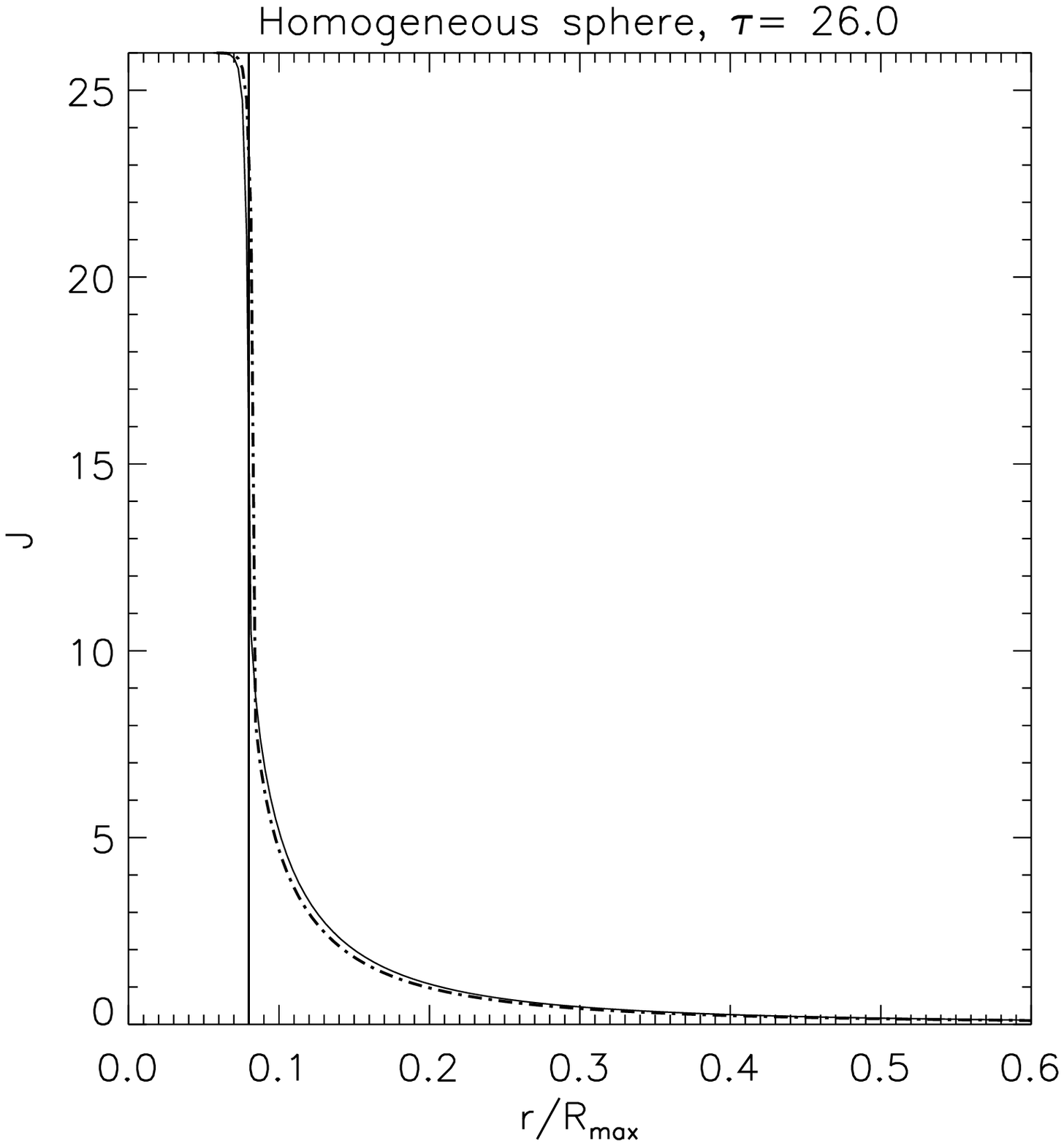}
\vspace{-0.2cm} \\
\put(0.9,0.3){{\Large\bf c}}
\epsfxsize=8.8cm \epsfclipon \epsffile{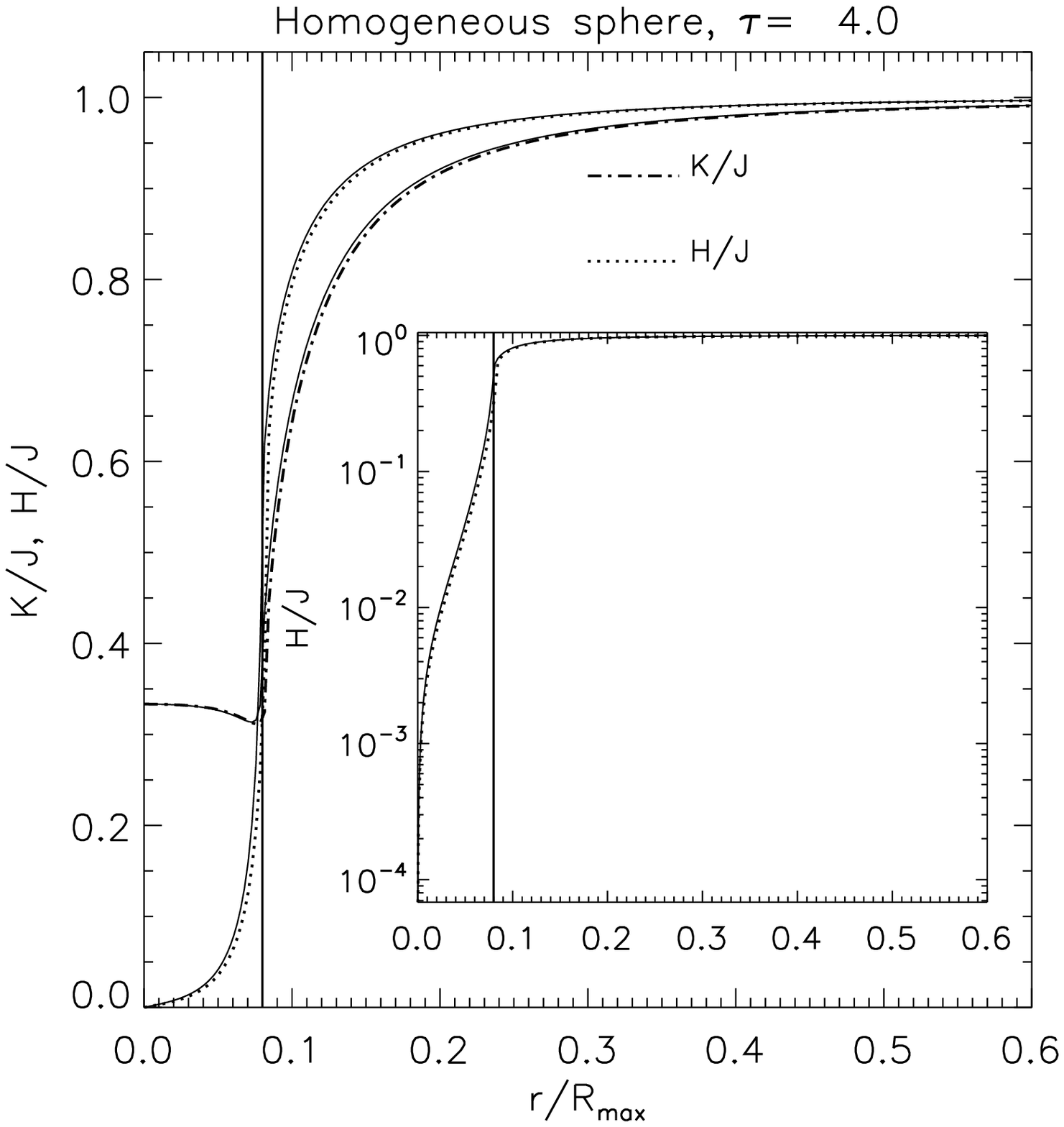} &
\put(0.9,0.3){{\Large\bf d}}
\epsfxsize=8.8cm \epsfclipon \epsffile{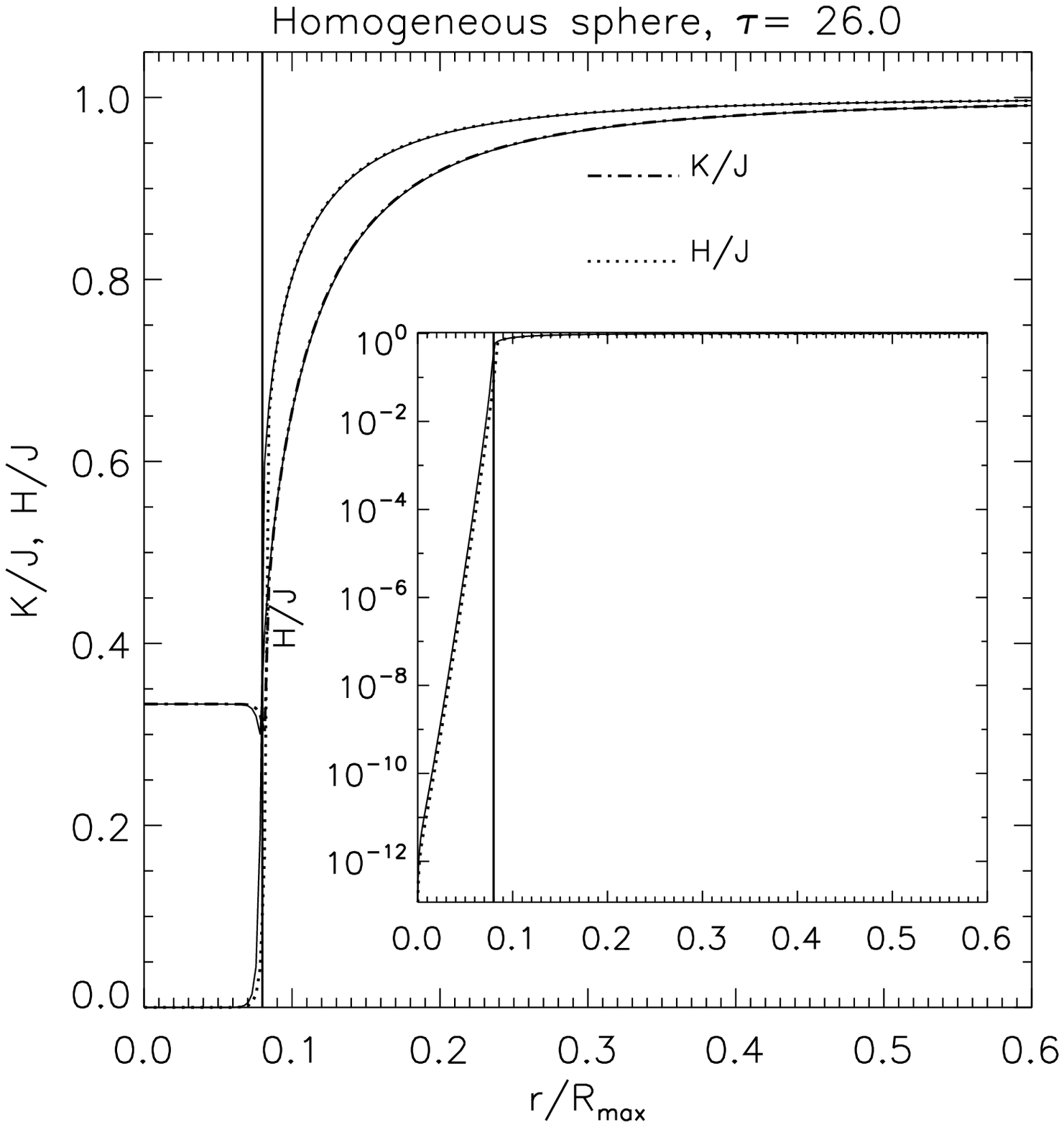}
 \end{tabular}
\caption[]{Numerical (dotted and dash-dotted lines) and analytical
  (solid lines) stationary-state solutions vs.~radius (normalized to
  the radius $R_\mathrm{max}$ of the outer boundary) of 
  homogeneous sphere problems with  
  ``low'' ($\tau=4$, left column, Panels {\bf a}, {\bf c}) and
  ``high'' opacity ($\tau=26$, right column, 
  Panels {\bf b}, {\bf d}). 
  Panels {\bf a} and {\bf b} show the zeroth moment of the
   intensity. In Panels {\bf c} and {\bf d} the Eddington
  factor $K/J$ (dash-dotted) and the flux factor $H/J$ (dotted) are
  displayed. The inserted panels show the flux factor on a logarithmic
  scale. Thin vertical lines mark the surface 
  $R$ of the homogeneous sphere.} 
\label{fig:hs}
\end{figure*}
%
The so-called ``homogeneous sphere'' is a test problem frequently
employed for radiative transfer 
calculations \cite[e.g.,][]{bru85,sinblu89,smicer97}.
Physically, one can think of a static, homogeneous and isothermal sphere
of radius $R$ radiating into vacuum. Inside the sphere, the only
interactions of the radiation with the background medium are isotropic
absorption and thermal emission of radiation. 

Despite of its simplification, the model has some important 
numerical and physical properties which are typically found 
in practical applications:
Finite difference methods are challenged by the
discontinuity at the surface of the sphere.
Moreover, the sudden transition from
radiation diffusion inside an optically thick sphere to free streaming
in the ambient 
vacuum --- a similar but less extreme situation arises in the 
neutrino-heating region of a core-collapse supernova --- is a major 
source of inaccuracies for approximate radiative transfer 
methods like, e.g., flux-limited diffusion.

The model is defined by setting $C=\kappa_\mathrm{a}(b-{\cal I})$ with
$b=\mathrm{const.}$, $\kappa_\mathrm{a}=\mathrm{const.}$ for
$0\le r \le R$, and $\kappa_\mathrm{a}=b\equiv 0$ for $r > R$.
Since for this choice of parameters, the right hand side of the
Boltzmann equation does not contain terms that depend 
explicitly on the angular moments of the radiation field, the solution 
of Eq.~(\ref{eq:BTE_sr}), with $\beta\equiv 0$, can be obtained 
analytically by computing a formal solution. 
With the boundary conditions 
${\cal I}(r=0,\mu=1) = {\cal I}(r=0,\mu=-1)$ and
${\cal I}(r=R_\mathrm{max},-1\le \mu \le 0) =  0$, which account for
isotropy (due to spherical symmetry) of the radiation field at
the center and no incoming radiation at the outer boundary,
respectively, the result for the stationary state 
is \cite[see e.g.,][]{smicer97}:
\begin{equation}\label{eq:hs1}
{\cal I}(r,\mu)=b\,(1-\mathrm{e}^{-\kappa_\mathrm{a}\,s(r,\mu)})
\,.
\end{equation}
Here $s$ is defined as
\begin{gather}\label{eq:hs3}
s(r,\mu) \equ
  \begin{cases}
     r\mu+R g(r,\mu)&\text{if $r < R, -1 \le \mu \le +1$}, \\
     2R g(r,\mu)&\text{if $r \ge R, \hspace{0.7em}x \le\mu \le +1$},          \\
     0&\text{else,}                                  
  \end{cases}   \nonumber \\
\text{with}\qquad
     g(r,\mu) \equ \sqrt{1-\left(\frac{r}{R}\right)^2(1-\mu^2)} \hspace{10cm}
     \\
\text{and}\qquad  x \equ \sqrt{1-\left(\frac{R}{r}\right)^2}
\,.\hspace{10cm}\nonumber 
\end{gather} 
The solution given by Eqs.~(\ref{eq:hs1}, \ref{eq:hs3}) depends on 
only three independent parameters, namely the radial position relative
to the radius of the sphere, $r/R$, its total optical depth $\tau=\kappa_\mathrm{a} R$,
and the equilibrium intensity $b$. The latter merely acts as a scale
factor of the solution.

We employ a radial grid consisting of 213 radial zones to cover the range
between $r=0$ and $r=R_\mathrm{max} \approx 12 R$.
Approximately 200 zones were distributed logarithmically between
$r\approx0.0006 R_\mathrm{max}$ and $r=R_\mathrm{max}$,  
about two thirds of which were spent to resolve the sphere.
Initially we set ${\cal I}\approx 0$ everywhere and evolved the
radiation field until a stationary state was reached.

In \reffigN{fig:hs} we display the stationary-state solutions for two
models, one with  $\tau=b=4$ (Figs.~\ref{fig:hs}a,c) and another with
$\tau=b=26$ (Figs.~\ref{fig:hs}b,d), representative of spheres with low
and high optical depth, respectively.   
In general, our numerical results agree very well
with the analytical solutions.
The Eddington factor $K/J$ (as calculated from the
Boltzmann equation) and the flux factor $H/J$ (as obtained from the
moment equations) both follow the exact values very 
closely (Figs.~\ref{fig:hs}c,d), in case of the flux factor over many
orders of magnitude. 
In both models, the transition to 
free streaming is reproduced very accurately (Figs.~\ref{fig:hs}c,d).
This region usually causes serious problems for flux-limited
diffusion methods \cite[see e.g.,][]{bru85,smicer97}.
Also the forward peaking of the radiation distribution at large
distances from the
surface of the sphere (cf.~Eqs.~\ref{eq:hs1}, \ref{eq:hs3}) is
described excellently by our method:   
At $r=R_\mathrm{max}$, the tangent ray grid 
yields approximately 150 angular grid points to resolve the radiation
field from the central source, which has an opening half angle of 
only $\Theta\approx 5^\circ$ ($\cos\Theta\approx 0.996$).
These numbers should be compared with calculations employing the 
discrete angles ($\mathrm{S}_\mathrm{N}$) method: In time dependent neutrino
transport calculations, typically less than 10 angular
grid points can be afforded to equidistantly cover the range
$0\le \cos\Theta \le 1$. 

Although we have used a geometrical  
radial zoning for our tangent-ray grid, we do not find any systematic
effects caused by a ``bias'' in angular binning as described by
\cite{buryou00}. 
Far from a central source, they report Eddington factors and flux
factors that asymptote to between 0.96 and 0.98 instead of 1.0 in this
case. 
For example, we find  values of $K/J=0.996602$ 
for the Eddington factor and 
$H/J=0.999361$ for the flux factor 
at $r=R_\mathrm{max}$ (in the model with $\tau=4$), to be 
compared with $K/J=0.996636$ and $H/J=0.998626$ obtained from the
analytical solution~(Eqs.~\ref{eq:hs1}, \ref{eq:hs3}).

In case of the mean intensity $J$ we observe a systematic trend towards 
larger deviations from the true solution for spheres with larger total
optical depth $\tau$.
In our ``low opacity'' case \reffigA[a]{fig:hs}{$\tau=4$, see}, there
is hardly any difference visible, whereas in the ``high opacity'' case 
\reffigA[b]{fig:hs}{$\tau=26$, see} the numerical solution slightly
overestimates the true solution in a narrow region beneath the surface
of the sphere and underestimates it in the ambient vacuum region. A
detailed analysis of the data yields values of $6\%$ and $10\%$ for
the relative deviations from the analytical solutions in the ``low
opacity'' and the ``high opacity'' case, respectively.
\begin{figure*}[!ht]
 \begin{tabular}{lr}
\put(0.9,0.3){{\Large\bf a}}
\epsfxsize=8.8cm \epsfclipon \epsffile{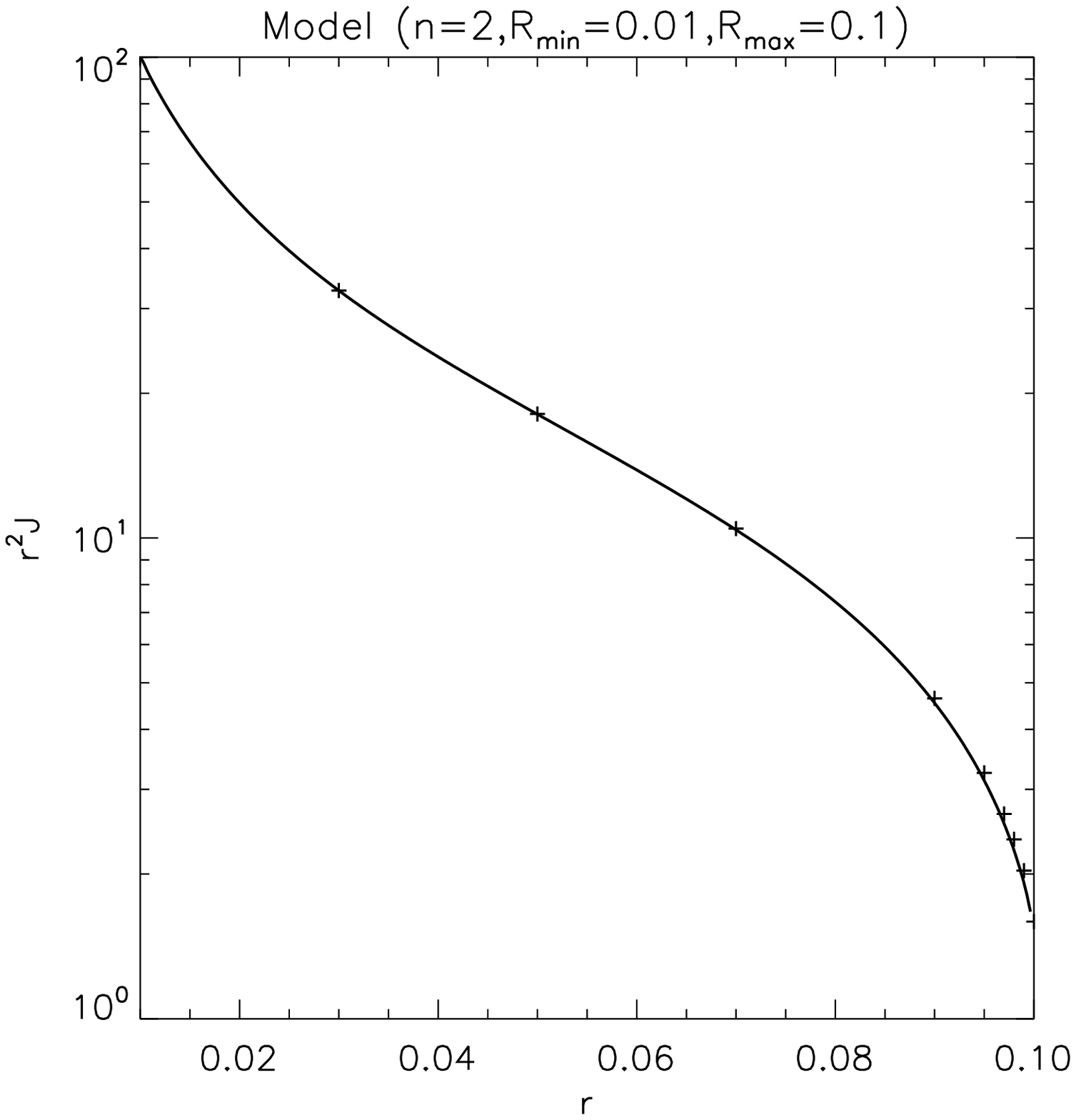} &
\put(0.9,0.3){{\Large\bf b}}
\epsfxsize=8.8cm \epsfclipon \epsffile{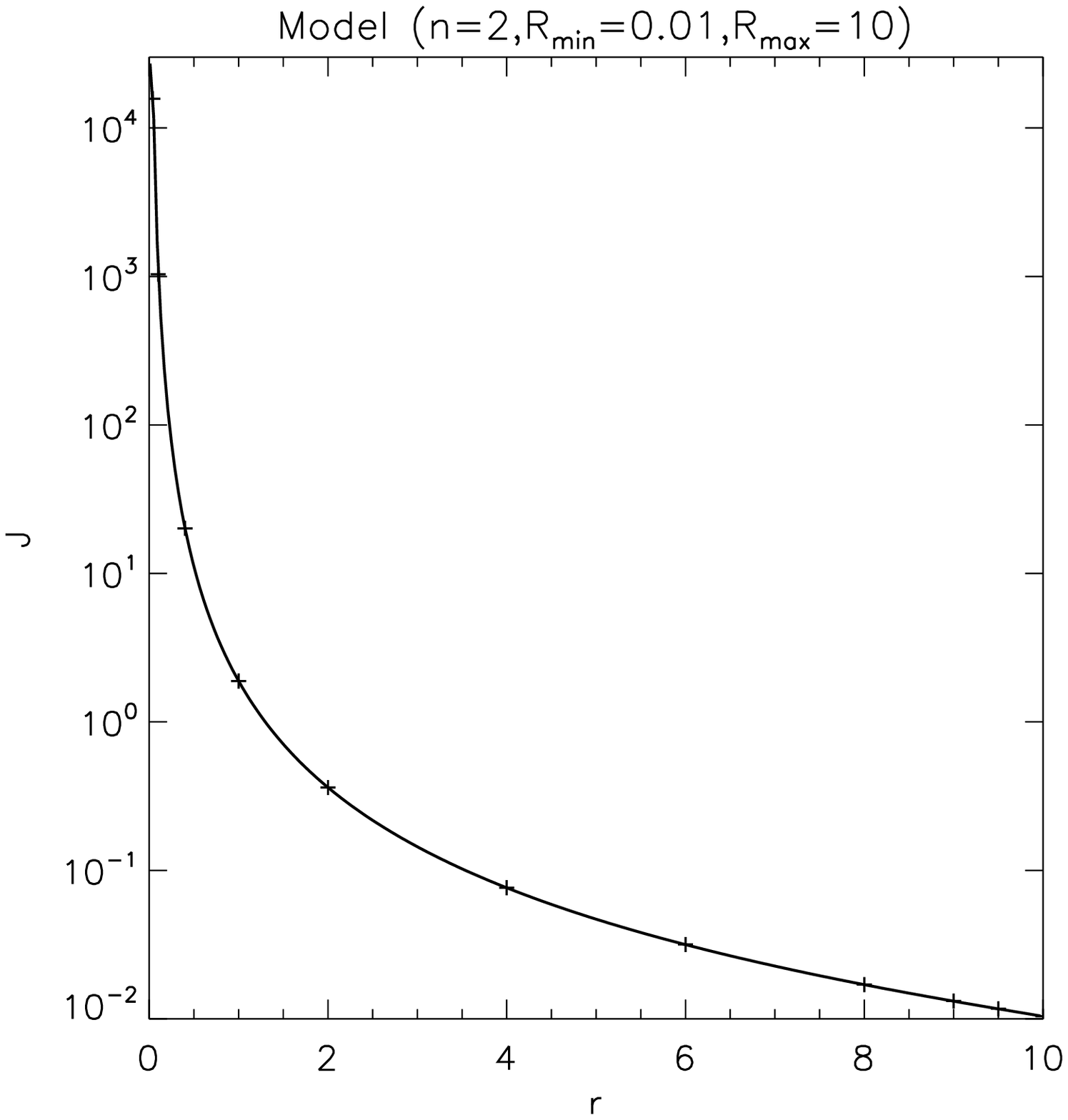} 
\vspace{-0.2cm} \\
\put(0.9,0.3){{\Large\bf c}}
\epsfxsize=8.8cm \epsfclipon \epsffile{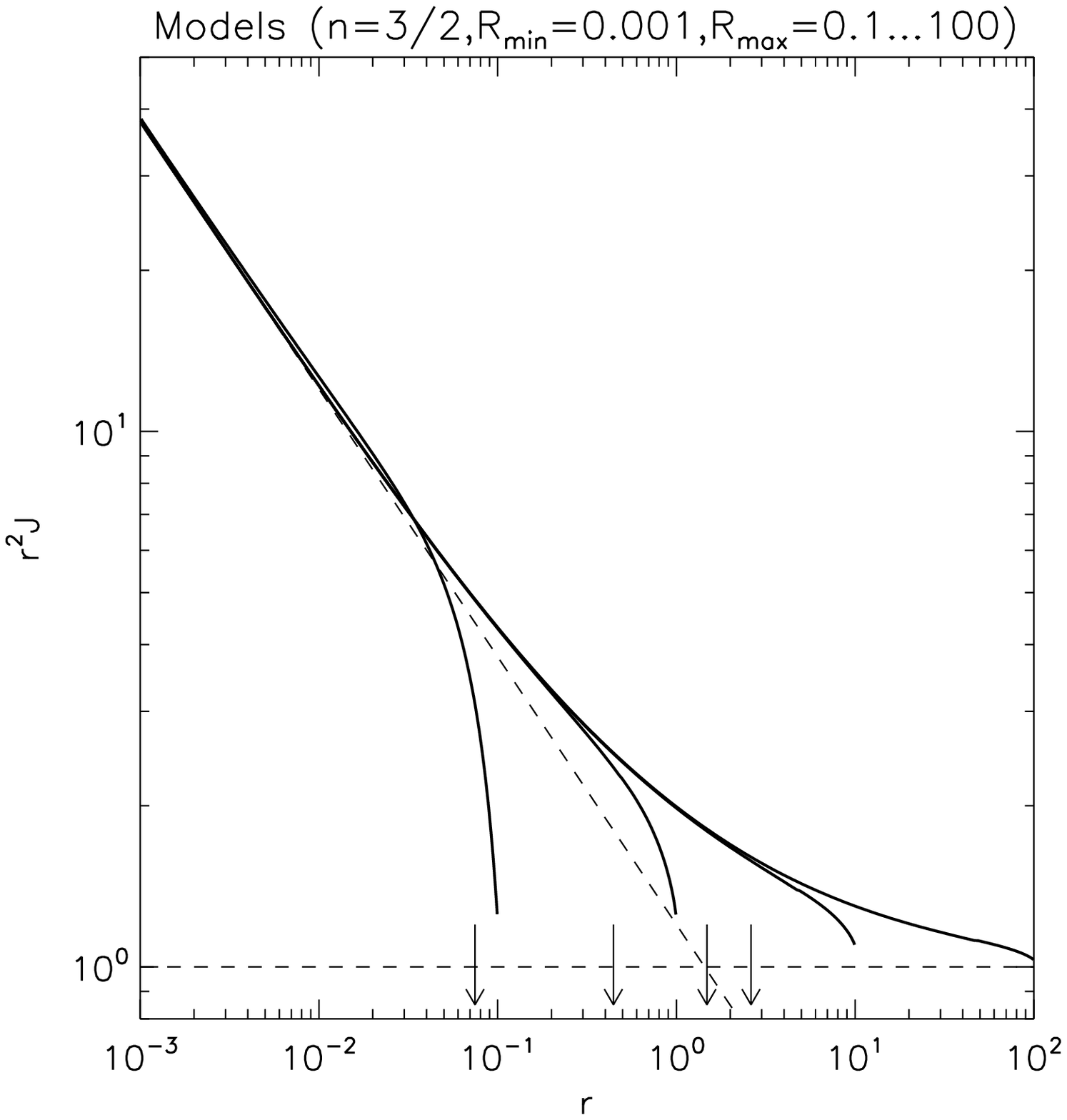} &
\put(0.9,0.3){{\Large\bf d}}
\epsfxsize=8.8cm \epsfclipon \epsffile{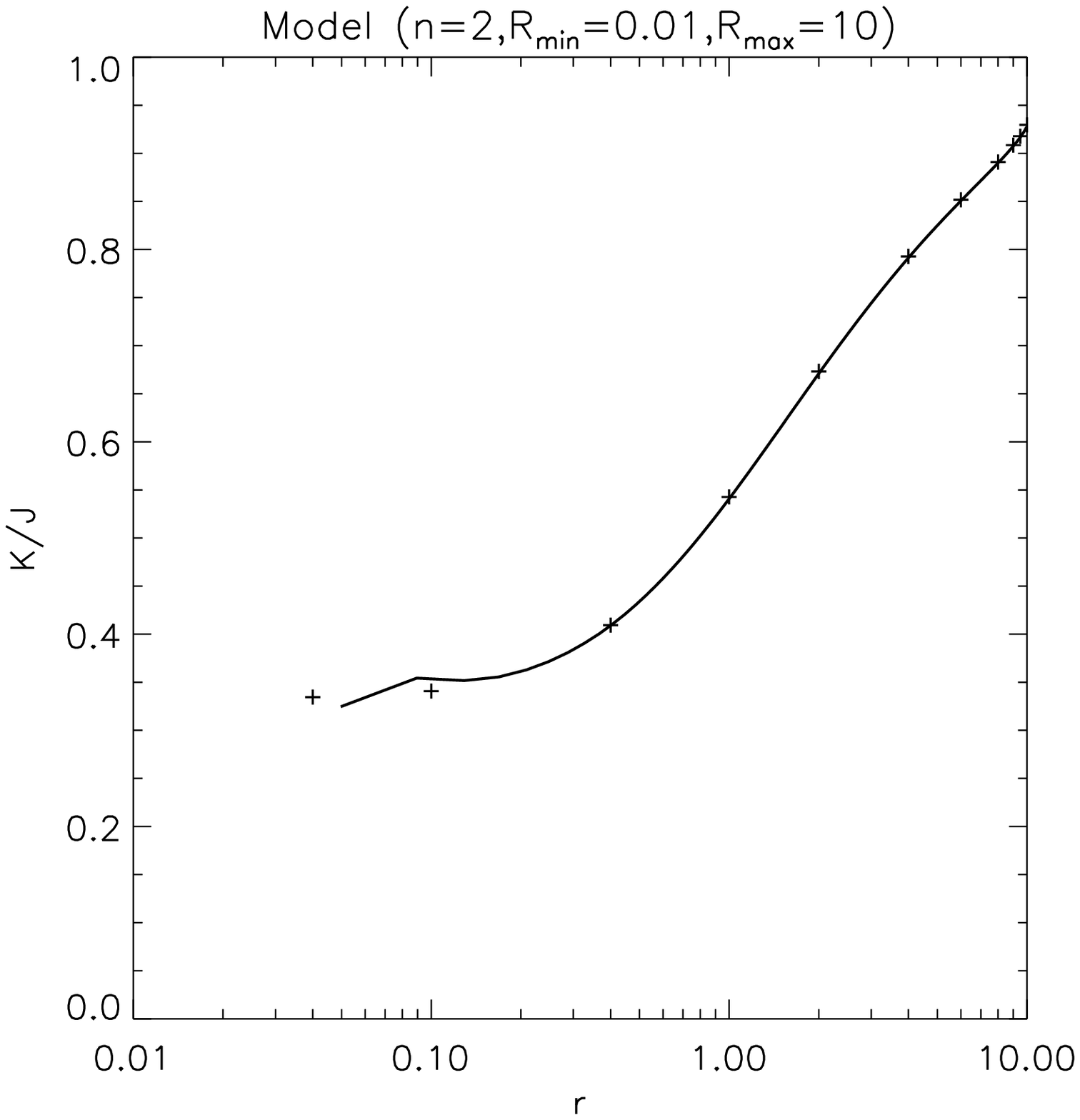}
 \end{tabular}
\caption[]{Stationary state solutions for selected radiation
  quantities as a function of radius obtained 
  with our radiative transfer code (solid lines). Our results are
  compared with the reference solutions (crosses) and asymptotic
  solutions (dashed lines) of \cite{humryb71}. 
  Panel {\bf a}: mean intensity $J$ (times $r^2$) for the combination
  of model parameters $(n=2,R_\mathrm{min}=0.01,R_\mathrm{max}=0.1)$;  
  Panel {\bf b}: mean intensity $J$ for $(n=2,R_\mathrm{min}=0.01, 
  R_\mathrm{max}=10)$; 
  Panel {\bf c}: $r^2 J$ for models with $(n=3/2,R_\mathrm{min}=0.001, 
  R_\mathrm{max}\in\{0.1,1,10,100\})$. The vertical arrows
  indicate the radial positions, where $\tau=1$ is reached in
  the different atmospheres;  
  Panel {\bf d}: Eddington factor $K/J$ for the test case with a
  purely radial distribution of the radiation intensity at the
  inner boundary and the model parameters
  $(n=2,R_\mathrm{min}=0.01,R_\mathrm{max}=10)$. Note that
  the solid line connects values which are evaluated at the 
  centers of the radial zones of our computational grid, the
  first one being located at $r_{\hlf}=0.05$.   
}
\label{fig:HUM}
\end{figure*}

This finding is caused by the steep opacity gradient near the surface
of the radiating sphere. Since we keep our radial grid to be the same
in all models, the transition from high optical depth
($\tau(r)=\kappa_\mathrm{a}\,(R-r)\gtrsim 1$, if $r<R$) to transparency  
($\tau(r)=0$, if $r>R$) occurs in a boundary layer beneath the surface
of the sphere, which is less well resolved when the opacity is very
large. 
In the ``high opacity''
case, the semi-transparent layer is geometrically much thinner than
in the ``low opacity'' case and therefore the radial gradients of the
radiation density are steeper and would require more radial zones for
a proper description.
In fact, in the ``low opacity'' case the radial zones of our grid
near the surface of the sphere are optically thin ($\Delta\tau\approx
0.15$), whereas in the ``high opacity'' case the outermost
zone has already an optical depth of $\Delta\tau\approx 2.3$.
A better quality of the numerical results could be achieved by
increasing the spatial resolution in
the vicinity of large opacity or emissivity gradients and/or by
using higher-order difference schemes. 

Due to the absence of scattering in the discussion of the homogeneous
sphere problem, the solution of the Boltzmann equation does not
require any information from the moment equations.
No iteration between both parts of the code is necessary. Therefore
this problem allows one to test the algorithm which solves the
Boltzmann equation for the radiation intensity independently from the 
numerical solution of the moment equations. 
For the stationary state, we find that the radiation moments calculated by a
numerical integration of the intensities are consistent with the
moments directly obtained from the moment equations to within an
accuracy of less than a percent. 

\paragraph{Static scattering atmospheres:}\label{subsec:test.scatt}

\cite{humryb71} published stationary-state solutions for the
spherical analogue of the classical Milne problem.
The model comprises a static, spherically symmetric, pure scattering
atmosphere of some radius $R_\mathrm{max}$, with a central point source
that is emitting radiation isotropically with a constant luminosity
$L_0$.  
Due to the presence of scattering, the problem is of
integro-differential nature and defies solution by
simply computing the formal solution of the Boltzmann equation.

The opacity of the atmosphere is assumed to be solely caused by isotropic
scattering with a simple power-law dependence on radius:
\begin{equation}\label{eq:milne1}
  \chi(t,r)\equiv\kappa_\mathrm{s}(r)=r^{-n}, \quad 0 < r\le R_\mathrm{max}, 
  \quad (n > 1)
\,.
\end{equation}
By a redefinition of the unit of length we have set a 
scale factor $\alpha>0$ to unity. This factor appears in the original
form  
$\kappa_\mathrm{s}(r,t)=\alpha r^{-n}$ used by
\citet[][ Eq.~1.1.]{humryb71}.
The emissivity vanishes within the atmosphere ($\eta\equiv 0$) and no
radiation is entering from outside (${\cal I}(t,R_\mathrm{max},\mu)=0$, for
$-1\le \mu<0$).   

For a number of atmospheres defined by various combinations of the
parameters  
$n\in\{1.5,2,3\}$, $R_\mathrm{max}\in\{0.1,\dots,100\}$ and $L_0=(4\pi)^2$,
\cite{humryb71} computed numerically the zeroth moment $J$ 
and the Eddington factor $K/J$ of the stationary-state solution as a
function of radius.  
They found values of better than one percent for the accuracy of their
results. 
The stationary-state solution for the first moment $H$ as a function
of radius can easily be derived analytically from the zeroth order
moment equation~(Eq.~\ref{eq:J_sr}, with $\beta\equiv 0$): 
\begin{equation}\label{eq:HUM:H}
H(r)=\frac{L_0}{(4\pi)^2}\,\frac{1}{r^2}
\,.
\end{equation}
This means that the luminosity is conserved ($L(r)\equ 4\pi r^2\cdot 4\pi
H(r)=L_0$). 
\cite{humryb71} also gave useful asymptotic expressions applicable
to regions of very low and very high optical depth, respectively. 

The idealized concept of a central point source with a given luminosity
$L_0$ is in practice modeled by imposing a suitable inner boundary
condition at some finite radius $R_\mathrm{min}$, which bounds the
atmosphere from below.
Since all atmospheres with a scattering opacity according to 
Eq.~(\ref{eq:milne1}) become optically thick at sufficiently small
radius, it is reasonable to employ the diffusion ansatz 
\begin{equation}\label{eq:HUM:ibc}
{\cal I}(t,R_\mathrm{min},\mu)={\cal
  I}_0+\mu {\cal I}_1
\end{equation}
for the radiation field at the inner boundary. 
In our program, only the quantity $h(t,R_\mathrm{min},\mu)=\mu {\cal I}_1$
needs to be specified.
Using Eq.~(\ref{eq:HUM:H}) and $H=\int_0^1\dlin{\mu}\mu h$
(cf.~Eq.~\ref{eq:numq.moments1}), the parameter ${\cal I}_1$ 
is easily verified to be ${\cal I}_1=3H(R_\mathrm{min})=3\,L_0/(4\pi
R_\mathrm{min})^2$.

We evolved the transport to stationary-state solutions for the two
sets of parameter combinations 
($n=1.5$, $R_\mathrm{min}=0.001$, $R_\mathrm{max}\in\{0.1,1,10,100\}$), 
and ($n=2$, $R_\mathrm{min}=0.01$, $R_\mathrm{max}\in\{0.1,1,10,20\}$).
The total optical depth at $r=R_\mathrm{min}$ is larger than
55 for all models of the former class ($n=1.5$) and larger than
90 for all of  
the models with $n=2$. Hence, we verify that the inner boundary is
placed at 
a radius which is sufficiently small to justify the use of the
diffusion approximation (Eq.~\ref{eq:HUM:ibc}) for the inner boundary
condition. 
In order to test the sensitivity of the results to the corresponding angular dependence of the
 intensity  (Eq.~\ref{eq:HUM:ibc}), we have calculated a model
with a different angular dependence: ${\cal I}(t,R_{\rm
  min},\mu)=\delta(\mu-1)\cdot{\cal I}_\mathrm{in}$. 
The parameter ${\cal I}_\mathrm{in}$ is chosen such that
$L(R_\mathrm{min})=L_0$. This prescription is appropriate for an
atmosphere around a central hollow sphere with radius $R_\mathrm{min}$,
which is irradiated from below by the central point source.
\reffigLN[d]{fig:HUM} shows the Eddington factor of the stationary
state for this model. 
The numerical solution deviates visibly from the reference solution
only in the innermost radial zones.  

In the models with $n=1.5$ the standard radial grid consisted of 200
logarithmically distributed zones with some additional zones giving
higher resolution near the
surface of the atmosphere, whereas 250 equidistant
radial zones were used for the models with $n=2$. Additional tests
with 500 radial grid points were performed for some selected models
without finding significant changes. 
Figure~\ref{fig:HUM} shows some results of our tests
together with the data of \cite{humryb71}. The agreement is excellent.
In all models we were able to reproduce the analytical solution for the
flux (Eq.~\ref{eq:HUM:H}) with a relative accuracy of
at least $10^{-4}$ throughout the atmosphere.

\paragraph{Radiative transfer in general relativity:}

\begin{figure*}[!ht]
\begin{tabular}{lr}
\put(0.9,0.3){{\Large\bf a}}
\epsfxsize=8.8cm \epsfclipon \epsffile{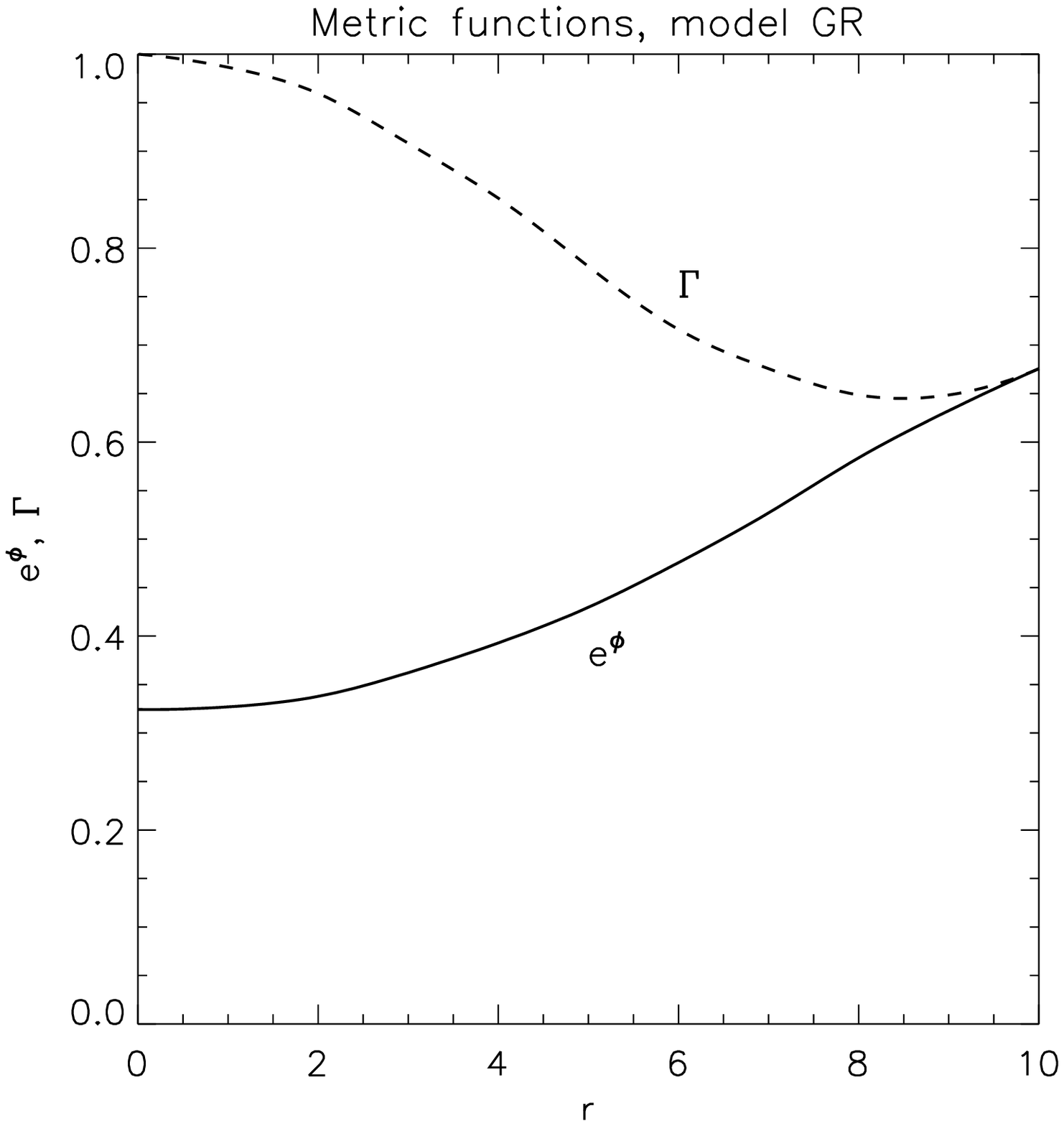} &
\raisebox{8.8cm}{\parbox[t]{8cm}{
\caption[]{Metric functions $\Gamma$ (dashed line) and $\mathrm{e}^\Phi$
  (solid line) of the background model (Panel~{\bf a})
  and stationary-state solutions for selected radiation quantities
  ((Panels~{\bf b}, {\bf c}) versus radius.
  Panel~{\bf b} shows the zeroth moment $J$ (solid line) and first
  moment $H$ (dashed line) of the  intensity normalized to the
  non-relativistic solutions. The reference solutions of
  \cite{sinblu89} are given by the symbols ($J$: triangles; $H$:
  diamonds). Error bars indicate the
  estimated uncertainties of reading off the reference solutions from
  the plots in \cite{sinblu89}.  
  The Eddington factor vs.~radius is shown as a solid line in
  Panel~{\bf c} together with the solution of
  \citet[][ Fig.~13; diamonds]{sinblu89}. }\label{fig:SCHIND} 
}} \\
\put(0.9,0.3){{\Large\bf b}}
\epsfxsize=8.8cm \epsfclipon \epsffile{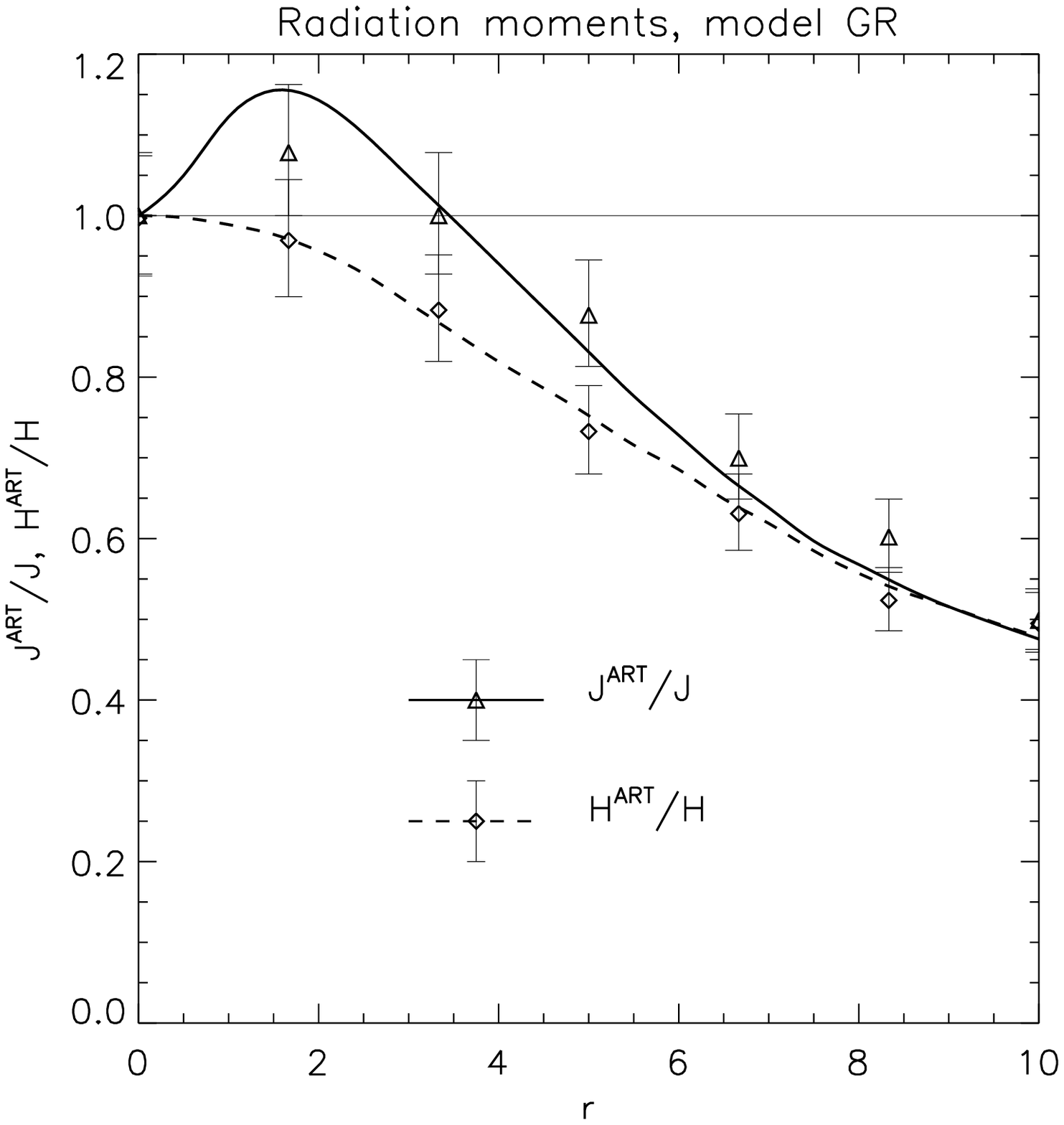} &
\put(0.9,0.3){{\Large\bf c}}
\epsfxsize=8.8cm \epsfclipon \epsffile{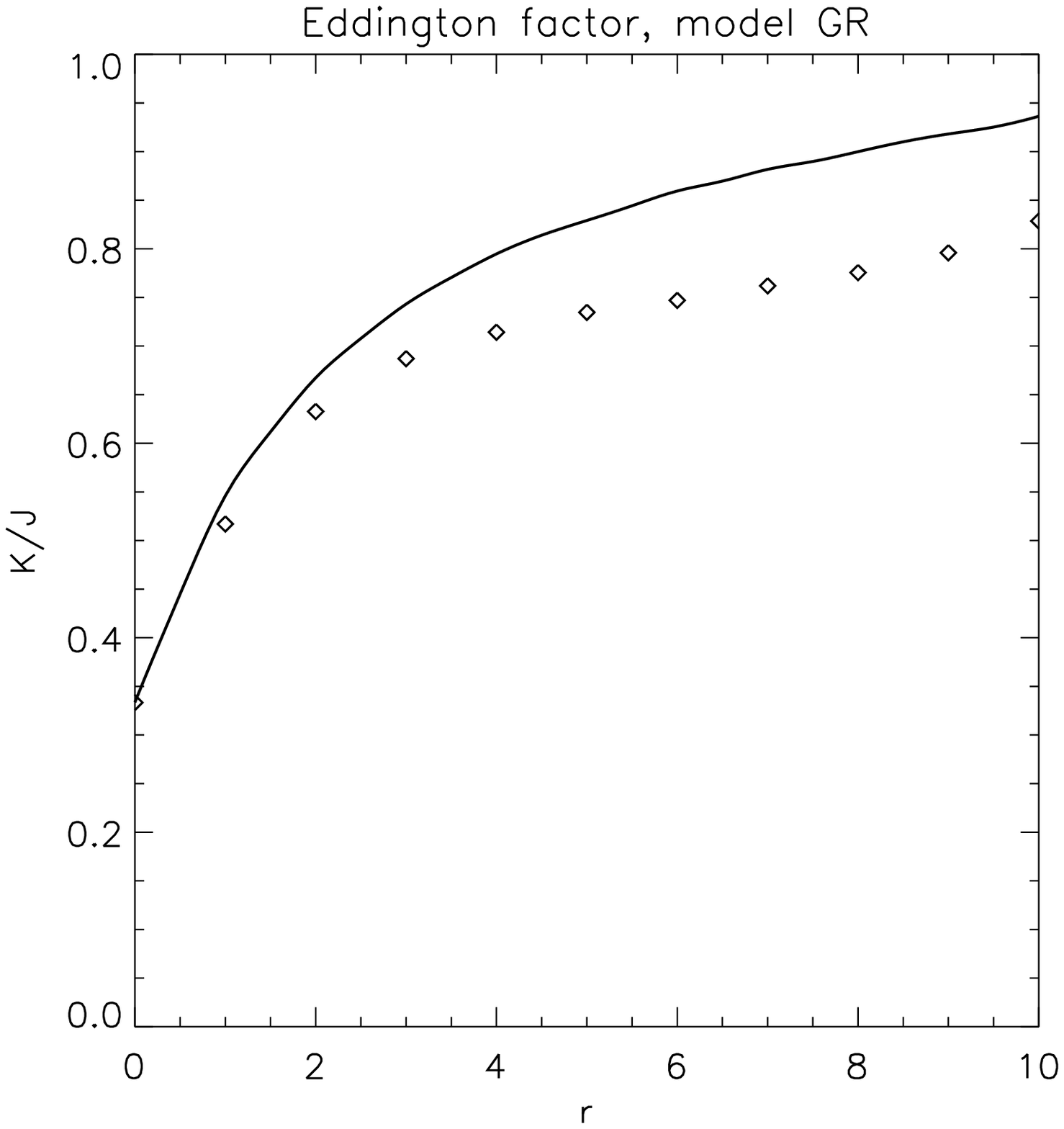} 
\end{tabular}
\end{figure*}

\cite{sinblu89} considered the effects of a strong gravitational field
in some of the radiative transfer problems discussed above, in
particular in the cases of a homogeneous sphere and a static
scattering atmosphere. 
They computed stationary-state solutions numerically for the  
general relativistic equations of radiative transfer and compared the
results to the weak field limit. 
For a static background ($\partial_t R=\partial_t \Lambda\equiv 0$), 
and by application of a change of coordinates
$(R,\epsilon)\mapsto(R,\epsilon_\infty\equ \mathrm{e}^{\Phi}\epsilon)$,
\cite{sinblu89} simplified the general relativistic moment equations
(Eqs.~\ref{eq:J_gr},\ref{eq:H_gr}) to  
\begin{eqnarray}
\frac{1}{c}\frac{\mathrm{D}}{\mathrm{D}t} J &+& 
\frac{\Gamma}{R^2}\frac{\partial}{\partial R}(R^2H\mathrm{e}^{\Phi}) 
=\mathrm{e}^{\Phi}\,C^{(0)} \,,     \label{eq:sinblu_J} \\
\frac{1}{c}\frac{\mathrm{D}}{\mathrm{D}t} H &+& 
\frac{\Gamma}{R^2}\frac{\partial}{\partial R}(R^2K\mathrm{e}^{\Phi})
+\mathrm{e}^{\Phi}\Gamma\frac{K-J}{R} \nonumber \\
&& +\Gamma\frac{\partial \mathrm{e}^{\Phi}}{\partial R}\,(J-K)
= -\mathrm{e}^{\Phi}\,C^{(1)} \,.   \label{eq:sinblu_H}
\end{eqnarray}
The new coordinates have the advantage that the moment equations
decouple in energy space (if energy-changing neutrino-matter
interactions are ignored). 
Hence, all radiation quantities depend only {\em parametrically} on
$\epsilon_\infty$.

\medskip

Different from the moment equations, which we treat in their most
general form (Eqs.~\ref{eq:J_gr}, \ref{eq:H_gr}), our
implementation of the tangent ray scheme for 
computing the variable Eddington factors employs a number of
approximations. In particular, by using straight lines for the tangent
rays, general relativistic ray bending is not included. 
The current tests therefore serve the purpose of
clarifying the influence of these approximations on the quality of the
solutions of the moment equations.
\cite{sinblu89} found significant differences of the Eddington factors
for the general relativistic and Newtonian simulations of the
scattering atmosphere (while in case of the homogeneous sphere the
differences were minor). The scattering atmospheres therefore seems 
to be an ideal test case for our purposes.

Following \cite{sinblu89},
we choose the parameters $R_\mathrm{max}=10$ and
$n=2$. The variation of the metric functions $\mathrm{e}^{\Phi}$ and 
$\Gamma$ with radius is depicted in \reffigN[a]{fig:SCHIND}.
All other model parameters like boundary conditions and initial
conditions are the same as in the ``weak-field''-case which is defined
by $\mathrm{e}^{\Phi}=\Gamma\equiv 1$ (see above).
The stationary-state Eddington factor as computed from the Boltzmann
equation in our program is displayed as a function of 
radius in \reffigN[c]{fig:SCHIND}. The deviation from the
relativistically correct results of \cite{sinblu89} is significant.
On the other hand, our result is very close to the corresponding
``weak-field'' case, which is not shown in Fig.~\ref{fig:SCHIND}. 
This indicates that general relativistic ray bending, which is not
correctly taken into account in our calculation, has a determining
influence on the Eddington factor, while   
the contraction of rods and time dilation --- both included in our
calculation of the Eddington factor --- are of minor importance.
Indeed, the relativistically correct characteristic curves
deviate considerably from straight rays, as can be seen 
in \citet[][ Fig.~9]{sinblu89}. 

The analytical stationary-state solution for the radiation flux density,
which is easily verified\footnote{Note that the more familiar result 
  $R^2\mathrm{e}^{2\Phi}H(R)=\mathrm{const.}$ only holds for the total
  (i.e.~energy integrated) ``flux density'' 
  $H(R)\equ\int H(R,\epsilon)\,\mathrm{d}\epsilon=
  \mathrm{e}^{-\Phi}\int H(R,\epsilon_\infty)\,\mathrm{d}\epsilon_\infty$.}
to read $R^2\mathrm{e}^{\Phi}H(R,\epsilon_\infty)=A(\epsilon_\infty)$,
(with $A(\epsilon_\infty)\in\Real$), is reproduced with an
accuracy of $10^{-7}$. 
This, of course, was to be expected for the atmosphere with
isoenergetic scattering, 
since in the stationary state, the solution $H(r)$ is
solely determined by the zeroth moment equation (Eq.~\ref{eq:sinblu_J})
with no reference to the (incorrect) Eddington factor.

More remarkably, the quality of our solution for
the mean intensity $J(r)$, which is governed in the stationary
state by Eq.~(\ref{eq:sinblu_H}) and thus 
is directly influenced by the Eddington factor, appears to be rather
good, too. 
As can be seen from \reffigN[b]{fig:SCHIND},
the slope of the general relativistic result is reproduced correctly
throughout the atmosphere. 
The quantitative agreement is quite satisfactory as well.
In judging the accuracy of our results one has to keep in mind that
the model computed here is a rather extreme case in two respects:
Firstly, the deviations of the metric functions $\mathrm{e}^{\Phi}$ and 
$\Gamma$ from unity \reffigA[a]{fig:SCHIND}{see} are larger 
than those typically encountered in supernova simulations when the
neutron star is not going to collapse to a black hole. 
Secondly, as stated in the beginning, the effect of a curved spacetime
on the radiation field was observed by \cite{sinblu89} to be much
larger for the scattering atmosphere than in a case dominated by
absorption (and emission). We expect real situations to be somewhere
in between these two extremes. 

\subsection{Radiative transfer in stationary background media}\label{chap:test.stationary}

\begin{figure*}[!ht]
 \begin{tabular}{cc}
\put(0.9,0.3){{\Large\bf a}}
\epsfxsize=8.8cm \epsfclipon \epsffile{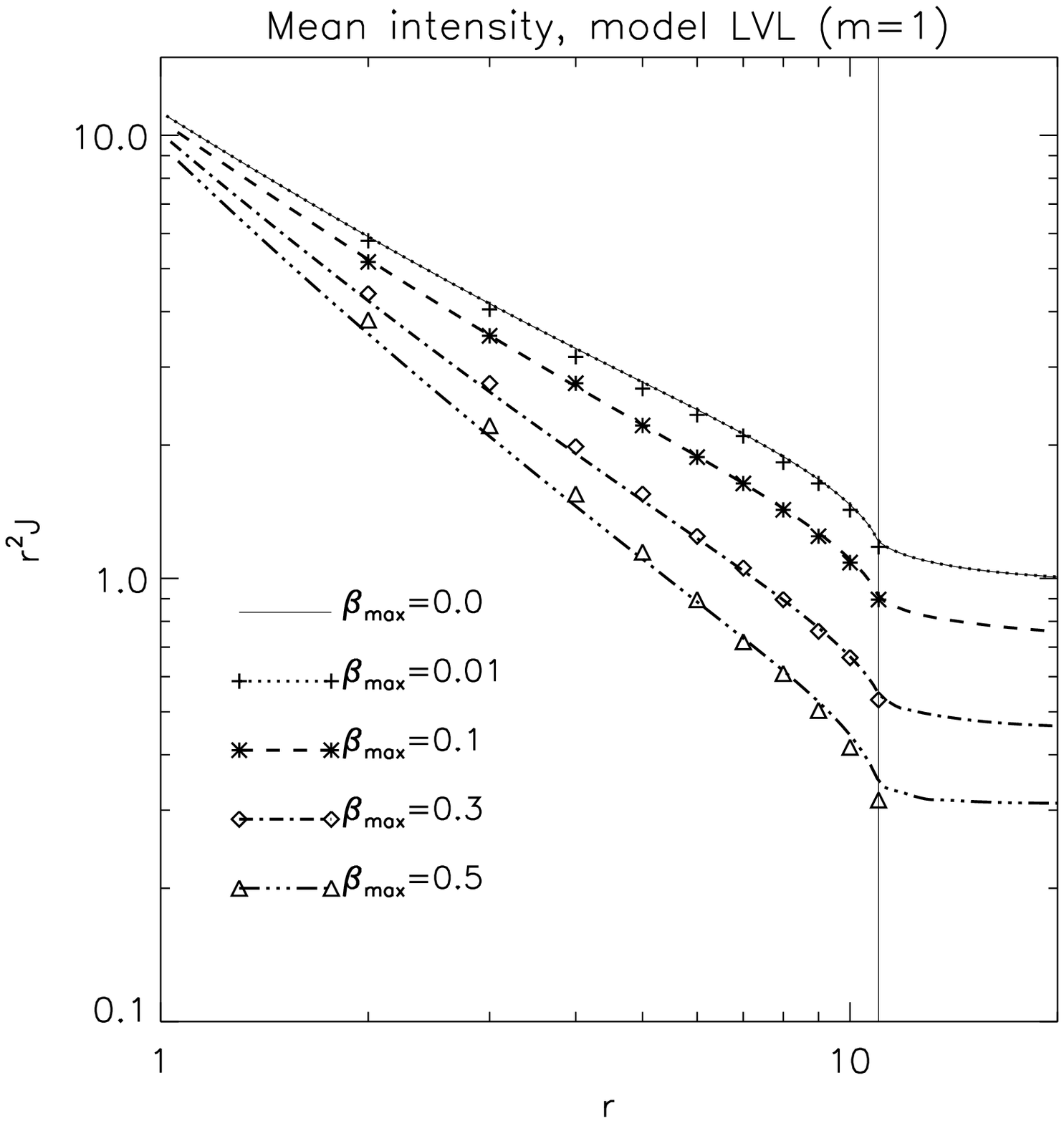} &
\put(0.9,0.3){{\Large\bf b}}
\epsfxsize=8.8cm \epsfclipon \epsffile{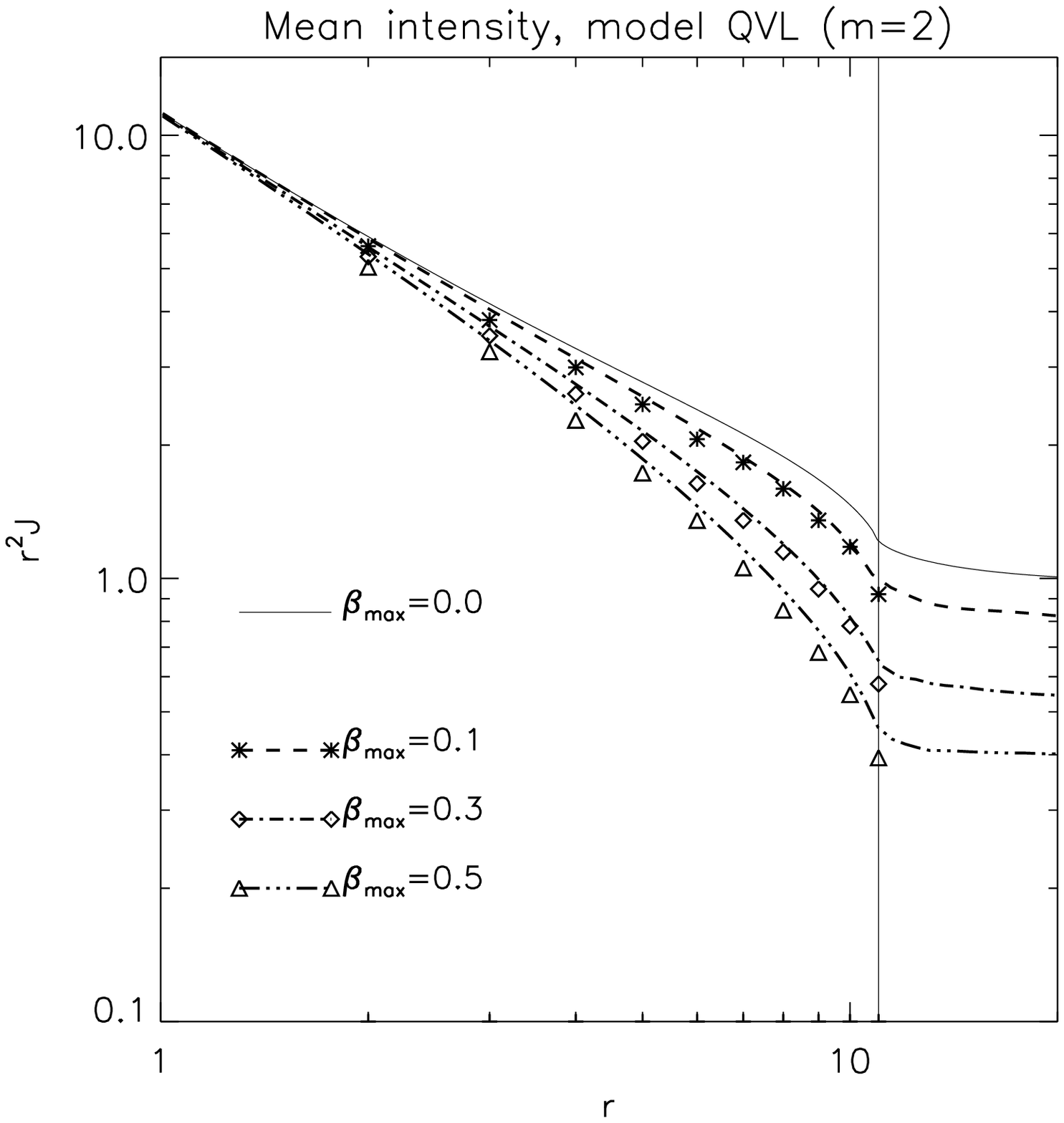} 
 \end{tabular}
 \caption[]{Stationary-state angular mean $J$ (times
   $r^2$) of the  intensity  as measured in the comoving frame of
   reference. The abscissa gives the radial position. Results are
   shown for gray scattering
   atmospheres that expand according to a linear (Model ``LVL'';
   Panel~\textbf{a}) and the quadratic (Model ``QVL'';
   Panel~\textbf{b}) velocity law. 
   The thin vertical line marks the upper boundary of the atmospheres.
   Different line styles of the curves correspond to different values
   of the parameter
   $\beta_\mathrm{max}$ which gives the maximum expansion velocity in
   units of $c$ reached at the surface ($r=11$) of the model
   atmospheres. The thin solid curve corresponds to the static
   case. Reference solutions (symbols) were taken from \citet[
   Fig.~2]{mih80}.}\label{fig:MIH_grey}  
\end{figure*}

In the following section we consider radiative transfer problems in moving,
yet stationary background media. This means that in addition to 
time-independent radial profiles of the opacity and emissivity, 
a time-independent velocity field as a function of radius is prescribed.
Stationary-state solutions for the radiation field are expected to
exist is such cases and have been computed to high accuracy for
some test problems including differentially moving atmospheres with
relativistic velocities.
By comparison with fully (special) relativistic calculations, we are not
only able to 
test our implementation of the velocity dependent terms but can also
judge the quality of the employed ${\cal O}(v/c)$-approximation of
special relativistic effects in the equations of radiative transfer.

Upon introducing a non-vanishing velocity field a particular frame has
to be specified, where physical quantities are measured. In all cases
considered in this work, the latter is chosen to be the Lagrangian or
comoving \emph{frame of reference}. This has to be distinguished from
the (numerical) \emph{coordinate grid} that is used to simply label the
events in spacetime. Although the
transformation between different coordinate grids is trivial from an
analytical point of view (e.g., the simple replacement $\partial/\partial
t+v\partial/\partial r\to \mathrm{D}/\mathrm{D}t$ transforms from Eulerian
to Lagrangian coordinates), the numerical
treatment can be involved (cf. Sect.~\ref{chap:transp.BTE}, where our
algorithm for computing a formal solution of the radiative transfer
equation in Eulerian coordinates is described).  
Therefore we present test results obtained with two different
versions of our radiative-transfer code, one which uses 
Lagrangian coordinates and another one which employs an Eulerian
coordinate grid. 
\begin{figure*}[t]
 \begin{tabular}{cc}
\put(0.9,0.3){{\Large\bf a}}
\epsfxsize=8.8cm \epsfclipon \epsffile{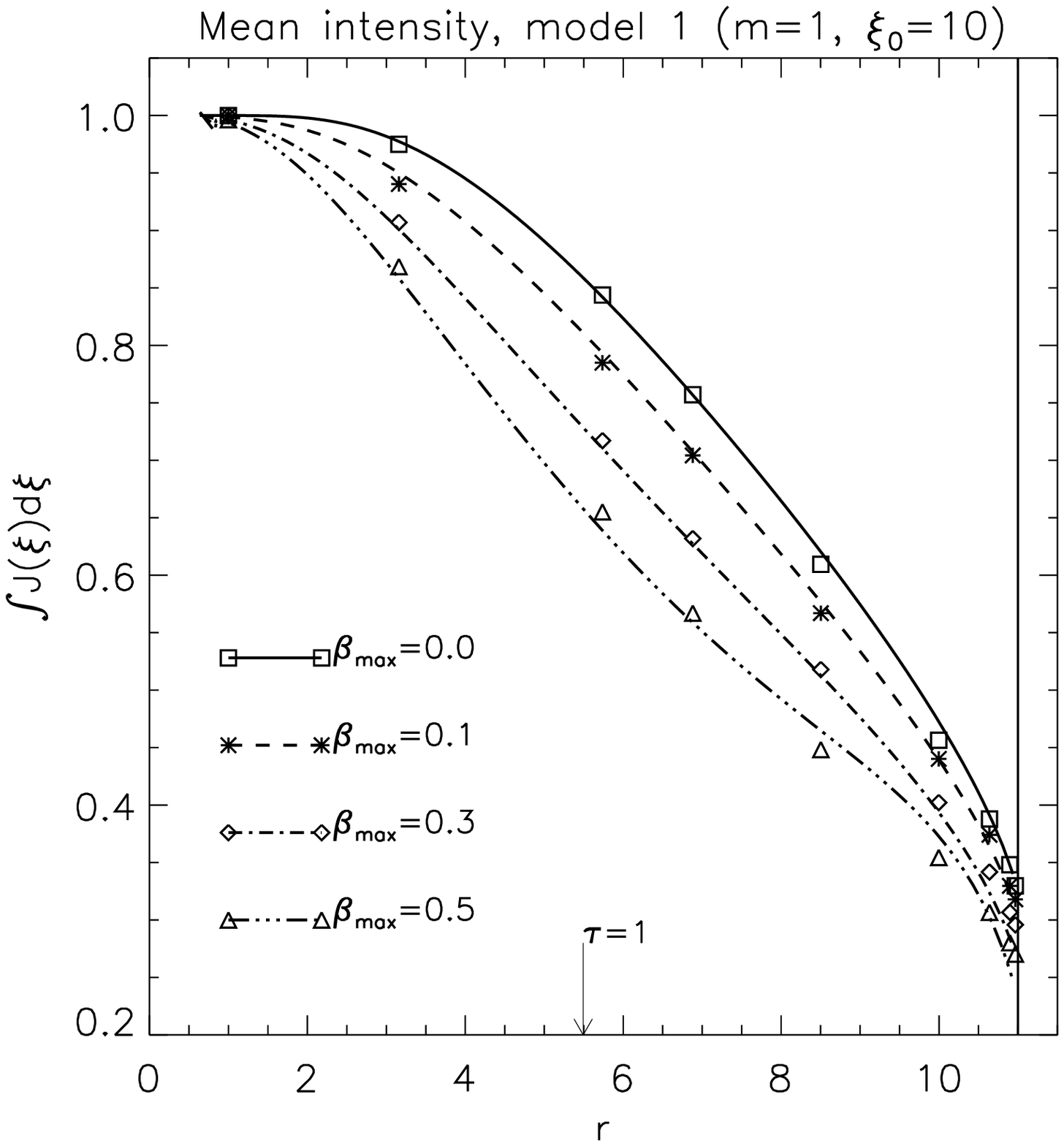} &
\put(0.9,0.3){{\Large\bf b}}
\epsfxsize=8.8cm \epsfclipon \epsffile{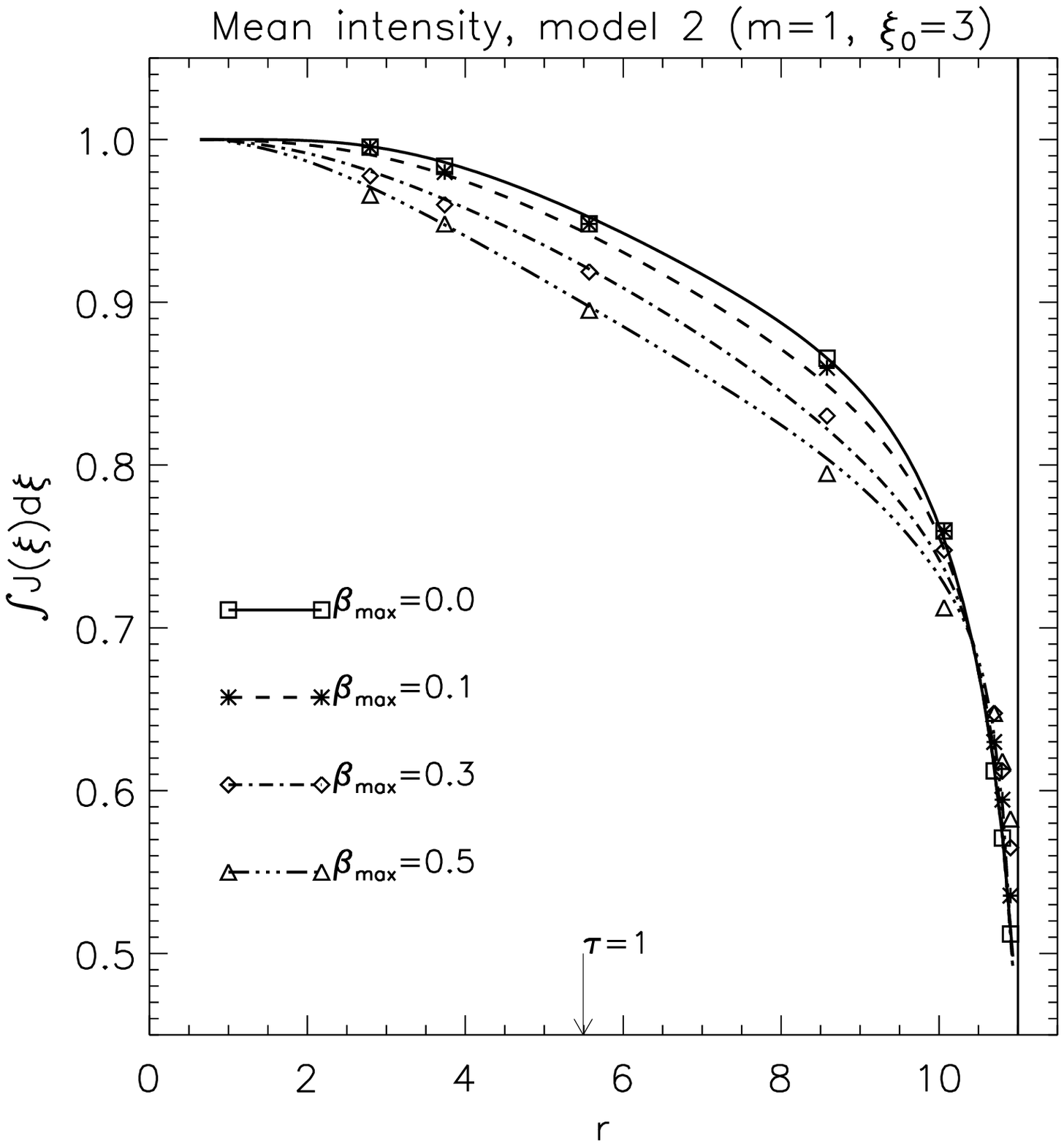} 
 \end{tabular}
\caption[]{Frequency integrated zeroth order angular moment of the 
  comoving frame
  intensity ($\int_0^\infty\dlin{\xi}J(\xi)$) as a function of 
  radius for our ${\cal O}(v/c)$ stationary-state solutions.
  The thin vertical line marks the outer boundary of the atmospheres.
  The small arrows point to the positions where an optical depth
  (which is calculated for $\xi<\xi_0$) of unity is reached.
  Reference solutions (symbols) obtained with a fully relativistic code 
  were taken from \citet[][ Figs.~3 and 5]{mih80}, where  
  the read-off error is approximately given by the size of the symbols. 
  Different maximum velocities $\beta_\mathrm{max}$
  of the atmospheres are indicated by different line styles
  and symbols.
  In Panel~{\bf a} the results are depicted for Model~(1) with the
  opacity step located at $\xi_0=10$  
 ($\Delta=0.25$). Panel~{\bf b} shows results for
  Model~(2) with  $\xi_0=3$ ($\Delta=0.2$).}\label{fig:MIH_jerg}
\end{figure*}

\paragraph{Differentially expanding gray scattering atmospheres:}

\cite{mih80} presents stationary-state solutions for the same
type of purely scattering atmospheres considered above.
In addition to the static case, he investigated differentially expanding 
atmospheres with relativistic velocities. 

We refer to this paper and study two classes of atmospheres as test
problems, which differ in the functional dependence of the expansion
velocity ($v\equiv c\,\beta$) on the radius:
\begin{equation}
  \label{eq:expvel}
  \beta(r)=\beta_\mathrm{max}\,
  \left(\frac{r-r_\mathrm{min}}{r_\mathrm{max}-r_\mathrm{min}}\right)^m\,,
  \quad \text{with} \quad m \in \{1,2\}
\,.
\end{equation}
The case $m=1$ describes a ``linear velocity law'', the case
$m=2$ is called ``quadratic velocity law''. 
Different models are labeled by the parameter 
$0 \le \beta_\mathrm{max} < 1$, which is the maximum expansion velocity in
units of $c$ reached at the surface of the atmosphere ($r=r_{\max}$).

The scattering opacity of the expanding atmospheres is the same
as the one used for the static cases in
Sect.~\ref{subsec:test.scatt} except for the unit of length, which,
for the ease of comparison, is adopted from \cite{mih80}: 
\begin{equation}\label{eq:scattopac}
  \chi(t,r,\epsilon)\equiv\kappa_\mathrm{s}(r)=\alpha\,r^{-2},\quad 
  r_\mathrm{min} \le r \le r_\mathrm{max}
\,,
\end{equation}
with the parameters 
$r_\mathrm{min}=1$, $r_\mathrm{max}=11$ and $\alpha=10.989$.
This yields the same optical depth as the atmosphere with
$r_\mathrm{max}=1$ (and $r_\mathrm{min}=1/11$) in the 
units of \cite{humryb71}, who set $\alpha=1$ (cf.~Eq.~\ref{eq:milne1}).
For the profile of Eq.~(\ref{eq:scattopac}), an optical depth of unity is
reached at a radius of $r\approx 5.5$. 
Since the opacity does not depend on the frequency of the radiation, the
atmospheres are referred to as ``gray''.  
In this case, the transport solution does not depend on $\epsilon$ and
the $\partial/\partial\epsilon$-terms in the comoving-frame Boltzmann
equation can be dropped.

For our simulations we used a numerical grid with 200 radial
points, which were initially distributed logarithmically between
$r_\mathrm{min}$ and $r_\mathrm{max}$.
Following \cite{mih80} as closely as possible, 
we impose the boundary conditions
\begin{alignat}{3}
{\cal I}(t,r_\mathrm{min},\mu)&= 11+3\mu,\quad & 0 \le \mu \le 1, \nonumber\\
{\cal I}(t,R(t),\mu)       &= 0,      \quad & -1 \le \mu \le 0& , 
\end{alignat}
with $R(t)=r_\mathrm{max}+\beta_\mathrm{max}\,ct$. 
The numerical values in the diffusion ansatz at the inner
boundary are chosen according to the asymptotic solution of
\citet[][ Eq.~2.22]{humryb71}.  
We started the simulations using the stationary-state solution
of the corresponding static atmosphere (see
Sect.~\ref{subsec:test.scatt}) as an initial condition (at $t=0$).  
The results presented here were obtained by using a Lagrangian
coordinate grid.
Our calculations were terminated when the radiation field (as measured
in the comoving frame of reference) showed no more substantial
temporal variations at any given radial position. 
We refer to our results as ``stationary-state solutions'' in this 
sense. 
We also tested the Eulerian grid version of the code with
a less extensive set of models. The results were practically identical.

We compare our stationary-state solutions with results obtained by 
\cite{mih80}. In the latter investigation a numerical method that is
accurate to all orders in $v/c$ was 
employed to solve the stationary radiative transfer equation. 
We therefore do not only test the correct numerical implementation of
the ${\cal O}(v/c)$ equations~(Eqs.~\ref{eq:BTE_sr}, \ref{eq:J_sr},
\ref{eq:H_sr}),  
but we can also get a feeling for the quality of the 
${\cal O}(v/c)$-approximation to the special relativistic transfer
equation. 
All characteristic features 
of the solution, discussed in detail by \cite{mih80}, are reproduced
correctly by our implementation of the ${\cal O}(v/c)$ equations.
It is remarkable how accurately the fully relativistic solutions are
reproduced \reffig{fig:MIH_grey}. The differences are $\lesssim 10\%$, 
even for $\beta_\mathrm{max}=0.5$.

\begin{figure*}[t]
 \begin{tabular}{cc}
\put(0.9,0.3){{\Large\bf a}}
\epsfxsize=8.8cm \epsfclipon \epsffile{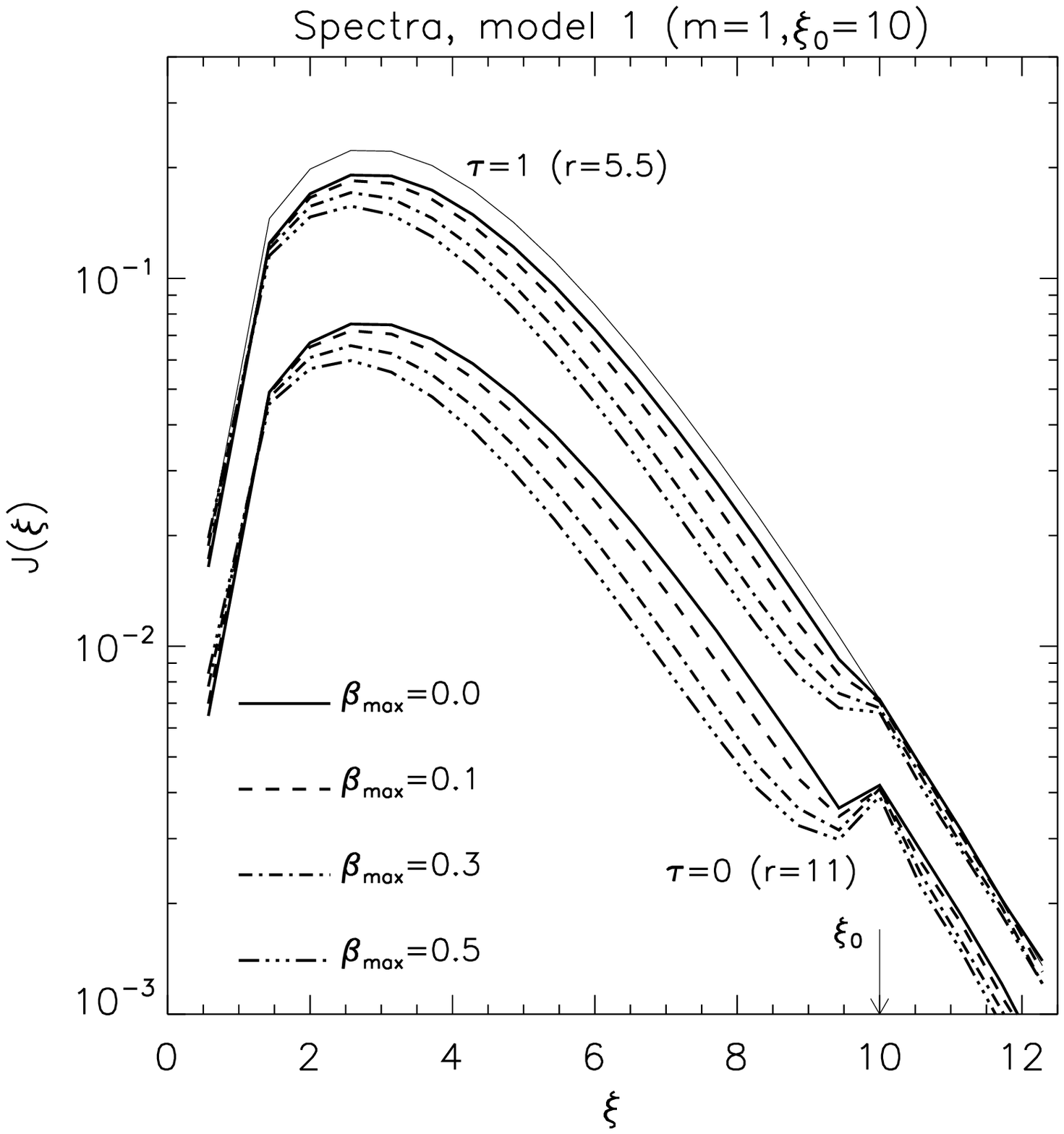} &
\put(0.9,0.3){{\Large\bf b}}
\epsfxsize=8.8cm \epsfclipon \epsffile{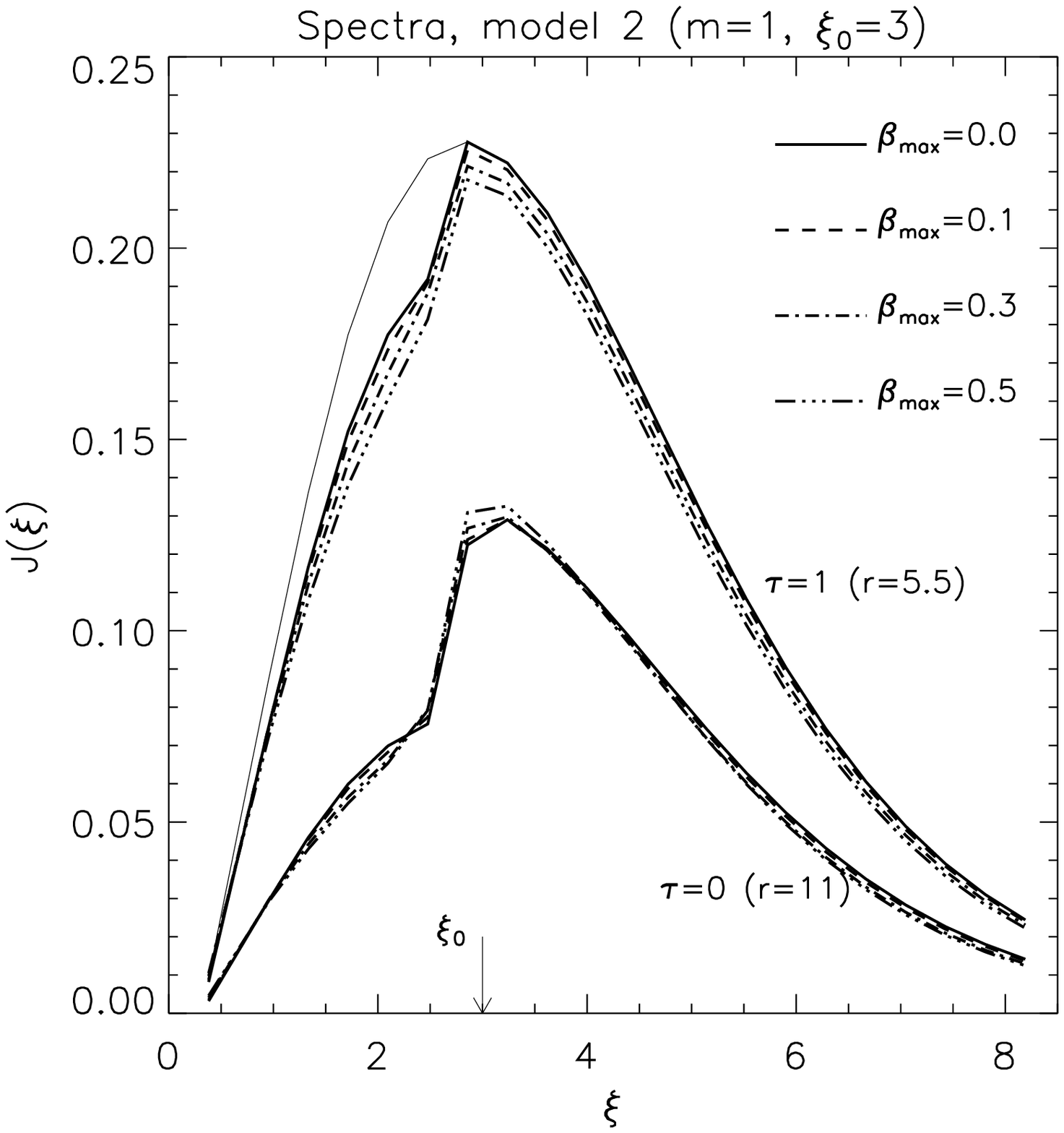} 
 \end{tabular}
\caption[]{Spectral distributions of the angular moment $J$ of the
  comoving frame intensity for our  
  stationary-state solutions of the ${\cal O}(v/c)$ radiative transfer
  equation.  Spectra
  are displayed for a radial position corresponding to
  an optical depth (which is calculated for $\xi<\xi_0$) of unity and
  for the surface of the atmosphere. For 
  comparison, the equilibrium intensity (Planck function) at
  an optical depth  of unity is plotted as a thin line. 
  Different maximum velocities ($\beta_\mathrm{max}$) of the expanding
  atmospheres are indicated by different line styles.  
  Panel~{\bf a} shows results for Model~(1), where the
  opacity step (vertical arrows in the plots) is located at $\xi_0=10$
  ($\Delta=0.25$).  
  Panel~{\bf b} shows Model~(2) with $\xi_0=3$ ($\Delta=0.2$).  
  Our results may be compared with \citet[][ Figs.~4 and 6]{mih80}.}
\label{fig:MIH_spec}
\end{figure*}

\paragraph{Differentially expanding, nongray, isothermal atmospheres:}

\cite{mih80} computed stationary-state solutions
also for relativistically expanding {\em nongray} atmospheres,
i.e., the opacity and emissivity depended explicitly on the radiation
frequency. Therefore the full frequency dependence of the comoving frame
Boltzmann equation had to be retained.
The atmospheres were assumed to be stationary, to emit a
thermal continuum of radiation, and to be isothermal, i.e., no radial
or temporal variations of the source function were present.
The radiation-matter interaction was solely via absorption and emission.

Introducing the dimensionless ``frequency'' $\xi=\epsilon/T$ (the
temperature $T$ is measured in units of the Boltzmann constant \kb),
the emissivity reads  
\begin{equation}\label{eq:mihemiss}
  \eta(\xi)=\kappa_\mathrm{a}(\xi)\,b(\xi)=
  \kappa_\mathrm{a}(\xi)\,\frac{\xi^3}{\exp(\xi)-1}
\,. 
\end{equation}
The absorption opacity $\kappa_\mathrm{a}$ was given a step-like 
functional dependence on frequency, characterized by 
the frequency $\xi_0$ where the step is located and some 
characteristic width $\Delta$ of the function
$s(\xi)\equ\exp[-(\xi-\xi_0)^2/\Delta^2]$, which mediates a
continuous variation of $\kappa_\mathrm{a}$ across the step:
\begin{equation}
  \label{eq:MIH_opa}
  \kappa_\mathrm{a}(r,\xi)= 
  \begin{cases}
    \chi_1(r)\cdot s(\xi)
    +\chi_2(r)\cdot(1-s(\xi)),& 
    \text{for $\xi \le \xi_0$,}   \\
    \chi_1(r),&                    \text{for $\xi > \xi_0$,}  
  \end{cases}
\end{equation}
with $\chi_1(r)\equ 10\alpha\,r^{-2}$, $\chi_2(r)\equ
  \alpha\,r^{-2}$, and the value $\alpha\equ 10.9989$.
Thus, the optical depth at a given radius is roughly 10 times 
smaller for ``low-frequency'' radiation ($\xi < \xi_0$) than for 
radiation with ``high'' frequency ($\xi > \xi_0$).

Following \cite{mih80},
we consider the two examples (1) $\xi_0=10$, $\Delta=0.25$, and (2)
  $\xi_0=3$, $\Delta=0.2$.
In the former case the frequency $\xi_0$,
where the opacity jump is located, is considerably larger than the
frequency of the maximum of the Planck function
($\xi_\mathrm{max}\approx 2.82$),
whereas it is close to the maximum in the latter case. 

The radial extent of the atmospheres was chosen to be
$0 \le r \le 11$ in both cases. For all models the linear 
law ($m=1$, cf.~Eq.~\ref{eq:expvel}) for  the expansion 
velocity as a function of radius was used. 
The results presented below were obtained using an Eulerian
coordinate grid. Our code version with a Lagrangian grid yields very
similar results.

\bigskip

We have used 200 logarithmically distributed radial zones for the
radial range $1 \le r \le 11$, supplemented by a few additional
zones between $0 \le r < 1$.
Boundary conditions were chosen as usual: Spherical symmetry at the
coordinate center requires
${\cal I}(t,r=0,\mu=1)={\cal I}(t,r=0,\mu=-1)$, and demanding that no
radiation is entering through the surface of the atmosphere translates to
the condition ${\cal I}(t,r=11,\mu \le 0)=0$.

The energy grid consisted of 21 bins (very similar results were
obtained using 8 and 12 bins instead) of equal width covering the 
range $0\le \xi \le 12$ for Model~(1) and $0\le \xi \le 8$ 
for Model~(2), respectively.
Details about the stationary-state solutions 
and their physical interpretation can be found in \citet[][ \S
IV]{mih80}. 
In our test calculations we were able to reproduce all qualitative
trends discussed there. 
The quantitative agreement of the zeroth order angular moment $J$ of the
specific intensity is within a few percent
\reffigA{fig:MIH_jerg}{see}, 
even for the case of $\beta_\mathrm{max}=0.5$. 
This, as previously stated, does not only demonstrate the accuracy of
our numerical implementation  
but even more importantly also justifies the physical 
approximations employed in the underlying finite difference equations.
Note that in applications of our code to supernova calculations, one
expects velocities that are not in excess of $v\approx 0.3c$.

\begin{figure*}[!ht]
 \begin{tabular}{cc}
\put(0.9,0.3){{\Large\bf a}}
\epsfxsize=8.8cm \epsfclipon \epsffile{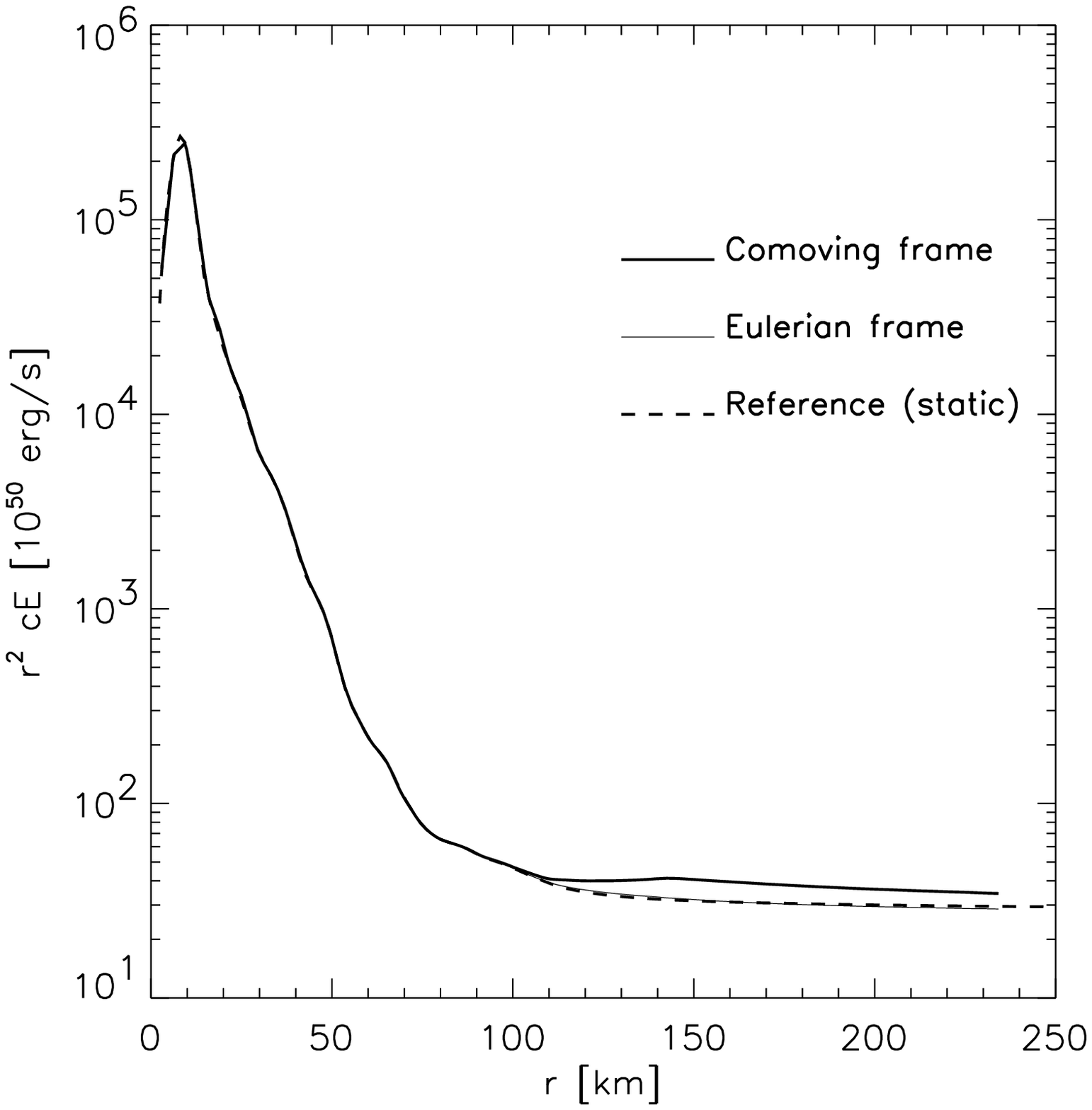} &
\put(0.9,0.3){{\Large\bf b}}
\epsfxsize=8.8cm \epsfclipon \epsffile{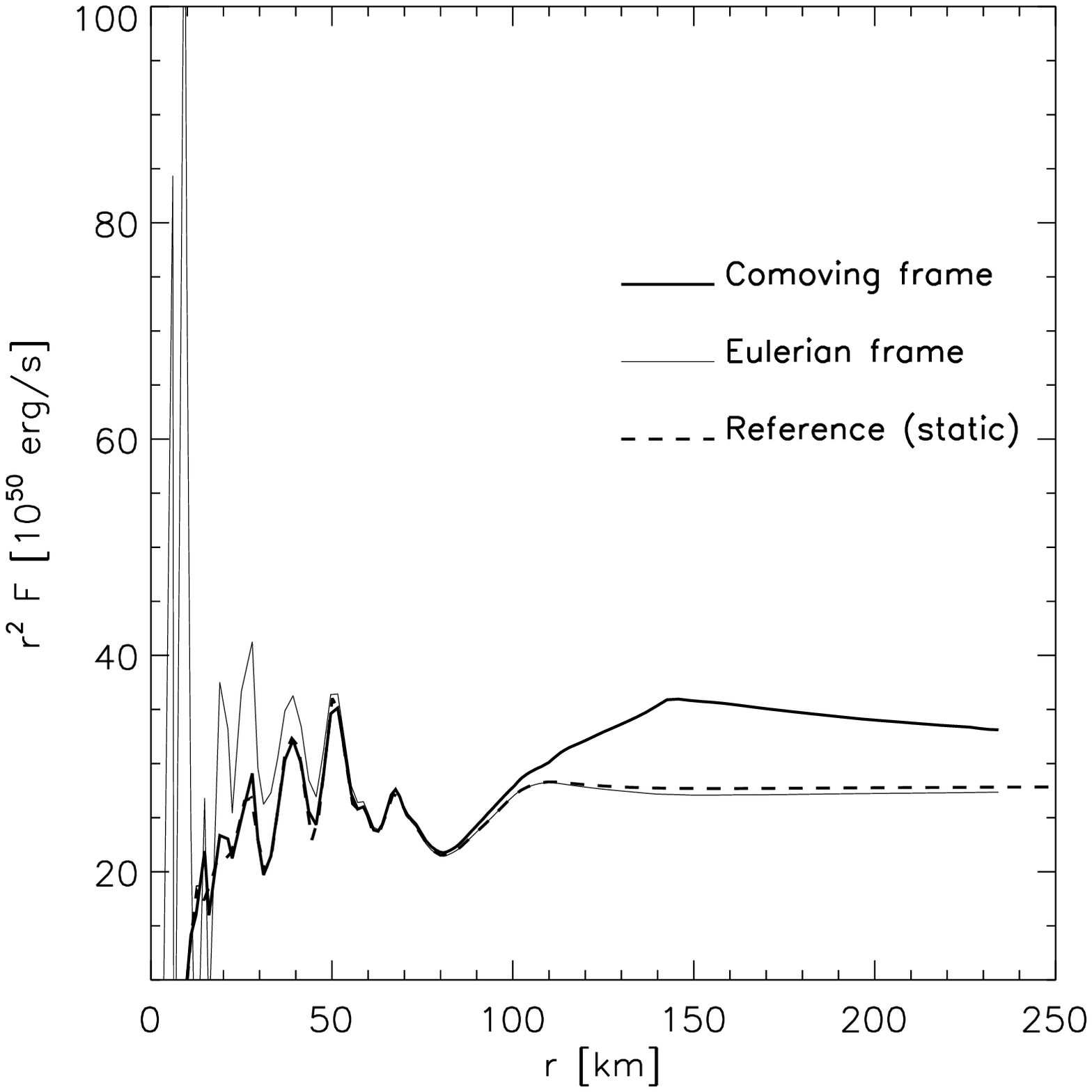} 
 \end{tabular}
\caption[]{Comparison of the stationary-state solutions of the comoving
  frame transfer equation (bold lines) with the solutions after
  transformation  to
  the Eulerian frame (thin lines) and a reference calculation for a
  static stellar background ($v\equiv 0$; dashed lines). Panel
  {\bf a} shows the total energy density of neutrinos (times
  $r^2\,c$), Panel~{\bf b} displays the total energy flux density (times
  $r^2$).}
  \label{fig:BRU}
\end{figure*}

A rigorous comparison of the spectral distribution of the 
angular moment $J$ as calculated by our radiative transfer code
\reffigA{fig:MIH_spec}{cf.} with the results of \cite{mih80} is
difficult, since he shows only results for the
static case and the rather extreme case $\beta_\mathrm{max}=0.8$.
The latter atmosphere obviously cannot be modeled reasonably well
with our ${\cal O}(\beta)$-code.
Nevertheless, by inspection of Figs.~3 and 4 of \cite{mih80} one can
infer that the shapes of our spectra as displayed in
\reffigN{fig:MIH_spec} exhibit the correct qualitative dependence on the
expansion velocity. 

\paragraph{Frame dependence?}

As an important consistency check we verified that, to order
${\cal O}(v/c)$, the numerical solution of our implementation of the
comoving frame equations shows the correct transformation properties of
the stress-energy tensor of radiation (see Eq.\ref{eq:momentstrafo})
and no unphysical artifacts are caused by the choice of a moving
reference frame \cite[cf.~the critique of][]{lowmor99}. 

For this purpose results are produced for neutrino transport in a
thermally and hydrodynamically frozen post-bounce model of 
a supernova calculation by \cite{bru93}. This model can be characterized
as follows:  
Neutrinos are emitted from a hot and dense hydrostatic inner core with
a radius of $r\approx 110$~km. 
This inner core is surrounded by a supersonically infalling outer core,
which is effectively transparent to the radiation. 
Velocities in this outer atmosphere range from
$v(r=140~\mathrm{km})\approx -0.12 c$ to $v(r=250~\mathrm{km})\approx -0.08 c$.

The stress-energy tensor as derived from the stationary-state solution
of the \emph{comoving frame} transport equation (Eq.~\ref{eq:BTE_sr}) is
transformed to the Eulerian frame (i.e.~the inertial frame in which the
center of the star is at rest) and can then
be compared with the result of an independent calculation which
solves the transport equation directly in the Eulerian frame.

\reffigLN{fig:BRU} shows the components 
$E=4\pi/c\int_0^\infty \mathrm{d}\epsilon\, J$ and 
$F=4\pi\int_0^\infty \mathrm{d}\epsilon\, H$ of the
stress-energy tensor (scaled by $r^2 c$ and $r^2$, respectively). 
Due to the large inwardly 
directed velocities in the outer atmosphere, the neutrinos emitted by
the inner core are blueshifted for observers which
are locally comoving with the fluid flow (see the bold
lines in Fig.~\ref{fig:BRU}). 
When the stress-energy tensor is transformed to the 
Eulerian frame (Eq.\ref{eq:momentstrafo}), this effect obviously
disappears \reffigA{fig:BRU}{thin lines in}. 
Comparison with the stress-energy tensor obtained by solving the
transport equation directly in the Eulerian frame
\reffigA{fig:BRU}{dashed lines in} reveals good agreement for both
the energy density $E$ and the energy 
flux density $F$ in the rapidly moving atmosphere. This demonstrates
that no unphysical frame dependence is present in the calculations.
The fluctuations of the transformed energy flux density near the
center are caused by the term $\beta (J + K )$ in the expression for
$H^\mathrm{Eul}$ in Eq.~(\ref{eq:momentstrafo}). Because $J$ (and $K$)
are large compared with $H$ in the dense interior, even small
velocities lead to a significant ``advective contribution'' to
$H^\mathrm{Eul}$. 

Note that an Eulerian-frame calculation in general requires complicated
velocity-dependent transformations of the source terms on the rhs. of
the transport equation \cite[see][]{mihmih84}. This, however, is
unnecessary in the specific situation of the discussed model:
Because the region where most of the neutrinos are produced 
moves with low velocities, the source terms there can be
evaluated in the rest frame of the fluid. In the outer atmosphere on
the other hand, where the large velocities would require a careful
transformation, the source terms nearly vanish. This allowed us to
simply drop the velocity-dependent terms on the lhs. of
Eq.~(\ref{eq:BTE_sr}) in order to perform a calculation with results
that are in agreement with the Eulerian frame solution in the
low-density part of the star. 

\subsection{Core collapse and supernova simulations}\label{chap:test.sn2}

\subsubsection{Newtonian gravity} \label{chap:test.sn2.newt}

Our new neutrino hydrodynamics code has already been applied to 
dynamical supernova simulations \cite[]{ramjan00}.
Here we report some results of a Newtonian calculation of the 
collapse phase of a stellar iron core with mass
$M_\mathrm{Fe}=1.28~\msol$  
(plus the innermost $0.1~\msol$ of the silicon shell).
It is the core of a $15~\msol$ progenitor 
star \cite[Model ``s15s7b2'', ][]{woo99,heg00}.
We compare the results to a calculation published by \cite{brumez97}. 
The input physics, in particular the stellar model and the
high-density equation of state, were adopted from 
model ``B'' of \cite{brumez97}, with the exception that we do
not include neutrinos other than $\nue$. This,
however, does not make a difference during the collapse phase where the
electron degeneracy is so high that only $\nue$ can be produced in
significant numbers. 
%
\begin{figure}[!t]
\epsfxsize=8.8cm \epsfclipon \epsffile{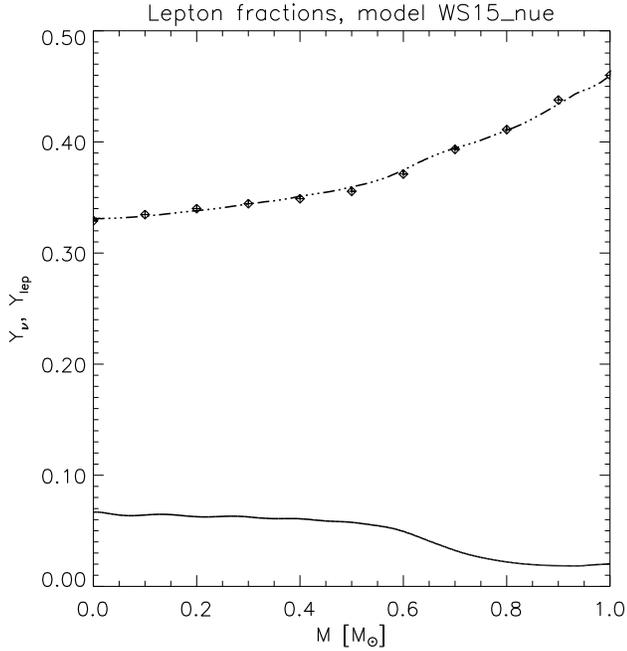} 
\caption[]{Profiles of the total lepton fraction $Y_\mathrm{lep}$
  (dash-dotted line) and the electron neutrino fraction $Y_\nue$
  (solid line) as a function of the enclosed mass at the time when the
  central density reaches $10^{14}$~\gcm. For comparison the results 
  of \citet[][ Fig.~5a]{brumez97} are plotted with
  diamonds.}\label{fig:N120_m} 
\end{figure}
%
The hydrodynamics was solved on a grid with 400 radial zones out 
to 20\,000~km, which were moved with the infalling matter during core
collapse. For the transport we used an Eulerian grid with 213
geometrically spaced radial zones, 233 tangent rays and 21 energy bins
geometrically distributed between 0 and 380 MeV the center of the
first zone being at 2 MeV. 

\begin{figure*}[!t]
 \begin{tabular}{ll}
\put(0.9,0.3){{\Large\bf a}}
\epsfxsize=8.8cm \epsfclipon \epsffile{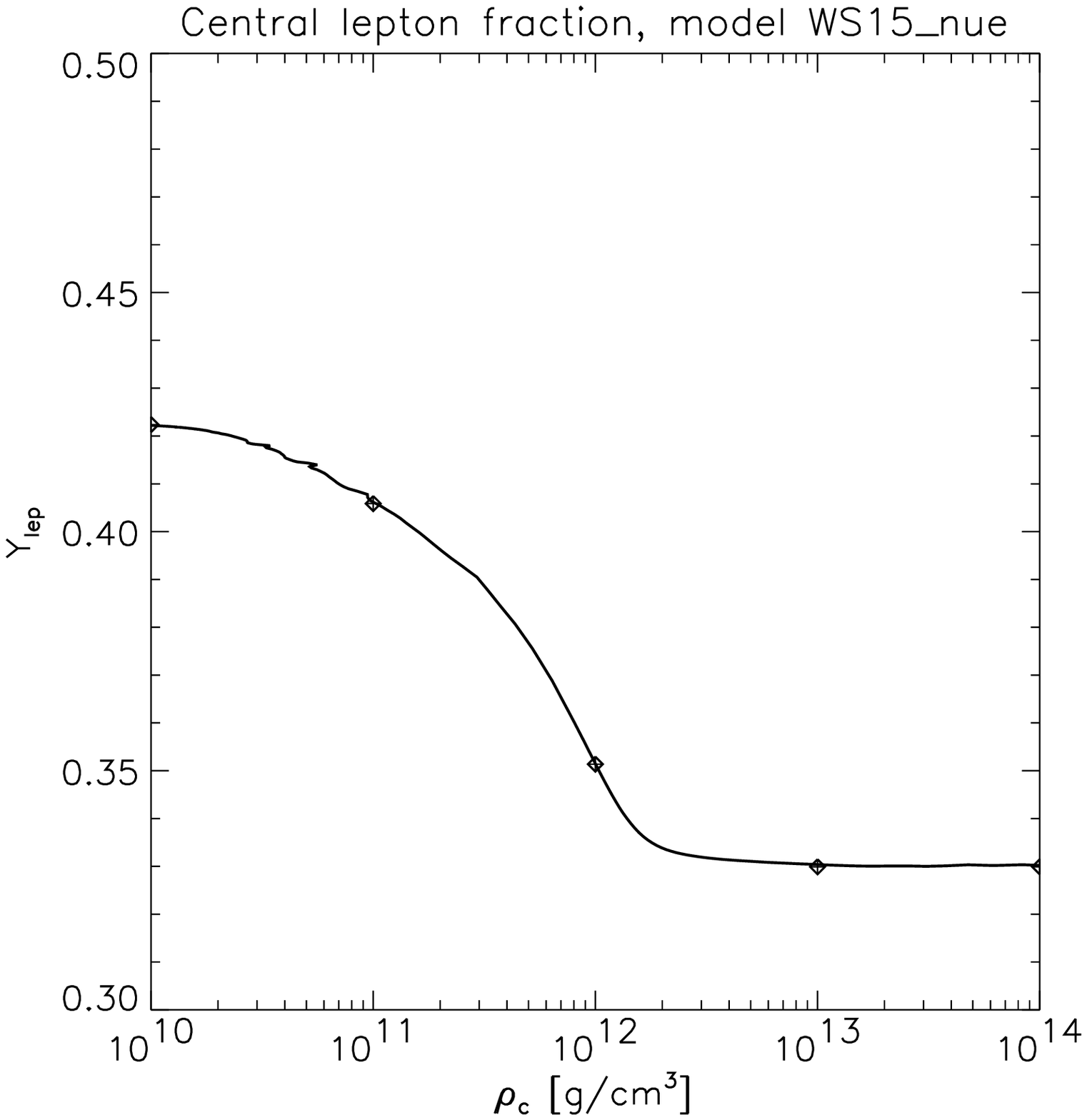}
\put(0.9,0.3){{\Large\bf b}}
\epsfxsize=8.8cm \epsfclipon \epsffile{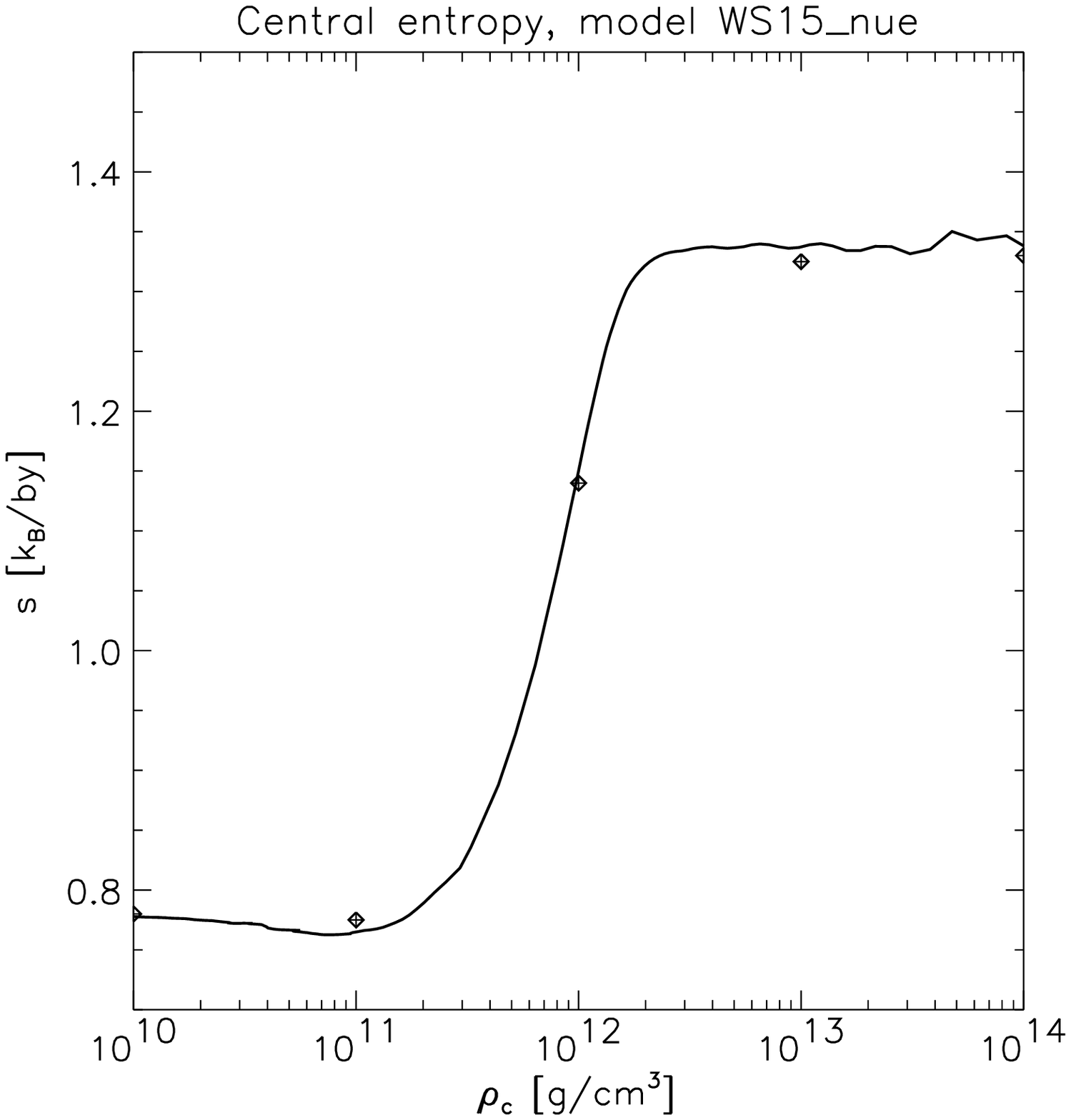} &
 \end{tabular}
\caption[]{Central lepton fraction 
  $Y_\mathrm{lep}\equ \ye+Y_\nue$ (Panel {\bf a}) and  
  central entropy per baryon ({\bf b}) as functions of the central
  density for our Newtonian core-collapse simulation of a $15~\msol$
  star (solid lines), compared with the
  results 
  (diamonds) obtained by \citet[][ Figs.~4a, 8a]{brumez97}.}
\label{fig:N120_t}
\end{figure*}

The total lepton fraction $Y_\mathrm{lep}\equ \ye+Y_\nue$
and the entropy per baryon at the center of the core as a function of
the central density are displayed in \reffigN{fig:N120_t}. 
The agreement with results of \cite{brumez97} is excellent.
Also the total lepton fraction as a function of enclosed mass matches
perfectly \reffig{fig:N120_m}.
Defining the shock formation point according to \cite{brumez97} as 
the mass coordinate where the entropy per baryon after 
core bounce first reaches a value of $s=3~\kb$ we find 
$M_\mathrm{S}\approx 0.62 M_\odot$.
This is about $0.03 M_\odot$ or 5\%
larger than the value given by \citet[][ Table~V]{brumez97}.

Judging our results, one should, of course, take into account 
that \cite{brumez97} employed
a multi-group flux-limited diffusion (MGFLD) approximation for treating
the neutrino transport, whereas our code solves the Boltzmann equation. 
\cite{mezbru93:nes,mezbru93:coll} performed a detailed comparison of
core-collapse simulations with MGFLD and with their Boltzmann solver
based on the $\mathrm{S}_\mathrm{N}$-method.
For the quantities presented here they also found good agreement of
their results throughout the inner core.

In the outer regions of the collapsing core, the situation is somewhat
different: 
\cite{mezbru93:nes,mezbru93:coll} demonstrated that a
number of quantities reveal significant deviations between
MGFLD and the Boltzmann transport ($\mathrm{S}_\mathrm{N}$-method). 
In particular, they obtained an electron neutrino
fraction which was by up to 20\% larger in the Boltzmann run.
For the total lepton fraction, however, the difference was only 1\%.
Our results \reffig{fig:N120_m} confirm the agreement of Boltzmann and
MGFLD results for the latter quantity.
A more detailed comparison between the variable
Eddington factor method and the $\mathrm{S}_\mathrm{N}$-method,
however, is desirable and currently underway.

\begin{figure*}[t]
 \begin{tabular}{lr}
\put(0.9,0.3){{\Large\bf a}}
\epsfxsize=8.8cm \epsfclipon \epsffile{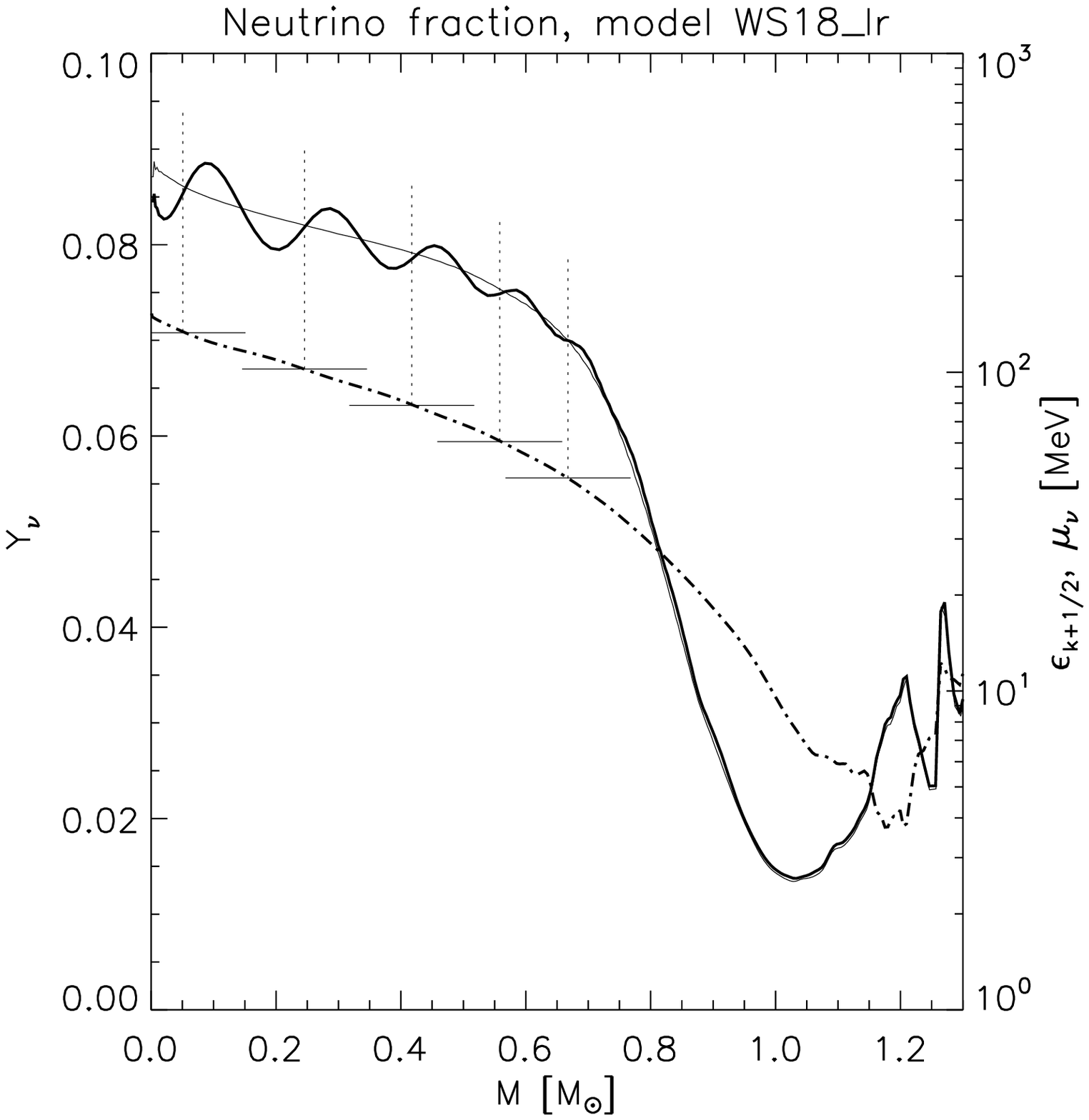} &
\put(0.9,0.3){{\Large\bf b}}
\epsfxsize=8,8cm \epsfclipon \epsffile{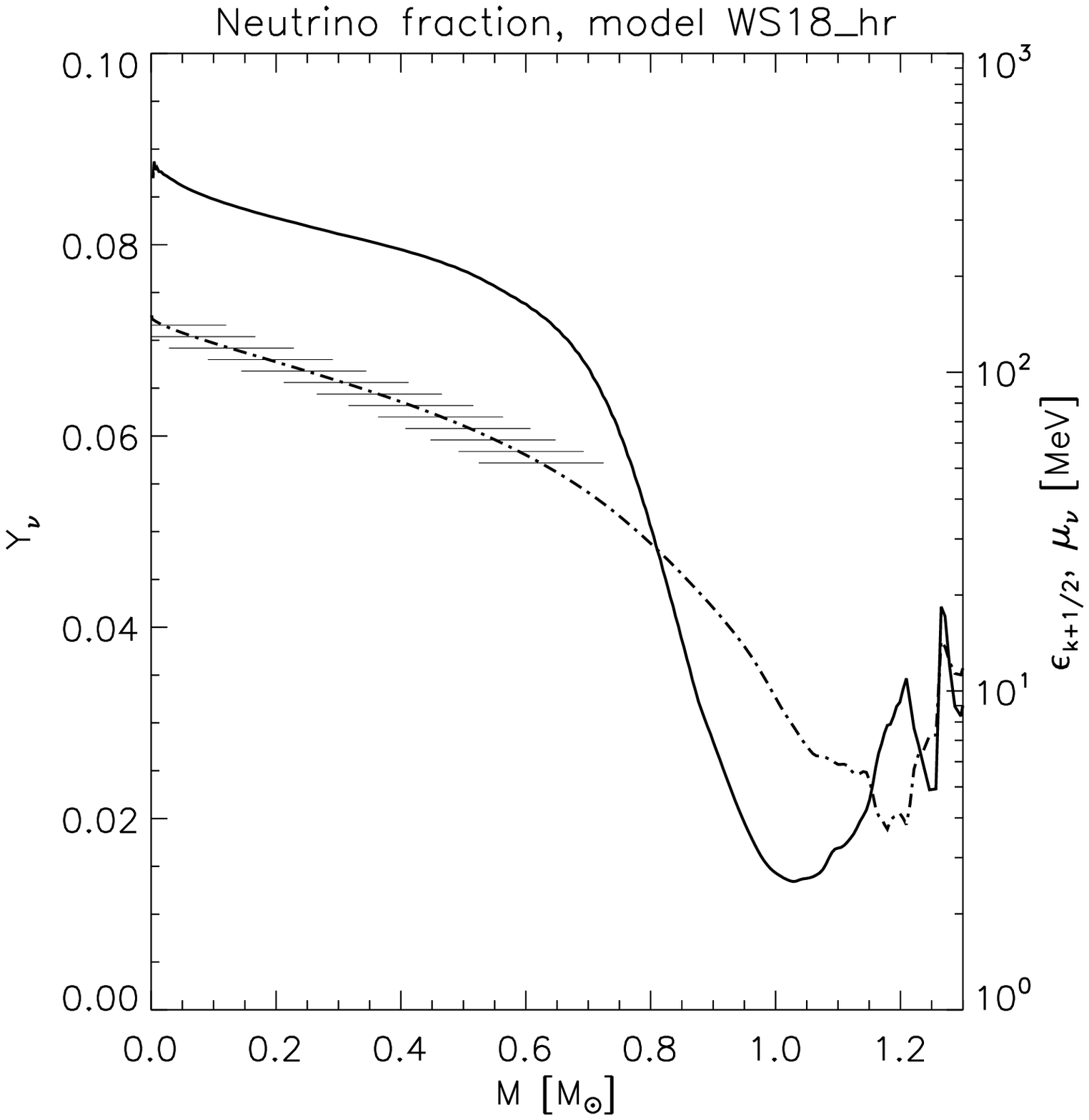} 
 \end{tabular}
\caption[]{Electron neutrino fraction $Y_\nue$ (bold solid line, scale
  given by the axis on the left side of the plots) 
  and electron neutrino equilibrium chemical potential
  $\mu_\nue^\mathrm{eq}$ (``Fermi surface'', 
  dash-dotted line, scale given by the axis on the right side) as 
  functions of enclosed mass at the time when the central density
  has reached $10^{14}$~\gcm (The plots show results of collapse
  simulations for a $18 M_\odot$ blue supergiant progenitor star;
  \citealp{wooheg97}). The spectral resolution of the energy
  grid is indicated by the spacing of the short horizontal lines
  (right ordinate), which correspond to the boundaries of the energy
  grid. 
  Panel {\bf a} displays results obtained with ``low'' spectral
  resolution (21 energy bins distributed geometrically between 0 and
  $380~\mev$), Panel {\bf b} shows the run with
  ``high'' spectral
  resolution (81 energy bins covering the same spectral range). For
  better comparison $Y_\nue$ of the high-resolution run is drawn as
  a thin solid line also in Panel {\bf a}.}
\label{fig:C640}
\end{figure*}

\paragraph{Spectral resolution:}

A few quantities exhibit spurious oscillations as a function of the
radial (mass-) coordinate and time in our core-collapse simulations.
Figure~\ref{fig:C640}a shows such features for 
the electron neutrino fraction in the mass shells $0 \le M \lesssim
0.6~\msol$ 
when the central density is $\rho_\mathrm{c} \simeq 10^{14}$~\gcm.
Similar structures are visible in the profile of the entropy per
baryon and in the neutrino luminosity as a function of radius (or
mass) once the central density becomes larger than $\rho_\mathrm{c}
\simeq 10^{13}$~\gcm. They can also be seen in the
central entropy as a function of the central density or time
\reffigA[b]{fig:N120_t}{cf.}.  
\cite{mezbru93:coll} found a similar effect in their 
core-collapse calculations. They identified the finite 
spectral resolution of the degenerate Fermi distribution of the
electron neutrinos 
to be the origin of these oscillations \cite[][ \S 3.3]{mezbru93:coll}.

For comparison, we have computed a collapse model with
a significantly improved spectral resolution: Instead of 21 bins we
used 81 energy zones to describe the energy spectrum between 0 and 380
MeV. 
A result of this simulation is displayed in \reffigN[b]{fig:C640}.
The neutrino fraction (and related quantities) are now perfectly
smooth functions of the mass coordinate.

A reasonable compromise between accuracy and computational
load could be achieved by using approximately 20--30 (geometrically
spaced) energy bins. This seems to be sufficient to adequately treat
the sharp Fermi surface of the degenerate distribution of electron
neutrinos that builds up during collapse. 
Even for less energy bins fluctuating
quantities follow the correct evolutionary trends \reffig[a]{fig:C640}.
Therefore we conclude with the remark that the presence of
oscillations in some transport quantities may be considered as
undesirable. The overall evolution of the collapsing core, its
deleptonization or the shock formation radius, however, were found not
to be sensitive to such ``noise''.

\bigskip

Also during the post-bounce evolution (for which the transport of
$\nuae$ was included), we determined no significant differences  
between the dynamical evolution of a model with 17 energy bins and
a model computed with 27 energy bins, using an otherwise identical
setup: At a 
time of 100~ms after bounce, the structure of the models, measured by
the radial positions of fluid elements with the same mass coordinate,
agreed within a relative error of $<1\%$. 
This is corroborated by the fact that 
for given snapshots from the dynamical evolution we found the neutrino
source terms
$Q_\mathrm{E}$ and also $Q_\mathrm{N}$ to be practically identical
in both calculations.

\paragraph{Radial resolution and size of the numerical time step:}

Varying the number of radial zones for the hydrodynamic and transport
parts of the code between 100 and 200 and using a comoving
hydrodynamic grid during the phase of stellar core collapse, we found
no significant changes of the results. The same holds true for the
post-bounce evolution where we switch to a fixed, Eulerian grid. 
The only exception to this insensitivity is the central density at
bounce, which shows variations of up to $10\%$, depending on the size
of the innermost zones of the hydrodynamic grid.

The choice of the radial grid, however, also influences the
size of the numerical time step. According to our experience the
latter has to be chosen very carefully. We give two examples here:
\begin{itemize}
\item[(a)] 
While neutrinos escape unhindered initially, the collapsing stellar
core becomes opaque when the density exceeds $\simeq 10^{12}~\gcm$,
and ``neutrino trapping'' sets in.
Since neutrinos are then carried along with the stellar plasma, the
subsequent, accelerating phase of the infall is adiabatic (the sum of
medium and neutrino entropy per baryon is constant:
$s+s_\nu=\text{const.}$) and the total lepton fraction of a fluid
element is conserved. Fulfilling the corresponding conditions, 
$\mathrm{D}Y_\mathrm{lep}/\mathrm{D}t=0$ and 
$\mathrm{D}(s+s_\nu)/\mathrm{D}t=0$, numerically on an Eulerian grid
requires a sufficiently accurate treatment of the advection of energy
and leptons. We found that treating the advection terms in the
neutrino transport at first order accuracy in time (setting $\zeta=1$
in Eq.~\ref{eq:time_interpolation}) is not sufficient and causes
deviations from $Y_\mathrm{lep}=\mathrm{const.}$ and 
$s+s_\nu=\mathrm{const.}$
at a level of $\gtrsim 10\%$. This problem can be diminished when the
transport time step is reduced significantly below the limits
mentioned in Sect.~\ref{chap:transp.rhd}. The time step reduction and
the corresponding increase
in computer time, however, can be avoided by handling the advection 
``second-order'' accurate in time, chosing $\zeta\approx 0.5$ in
Eq.~(\ref{eq:time_interpolation}). As visible from 
Fig.~\ref{fig:N120_t}, this yields  very good results (note that the
entropy of neutrinos, $s_\nu$, is small compared to the plasma entropy
$s$, and essentially constant after neutrino trapping).

\item[(b)]
Another problem occurred during the long-time evolution of the forming
neutron star after core bounce. For too large a value of the
hydrodynamic time step, we discovered an artificial inflation of the
mass shells that are located in the region of a steep density gradient
below the neutrinosphere. This effect was found not to be caused by
insufficient energy resolution of the neutrino transport nor does it
directly depend on the employed number of radial zones. Varying the
latter by a factor of two and testing $17$ and $25$ energy bins in the
interval between $0$ and $380~\mev$, we found that the agreement of
our transport results was better than $1\%$ and thus not worse than
the uncertainties due to the interpolation of the stellar
background. 
Since the post-bounce evolution is usually calculated 
with the same grids for the hydrodynamics and the transport, possible
errors associated with the mapping of quantities between the grids
can also be excluded as the source of the problem.
Moreover, even if different radial grids are used as in the model
published by \cite{ramjan00}, an expansion of the outer layers of the
forming neutron star is not necessarily a consequence.
The quality of our calculations, however, turned out to be dependent
on the size of
the time step of the hydrodynamics code. Our tests showed that a
reduction of the time step was necessary to prevent the artificial
expansion of the quasi-hydrostatic structure. We were so far unable to
exactly locate the origin of this numerical problem, but presume that
it might be connected with the accuracy of the implementation of the
neutrino momentum term in the hydrodynamics code. Good results,
however, were obtained, when the length of the time step was guided by
the typical time scale set by the neutrino momentum transfer to the
stellar medium, $\Delta t_\mathrm{Hyd}\simeq \rho\,v/Q_\mathrm{M}$ 
(cf.~Eq.~\ref{eq:hydro.v}).

\end{itemize}

\subsubsection{Approximate relativistic core collapse}

We also tried to assess the quality of our approximate treatment of
general 
relativity (GR) in the equations of hydrodynamics and neutrino transport
(cf.~Sect.~\ref{chap:transp.gr_rhd}).
For this purpose we performed core-collapse
simulations for the $15~\msol$ progenitor star ``s15s7b2'' and
compared characteristic quantities with the information given for a MGFLD
simulation of the same star in a recent paper by \cite{brunis01}, and
for a Boltzmann simulation by \cite{liemes01}.

For the hydrodynamics we used 400 radial zones out 
to 10\,000~km, which were moved with the infalling matter during core
collapse. The transport was solved on an Eulerian grid with 200
geometrically spaced radial zones, 220 tangent rays and 17 energy bins
geometrically distributed between 0 and 380 MeV the center of the
first zone being at 2 MeV. 

At bounce, which occurs at $t_\mathrm{b}^\mathrm{GR}=177.7~\ms$ 
($t_\mathrm{b}^\mathrm{Newt}=196.1~\ms$)
after the start of the calculation, the central density reaches a
maximum of $\rho^\mathrm{GR}_\mathrm{c}=4.8\times 10^{14}~\gcm$ 
($\rho^\mathrm{Newt}_\mathrm{c}=3.35\times 10^{14}~\gcm$) in our GR
model (for comparison, the corresponding values for the Newtonian runs
are given in brackets). The hydrodynamic shock forms at a mass
coordinate of $M_\mathrm{S}^\mathrm{GR}=0.51~\msol$ 
($M_\mathrm{S}^\mathrm{Newt}=0.62~\msol$). 
For the same stellar model, \cite{brunis01} found a bounce density of 
$\rho^\mathrm{GR}_\mathrm{c}=4.23\times 10^{14}~\gcm$  for the general
relativistic simulation
($\rho^\mathrm{Newt}_\mathrm{c}=3.20\times 10^{14}~\gcm$).
The corresponding time of the bounce was $t_\mathrm{b}^\mathrm{GR}=182.9~\ms$ ($t_\mathrm{b}^\mathrm{Newt}=201.3~\ms$). 
\cite{liemes01} located the shock formation at a mass coordinate of
$M_\mathrm{S}^\mathrm{GR}=0.53~\msol$ ($M_\mathrm{S}^\mathrm{Newt}=0.65~\msol$). 

Since the collapse time is determined by the slow initial phase of the
contraction, which is sensitive to the
initial conditions and the details of the numerical treatment at the
start of the calculation (Liebend\"orfer, personal communication),
it may be better to use the difference between the GR and the
Newtonian simulations, instead of comparing absolute times at bounce.
Our value of
$t_\mathrm{b}^\mathrm{Newt}-t_\mathrm{b}^\mathrm{GR}=18.4~\ms$ is very
close to the result of \cite{brunis01}:
$t_\mathrm{b}^\mathrm{Newt}-t_\mathrm{b}^\mathrm{GR}=21.6~\ms$.
A part of the remaining difference may be attributed to the fact
that \cite{brunis01} employed a MGFLD 
approximation of the neutrino transport with some smaller
differences also in the input physics (e.g., \citealp{brunis01} as 
well as
\citealp{liemes01} did not take into account ion-ion correlations for the
coherent scattering, but we did). 

At the level of comparison which is possible on grounds of
published numbers we conclude that the agreement between
our approximate treatment of GR and the relativistic core collapse
seems to be reasonably good.

\section{Summary and conclusions}\label{chap:concl}

We have presented a detailed description of the numerical
implementation of our new neutrino transport code and its
coupling to a hydrodynamics program which allows simulations
in spherical symmetry as well as in two or three dimensions.
The transport code solves the energy and time dependent 
Boltzmann equation by a variable
Eddington factor technique. To this end a model Boltzmann 
equation is discretized along tangent rays and integrated 
for determining the variable Eddington factors which provide
the closure relations for the moment equations of neutrino
energy and momentum. The latter yield updated values of the
neutrino energy density and neutrino flux, which are fed 
back into the Boltzmann equation to handle the collision 
integral on its right hand side.

The system of Boltzmann equation and moment
equations together with the operator-splitted terms of lepton
number and energy exchange with the stellar background
(which influence the evolution of the thermodynamical 
quantities and the composition of the stellar medium and
thus the neutrino-matter interactions)
are iterated to convergence. Lepton number conservation is 
ensured by solving additional moment equations for neutrino
number density and number flux. 

The integration is performed with implicit time stepping,
which avoids rigorous limitations by the CFL condition
and ensures that equilibrium between neutrinos and stellar
medium can be established accurately and without oscillations 
despite of stiff neutrino absorption and emission terms.
The coupling of energy bins due to Doppler shifts and 
energy changing scattering processes is implemented in a
fully implicit way. When coupling the transport part to an 
explicit hydrodynamics program, as done in this work,
computational efficiency requires to have the option of using
different time step lengths for transport and hydrodynamics.
In addition, we found it advantageous to have implemented
the possibility of choosing different radial grids for both
components of the program, and to switch between Lagrangian 
and Eulerian coordinates dependent on the considered physical 
problem (although we always measure physical
quantities of the transport in the comoving frame of 
reference).

We have suggested an approximation for applying the code to
multi-dimensional supernova simulations which can
account for the fact that hydrodynamically unstable layers
develop in the collapsed stellar core. The variable Eddington
factor method offers the advantage that this can be done
numerically rather efficiently (concerning programming
effort as well as computational load). Although the 
neutrino moment equations for the radial transport are solved
independently in all angular zones of the spherical grid of the 
multi-dimensional hydrodynamical simulation, 
the set of moment equations is closed by a variable Eddington factor 
which is computed only once for an angularly averaged stellar
background. 

This approximation saves a significant 
amount of computer time compared to truely multi-dimensional
transport, and yields a code structure which can easily be
implemented on parallel computers. The treatment is
constrained to radial transport and thus   
neglects lateral propagation of neutrinos.
Its applicability is therefore limited to physical
problems where no global asphericities occur. It must
also be considered just as a first, approximate step towards a
really multi-dimensional transport in convective layers
inside the neutrinosphere. The approach, however, should be 
perfectly suitable to treat anisotropies associated with
convective overturn in the
neutrino-heated region between the neutron star and the
supernova shock. In the latter region neutrinos are only
loosely coupled to the stellar medium (the optical depth
is typically only 0.05--0.2). Therefore the energy
and lepton number exchange between neutrinos and the stellar
medium depends on the presence of anisotropies and
inhomogeneities, whereas the transport of neutrinos is
essentially unaffected by non-radial motions of the stellar
plasma.

In preliminary multi-dimensional simulations of transport in
convecting neutron stars we have convinced ourselves that this method
guarantees good numerical convergence, because the variable
Eddington factor as a normalized quantity depends only
weakly on lateral variations in the stellar medium. The 
method ensures global energy conservation and enables local
thermodynamical equilibrium between stellar medium and 
neutrinos to be established when the optical depth is
sufficiently high. We therefore think that the proposed
approximation is practicable for neutrino transport
in multi-dimensional supernova simulations before fully
multi-dimensional transport becomes feasible. The latter
is certainly a major challenge for the years to come.

Our transport code exists in a ${\cal O}(v/c)$ version for
nonrelativistically moving media, and in a general relativistic
version. We have not yet coupled it to a hydrodynamics 
program in full general relativity. Instead, we performed
calculations of approximate relativistic core collapse, where
corrections to the Newtonian gravitational potential were included and 
only the redshift and time dilation effects were retained in the
transport equations (this means that space-time is considered 
to be flat). The results for stellar core collapse compared
with published fully relativistic calculations 
(with Boltzmann neutrino transport by Liebend\"orfer et 
al.~2001 and with multi-group flux-limited diffusion by 
Bruenn et al.~2001) are 
encouraging and suggest that this approximation works 
reasonably well and accounts for the most important effects
of relativistic gravity as long as one does not get very
close to the limit of black hole formation. 

We have presented results for a variety of idealized, partly
analytically solvable test calculations (in spherical
symmetry) which demonstrate the 
numerical efficiency and accuracy of our neutrino transport 
code. The ``neutrino radiation hydrodynamics'' was verified
by core-collapse and post-bounce simulations for cases where
published results were available and could reasonably well 
serve for a comparison 
(a direct and more detailed comparion with Boltzmann
calculations using the S$_{\mathrm N}$ method by
M.~Liebend\"orfer and A.~Mezzacappa is in progress).

Tests for a number of static model atmospheres showed that
the treatment of the angular dependence of the neutrino
transport and phase space distribution can be 
handled accurately by the variable Eddington factor method.
This holds for moderately strong general relativistic
problems, too, even if ray bending effects are neglected in
the determination of the variable Eddington factor. 
The numerical quality of the handling of the energy dependence
of the transport was also checked by considering background
models with stationary fluid flow. We included a model with
mildly relativistic motion (up to $v/c = 0.5$) and found 
that the code produces accurate and physically 
consistent results although we employ
an ${\cal O}(v/c)$ approximation to the special relativistic,
comoving frame radiative transfer equation. Also the omission
of some velocity-dependent terms in the model Boltzmann 
equation for determining the variable Eddington factor 
turned out not to be harmful in this respect. Closing the 
radiation moment equations by a variable Eddington factor
seems to be remarkably robust against approximations in the 
treatment of the Boltzmann equation.

The tests have also demonstrated that our code performs very 
efficiently.
Only a few iterations (typically between 3 and 7) are 
needed for obtaining a converged solution of the system of
Boltzmann equation and moment equations even in cases where
scattering rates dominate neutrino absorption. The calculation 
of the formal solution of the Boltzmann equation, from which the
variable Eddington factor is derived, turned out to consume
only 10--20\% of the computer time. The major computational 
load comes from the inversion of the set of moment equations.
The latter have a reduced dimensionality relative to the 
Boltzmann equation, because the dependence on the angular
direction of the radiation propagation was removed by
carrying out the angular integration. This advantage, however, 
has to be payed for by the iteration procedure with the 
Boltzmann equation. 

Using Boltzmann solvers for the neutrino transport means
a new level of sophistication 
in hydrodynamical simulations of supernova core
collapse and post-bounce evolution. It permits a
technical handling of the transport which for the first 
time is more accurate than the standard treatment of 
neutrino-matter interactions. The latter includes a number 
of approximations and simplifications which should be 
replaced by a better description, for example the detailed
reaction kinematics of neutrino-nucleon interactions,
phase space blocking of nucleons, effects due to 
weak magnetism in neutrino-nucleon reactions, or 
nucleon correlations in the dense medium of the neutron
star \cite[]{redpra98,redpra99,bursaw99,hor02}.

Boltzmann calculations will not only remove 
imponderabilities associated with the use of approximate
methods for the neutrino transport in hydrodynamical supernova
models. They will also allow for much more reliable predictions
of the temporal, spectral and flavour properties of the 
neutrino signal from supernovae and newly formed neutron
stars. This is an indispensable requirement for the 
interpretation of a future measurement of neutrinos from
a Galactic supernova.

\begin{acknowledgements}
We thank our referee, A.~Mezzacappa, for insightful comments which 
helped us improving the original manuscript.
We are grateful to K.~Kifonidis, T.~Plewa and E.~M\"uller 
for the latest version of the PROMETHEUS code, to K.~Kifonidis
for contributing a matrix solver for the transport on parallel
computer platforms, and to R.~Buras for
implementing the three-flavour version of the code and
coining subroutines to calculate the neutrino pair and
bremsstrahlung rates.
We thank S.~Bruenn, R.~Fischer, W.~Keil, M.~Liebend\"orfer and 
G.~Raffelt for helpful conversations, and M.~Liebend\"orfer for 
providing us with data of his simulations.
The Institute for Nuclear Theory at the University 
of Washington is gratefully acknowledged for its hospitality and the
Department of Energy for support during a visit of the Summer Program
on Neutron Stars. 
This work was also supported by the Sonderforschungsbereich~375 on
``Astroparticle Physics'' of the Deutsche Forschungsgemeinschaft.
The computations were performed on the NEC SX-5/3C of
the Rechenzentrum Garching and on the CRAY T90 of the John 
von Neumann Institute for Computing (NIC) in J\"ulich.
\end{acknowledgements}

\begin{appendix}

\section{Neutrino opacities}\label{appx:opa}

In this appendix we shall describe the neutrino-matter interactions
that are included in the current version of our neutrino transport
code for supernova simulations. We shall focus on aspects which are
specific and important for their numerical handling and will present
final rate expressions in the form used in our code and with a
consistent notation.

\begin{table*}[!t]
\centerline{
\begin{tabular}{|rclll|}
\hline 
Reaction  & & & Rate described in & Reference \\
\hline\hline
$\nu  \mathrm{e}^{\pm}$ &$ \rightleftharpoons $ & $\nu \mathrm{e}^{\pm}$   &
Sects.~\ref{appx:opa.scatt}, \ref{appx:opa.specifics.nes}                & 
\cite{mezbru93:nes,cer94}  \\
%
$\nu  A$       &$ \rightleftharpoons $ & $\nu A $        & 
Sects.~\ref{appx:opa.scatt}, \ref{appx:opa.specifics.scat_nuclei}        &
 \cite{hor97,brumez97}               \\
%
$\nu  \mathrm{N}$     &$ \rightleftharpoons $ & $\nu \mathrm{N} $      & 
Sects.~\ref{appx:opa.scatt}, \ref{appx:opa.specifics.abs_free_nuc}       & 
\cite{bru85,mezbru93:coll}                              \\
%
$\nue \mathrm{n}$       &$ \rightleftharpoons $ & $\mathrm{e}^- \mathrm{p} $        & 
Sects.~\ref{appx:opa.absemm}, \ref{appx:opa.specifics.abs_free_nuc}      & 
\cite{bru85,mezbru93:coll}          \\
%
$\nuae \mathrm{p}$       &$ \rightleftharpoons $ & $\mathrm{e}^+ \mathrm{n} $        & 
Sects.~\ref{appx:opa.absemm}, \ref{appx:opa.specifics.abs_free_nuc}      &
 \cite{bru85,mezbru93:coll}          \\
%
$\nue A'$       &$ \rightleftharpoons $ & $\mathrm{e}^- A $      & 
Sects.~\ref{appx:opa.absemm}, \ref{appx:opa.specifics.abs_nuclei}        & 
\cite{bru85,mezbru93:coll}                              \\
%
$\nu\bar\nu$   &$ \rightleftharpoons $ & $\mathrm{e}^- \mathrm{e}^+ $      & 
Sects.~\ref{appx:opa.pairs}, \ref{appx:opa.specifics.pairs}              &
 \cite{bru85,ponmir98}                              \\
%
$\nu\bar\nu \,\mathrm{N}\mathrm{N}$  &$ \rightleftharpoons $ & $\mathrm{N}\mathrm{N} $      & 
Sects.~\ref{appx:opa.pairs}, \ref{appx:opa.specifics.brems}  &
\cite{hanraf98}                              \\
\hline
\end{tabular}
}
\caption[]{Overview of all neutrino-matter interactions currently
  implemented in the code. 
  For each process we list the sections where we summarize fundamental
  aspects of the
  calculation of the corresponding rate and give details of its
  numerical implementation.
  The references point to papers where more information can be found
  about the approximations employed in the rate calculations.
  In the first column the symbol $\nu$ represents any of the neutrinos
  $\nue,\nuae,\nu_\mu,\bar\nu_\mu,\nu_\tau,\bar\nu_\tau$, the
  symbols $\mathrm{e}^-$, $\mathrm{e}^+$, $\mathrm{n}$, $\mathrm{p}$ and $A$ denote electrons, positrons,
  free neutrons and protons, and heavy nuclei, respectively, the
  symbol $\mathrm{N}$ means $\mathrm{n}$ or $\mathrm{p}$.}
\label{tab:reactions}
\end{table*}

\subsection{Basic considerations}

In the following we discuss the different  neutrino-matter interaction
processes which contribute to the source term 
(``collision integral'') on the rhs. of the Boltzmann
equation: 
\begin{equation}\label{eq:nolabel99}
\frac{1}{c}\frac{\partial}{\partial t}\,f + 
\mu\frac{\partial}{\partial r}\,f + 
\frac{1-\mu^2}{r}\frac{\partial}{\partial \mu}\,f  = 
\sum_{\{I\}} B_I
\,.
\end{equation}
The sum on the rhs.~of this equation runs over all
considered interaction processes 
that can change the distribution function of a particular type of
neutrino. For simplicity we have written down the Boltzmann
equation for static media, here.
In the general case of a moving stellar fluid, velocity-dependent
terms have to appear on the lhs. of Eq.~(\ref{eq:nolabel99}) when the
neutrino quantities and the neutrino-matter interaction rates are 
evaluated in the local rest frame of the fluid.
The phase-space distribution function $f$ is
related to the 
specific intensity ${\cal I}$ used in the main body of this paper by 
${\cal I}=c/(2\pi\hbar c)^3\cdot\epsilon^3\,f$.
Accordingly, the source term on the rhs. of the Boltzmann equation,
Eq.~(\ref{eq:BTE_sr}), and the corresponding moment equations for
neutrino number and 
neutrino energy or momentum (Eqs.~\ref{eq:JN_sr}, \ref{eq:HN_sr},
\ref{eq:J_sr}, \ref{eq:H_sr}) can be calculated from $B_I$ as
\begin{eqnarray}
{C}(\epsilon,\mu)&\equ&
\frac{c}{(2\pi\hbar c)^3}\epsilon^3 \cdot
\sum_{\{I\}} B_{I}(\epsilon,\mu) \,, 
\nonumber \\
{C^{(k)}}(\epsilon)&\equ&
\frac{c}{(2\pi\hbar c)^3}\epsilon^3 \cdot 
\sum_{\{I\}} B^{(k)}_{I}(\epsilon) \,, 
\label{eq:sourcet_I} \\
{\mathcal{C}^{(k)}}(\epsilon)&\equ&
\frac{c}{(2\pi\hbar c)^3} \epsilon^2 \cdot
\sum_{\{I\}} B^{(k)}_{I}(\epsilon) \,,\nonumber
\end{eqnarray}
where $B^{(k)}_{I}(\epsilon)\equ
1/2\int_{-1}^{+1}\mathrm{d}\mu\,\mu^k B_I(\epsilon,\mu)$.
Here and in the following, the dependence of quantities on the 
space-time coordinates $(t,r)$ is suppressed in the notation.
Throughout Appendix~\ref{appx:opa} the temperature $T$ is measured
in units of energy. 

\subsubsection{Neutrino absorption and emission}\label{appx:opa.absemm}

The rate of change (modulo a factor of $1/c$) of the neutrino
distribution function due to absorption and emission processes
is given by \cite[see][]{bru85}
\begin{equation}\label{eq:B_AE_bruenn}
B_\mathrm{AE}(\epsilon,\mu)=
j(\epsilon)[1-f(\epsilon,\mu)]-f(\epsilon,\mu)/\lambda(\epsilon)
\,,
\end{equation}
where $j$ denotes the emissivity and $\lambda$ is the  mean free
path for neutrino absorption. 
The factor $(1-f)$ accounts for fermion phase space
blocking effects of neutrinos.
Using the Kirchhoff-Planck relation (``detailed balance''), and
introducing the absorption opacity corrected for stimulated absorption,
\begin{equation}\label{eq:kapstar}
\kappa_\mathrm{a}^*\equ \frac{1}{1-f^\mathrm{eq}}\cdot \frac{1}{\lambda}=
j+\frac{1}{\lambda}
\,,
\end{equation}
Eq.~(\ref{eq:B_AE_bruenn}) can be rewritten as
\begin{equation}\label{eq:B_AE}
B_\mathrm{AE}(\epsilon,\mu)=
\kappa_\mathrm{a}^*(\epsilon)[f^\mathrm{eq}(\epsilon)-f(\epsilon,\mu)]
\,.
\end{equation}
In Eq.~(\ref{eq:B_AE}) it becomes evident that the source 
term drives the neutrino distribution function $f(\epsilon,\mu)$
towards its equilibrium value  
$f^\mathrm{eq}(\epsilon)=(1+\exp{[(\epsilon-\mu_\nu^\mathrm{eq})/T]})^{-1}$,
where $\mu_\nu^\mathrm{eq}$ denotes the chemical potential for
neutrinos in thermodynamic equilibrium with the stellar medium. 
In equilibrium, the source term vanishes in accordance with the
requirement of detailed balance.

\subsubsection{Scattering}\label{appx:opa.scatt}

\paragraph{Reduction of the collision integral:}

The rate of change of the neutrino
distribution function due to 
scattering of neutrinos off some target particles is 
given by the collision integral \cite[cf.][ Eqs.~2.1, 2.3]{cer94}
\begin{eqnarray}
B_\mathrm{S}(\vec{q})&=&
\int
\frac{\dvol{\vec{q'}}}{(2\pi)^3}\cdot\nonumber \\
&&[f'_\nu(1-f_\nu)\widetilde{R}^\mathrm{in}(\vec{q},\vec{q'})
  -f_\nu(1-f'_\nu)\widetilde{R}^\mathrm{out}(\vec{q},\vec{q'})]
\,,\nonumber \\
\label{eq:S_collkernel}
\end{eqnarray}
where $f'_\nu$ depends on $\vec{q'}$ and
the scattering kernels $\widetilde{R}^\mathrm{in}$ and
$\widetilde{R}^\mathrm{out}$ are defined as the following phase space
integrals over products of the transition rate ${\cal R}$ and
the Fermi distribution functions $F_\mathrm{T}$ of the target particles:
\begin{eqnarray}
\widetilde{R}^\mathrm{in}(\vec{q},\vec{q'})&=&\frac{2}{c\cdot(2\pi \hbar c)^3}\cdot
\nonumber \\
&&\int \frac{\dvol{p'}}{(2\pi)^3}\frac{\dvol{p}}{(2\pi)^3}
(1-F_\mathrm{T})F'_\mathrm{T}\,
     {\cal R}(\vec{p'},\vec{q'};\vec{p},\vec{q}) \,,\nonumber \\
\widetilde{R}^\mathrm{out}(\vec{q},\vec{q'})&=&\frac{2}{c\cdot(2\pi \hbar c)^3}\cdot
\nonumber \\
&&\int \frac{\dvol{p'}}{(2\pi)^3}\frac{\dvol{p}}{(2\pi)^3}
(1-F'_\mathrm{T})F_\mathrm{T}\,
     {\cal R}(\vec{p},\vec{q};\vec{p'},\vec{q'})
\,.\nonumber \\
\label{eq:S_inoutkernel}
\end{eqnarray}
The vectors $\vec{q}$ and $\vec{q'}$ 
are the momenta of the ingoing and outgoing neutrinos, respectively,
and $\vec{p}$ and $\vec{p'}$ those of their scattering targets.

The scattering kernels are usually given as functions of the
energies $\epsilon$ and $\epsilon'$ of the ingoing and outgoing neutrino 
and the cosine 
$\omega=\mu\mu'+\sqrt{(1-\mu^2)(1-{\mu'}^2)}\cos(\varphi-\varphi')$ of 
the scattering angle, where $(\mu,\varphi)$ and $(\mu',\varphi')$ are
the corresponding momentum space coordinates.  
Expanding the scattering kernels 
($R^\mathrm{in/out}\equ 1/(2\pi c)^3\cdot\widetilde{R}^\mathrm{in/out}$)
in a Legendre series
\begin{equation}
  \label{eq:S_legexp}
R^\mathrm{in/out}(\epsilon,\epsilon',\omega)=
\sum_{l=0}^{\infty} \frac{2l+1}{2}\,\phi_l^\mathrm{in/out}(\epsilon,\epsilon')P_l(\omega)
\,,
\end{equation}
with the Legendre coefficients
\begin{equation}\label{eq:legcoef}
\phi_l^\mathrm{in/out}(\epsilon,\epsilon')=
\int_{-1}^{+1}\mathrm{d}\omega\, P_l(\omega){R}^\mathrm{in/out}(\epsilon,\epsilon',\omega)
\,,
\end{equation}
and applying the addition theorem for Legendre polynomials
$P_l(\omega)=P_l(\mu)P_l(\mu')+2\sum_{m=1}^{l}\frac{(l-m)!}{(l+m)!}P_l^m(\mu)P_l^m(\mu')\cos[m(\varphi-\varphi')\big]$ 
\cite[e.g.,][ $P_l^m$ are the associated Legendre
polynomials]{brosem91},  
the integral over $\varphi'$ in the collision integral can
be performed analytically \cite[]{yuebuc77,schsha82:2}. 
For use in our Boltzmann transport code, it is necessary to truncate
the Legendre expansion at some level.
Note that the orthogonality relation 
 $\int_{-1}^{+1}\dlin{\omega}P_l(\omega)P_{l'}(\omega)=\frac{2}{2l+1}\delta_{ll'}$ 
\cite[e.g.,][]{brosem91} implies that any truncation of the Legendre
 series 
$R^\mathrm{in/out}(\epsilon,\epsilon',\omega)$ still gives the 
exact integral value
$\int_{-1}^{+1}\mathrm{d}\omega\,{R}^{\rm
  in/out}(\epsilon,\epsilon',\omega)$, 
independent of the level of truncation $l_\mathrm{max}>0$. 

The collision integral and its first two angular moments finally
read\footnote{We assume spherical symmetry, which implies that the
  neutrino 
  distribution functions entering Eq.~(\ref{eq:S_collkernel}) are
  independent of the angle $\varphi'$.}:
\begin{eqnarray}
\lefteqn{
B_\mathrm{S}(\epsilon,\mu) = 
2\pi\,\int_0^\infty \mathrm{d}\epsilon'\, {\epsilon'}^2  } \nonumber \\
\Big\{(1-f_\nu) &\cdot &
\sum_{l=0}^{\infty}(2l+1)P_l(\mu)
    \phi_l^\mathrm{in}(\epsilon,\epsilon')L'_l(\epsilon')  \nonumber \\
-f_\nu &\cdot &
\sum_{l=0}^{\infty}(2l+1)P_l(\mu)
    \phi_l^\mathrm{out}(\epsilon,\epsilon')(\delta_{l0}-L'_l(\epsilon'))\Big\}
\,, \nonumber \\
\label{eq:B_scatt}
\end{eqnarray}
\begin{eqnarray}
\lefteqn{
B^{(0)}_\mathrm{S}(\epsilon) =
2\pi\,\int_0^\infty \mathrm{d}\epsilon'\, {\epsilon'}^2  }  \nonumber \\
& &\Big\{\sum_{l=0}^{\infty}(2l+1)(\delta_{l0}-L_l(\epsilon))
    \phi_l^\mathrm{in}(\epsilon,\epsilon')L'_l(\epsilon')  \nonumber \\
&-&\sum_{l=0}^{\infty}(2l+1)L_l(\epsilon)
    \phi_l^{\rm
      out}(\epsilon,\epsilon')(\delta_{l0}-L'_l(\epsilon'))\Big\} \,,
    \label{eq:B0_scatt} 
\end{eqnarray}
\begin{eqnarray}
\lefteqn{
B^{(1)}_\mathrm{S}(\epsilon) =
2\pi\,\int_0^\infty \mathrm{d}\epsilon'\, {\epsilon'}^2  } \nonumber \\
& &\Big\{\sum_{l=0}^{\infty}
   [\delta_{l1}-(l+1)L_{l+1}(\epsilon)-lL_{l-1}(\epsilon)]
    \phi_l^\mathrm{in}(\epsilon,\epsilon')L'_l(\epsilon')  \nonumber \\
&-&\sum_{l=0}^{\infty}
   [(l+1)L_{l+1}(\epsilon)+lL_{l-1}(\epsilon)]
    \phi_l^{\rm
      out}(\epsilon,\epsilon')(\delta_{l0}-L'_l(\epsilon'))\Big\}
    \,,\nonumber  \\
    \label{eq:B1_scatt}
\end{eqnarray}
where the ``Legendre moment'' $L_l\equ\frac{1}{2}\int_{-1}^{+1}{\rm
  d}\mu\,\,P_l(\mu) f(\mu)$ of order $l$ 
of the distribution function has been introduced. 
These Legendre moments 
can obviously be written as linear combinations 
of the angular moments 
$M_m\equ\frac{1}{2}\int_{-1}^{+1}\mathrm{d}\mu\,\mu^m f(\mu)$, which
  occur in the formulation of the transport equations used in our
code.  
Detailed balance requires 
\begin{equation}
\phi_l^\mathrm{in}(\epsilon,\epsilon')=
\mathrm{e}^{-(\epsilon-\epsilon')/T}\,\phi_l^\mathrm{out}(\epsilon,\epsilon')
\,.
\end{equation}
Note that exploiting the in-out invariance of the transition rate 
(and therefore
$\phi_l^\mathrm{out}(\epsilon,\epsilon')=\phi_l^\mathrm{in}(\epsilon',\epsilon)$; 
e.g.~\citealp{cer94}) leads to
\begin{equation}\label{eq:numbconsv}
\int_0^\infty\mathrm{d}\epsilon\,\epsilon^2 B^{(0)}_\mathrm{S}(\epsilon)=0,\quad \text{but} \quad
\int_0^\infty\mathrm{d}\epsilon\,\epsilon^3 B^{(0)}_\mathrm{S}(\epsilon)\ne
0
\,,
\end{equation}
which means that neutrino number is conserved in the scattering process,
whereas there is a nonvanishing energy exchange between neutrinos and
matter due to scatterings.

\paragraph{Isoenergetic scattering:}

If for a particular 
scattering process the energy transfer between neutrinos and
target particles 
can be neglected, we have 
$\phi_l^\mathrm{out}(\epsilon,\epsilon')=
 \phi_l^\mathrm{in} (\epsilon,\epsilon')=:
 \phi_l(\epsilon)\cdot\delta(\epsilon-\epsilon')$, and
the source terms given by Eqs.~(\ref{eq:B_scatt}, \ref{eq:B0_scatt},
\ref{eq:B1_scatt}) 
simplify to
\begin{eqnarray}
B_\mathrm{IS}(\epsilon,\mu)&=&  
2\pi\sum_{l=0}^{\infty}(2l+1)P_l(\mu)\,
\epsilon^2\phi_l(\epsilon)L_l(\epsilon)\nonumber\\
&-&2\pi \epsilon^2\phi_0(\epsilon)\,f_\nu \,, \label{eq:B_ISscatt}\\
B^{(0)}_\mathrm{IS}(\epsilon) &=& 0 \,,\label{eq:B0_ISscatt} \\
B^{(1)}_\mathrm{IS}(\epsilon) &=& 2\pi \,\epsilon^2
L_1(\epsilon)\cdot\big(\phi_1(\epsilon)-\phi_0(\epsilon)\big) \label{eq:B1_ISscatt}
\,.
\end{eqnarray}

\subsubsection{Pair processes}\label{appx:opa.pairs}

Applying the procedures outlined in Sect.~\ref{appx:opa.scatt}
to the collision integral for the process of thermal emission and
absorption of neutrino-antineutrino pairs by electron positron
pairs and related reactions (e.g., nucleon-nucleon bremsstrahlung), the  
corresponding source terms read \cite[see][]{bru85,ponmir98}:
\begin{eqnarray}
\lefteqn{
B_\mathrm{TP}(\epsilon,\mu) =
2\pi\int_0^\infty \mathrm{d}\epsilon'\, {\epsilon'}^2 } \nonumber   \\ 
\Big\{
(1-f_\nu)&\cdot& \phi_0^\mathrm{p}(\epsilon,\epsilon')
-\sum_{l=0}^{\infty}(2l+1)P_l(\mu)
    \phi_l^\mathrm{p}(\epsilon,\epsilon')\bar{L}_l(\epsilon') \nonumber   \\
+f_\nu &\cdot&\sum_{l=0}^{\infty}(2l+1)P_l(\mu)
     \phi_l^\mathrm{a}(\epsilon,\epsilon')\bar{L}_l(\epsilon')) 
\Big\}
\,, \label{eq:B_TP}
\end{eqnarray}

\begin{eqnarray}
\lefteqn{B^{(0)}_\mathrm{TP}(\epsilon) =
2\pi\int_0^\infty \mathrm{d}\epsilon'\, {\epsilon'}^2  
\Big\{
\phi_0^\mathrm{p}(\epsilon,\epsilon')\cdot
(1-L_0(\epsilon)-\bar{L}_0(\epsilon'))  }\nonumber \\
&+&\sum_{l=0}^{\infty}(2l+1)\phi_l^\mathrm{a}(\epsilon,\epsilon')
L_l(\epsilon)\bar{L}_l(\epsilon')
\Big\}
\,, \label{eq:B0_TP}
\end{eqnarray}
\begin{equation}\label{eq:B1_TP}
\begin{split}
B^{(1)}_\mathrm{TP}(\epsilon) =&
2\pi\int_0^\infty \mathrm{d}\epsilon'\, {\epsilon'}^2 \Big\{ 
    -\phi_0^\mathrm{p}(\epsilon,\epsilon')L_1(\epsilon)
    -\phi_1^\mathrm{p}(\epsilon,\epsilon')\bar{L}_1(\epsilon)\\
+\sum_{l=0}^{\infty}&(2l+1)\phi_l^\mathrm{a}(\epsilon,\epsilon')
[
(l+1)L_{l+1}(\epsilon)+l L_{l-1}(\epsilon)
]\bar{L}_l(\epsilon')
\Big\}\,.
\end{split}
\end{equation}
Here bars indicate quantities for antineutrinos.
The absorption and production kernels are related by 
\begin{equation}
\phi_l^\mathrm{a}(\epsilon,\epsilon')=
\left(1-\mathrm{e}^{(\epsilon+\epsilon')/T}\right)\cdot\phi_l^\mathrm{p}(\epsilon,\epsilon')
\end{equation}
in accordance with the requirement of detailed balance.

\subsection{Numerical implementation of various interaction processes}
\label{appx:opa.specifics}

In the following we describe the implementation of the various
neutrino-matter interactions into our transport scheme.
The expressions are mainly taken from the works of
\cite{tubsra75}, \cite{schsha82:2}, \cite{bru85}, and \cite{mezbru93:nes,mezbru93:coll}.

All average values of the source terms within individual energy bins
(cf.~Eq.~\ref{eq:freqdis1}) are approximated to zeroth 
order by assuming the integrand to be a piecewise constant function of
the neutrino energy $\epsilon$, and hence:
\begin{equation}\label{eq:freqdis}
B^{(k)}_{j+1/2}\equ
 B^{(k)}(\epsilon_{j+1/2})
\,.
\end{equation}
With the rest-mass energy of the electron, $m_\mathrm{e}c^2=0.511~\mev$, and the
characteristic cross section of weak interactions $\sigma_0\equ
4(m_\mathrm{e}c^2 G_{\rm F})^2/(\pi(\hbar c)^4)=1.761\cdot 10^{-44}~\mathrm{cm}^2$
($G_{\rm F}$ is Fermi's constant), we define:
\begin{equation}\label{eq:G_const}
  {\cal G}\equ \frac{\sigma_0}{4\,{m_\mathrm{e}}^2c^4}
\,.
\end{equation}

\subsubsection{Neutrino absorption, emission and scattering by free
  nucleons}\label{appx:opa.specifics.abs_free_nuc} 

In the standard description of neutrino-nucleon interactions 
\cite[see][]{tubsra75,bru85,mezbru93:coll}, 
many-body effects for the nucleons in the dense medium, energy
transfer between leptons and nucleons as well as nucleon thermal
motions are ignored.
The corresponding simplifications allow for a straightforward
implementation of the processes, using Eq.~(\ref{eq:B_AE}) for the
charged-current reactions and
Eqs.~(\ref{eq:B_ISscatt}--\ref{eq:B1_ISscatt}) for the neutral-current
scatterings, respectively. Expressions for
$\kappa_\mathrm{a}^*$ and the 
Legendre coefficients $\phi_0(\epsilon)$, $\phi_1(\epsilon)$ can be 
computed from the rates given by \cite{bru85} and \cite{mezbru93:coll}.

For absorption of electron neutrinos by free neutrons
($\nue + \mathrm{n} \rightarrow \mathrm{e}^- + \mathrm{p}$)
the final result for the opacity is:
\begin{multline}\label{eq:kappaa_enu}
\kappa_{\rm a}^*(\epsilon)=
{\cal G}\cdot(g_\mathrm{V}^2+3g_\mathrm{A}^2) 
\frac{1-F_{\mathrm{e}^-}(\epsilon+Q)}{1-f^{\rm eq}_\nue(\epsilon)}
\;\eta_\mathrm{np}   \\
\cdot (\epsilon+Q)\sqrt{(\epsilon+Q)^2-m_\mathrm{e}^2c^4}
\,.
\end{multline}
The opacity of the absorption of
electron antineutrinos by free protons 
($\nuae + \mathrm{p} \rightarrow \mathrm{e}^+ + \mathrm{n}$) is
\begin{multline}\label{eq:kappaa_eanu}
\kappa_{\rm a}^*(\epsilon)=
{\cal G}\cdot(g_\mathrm{V}^2+3g_\mathrm{A}^2) 
\frac{1-F_{\mathrm{e}^+}(\epsilon-Q)}{1-f^{\rm eq}_\nuae(\epsilon)}
  \;\eta_\mathrm{pn} \\
  \cdot(\epsilon-Q)\sqrt{(\epsilon-Q)^2-m_\mathrm{e}^2c^4}\,,
\end{multline}
if the neutrino energy $\epsilon\ge m_\mathrm{e}c^2+Q$, and 
$\kappa_{\rm a}^*(\epsilon)=0$, else.
Here, $Q\equ m_\mathrm{n}c^2-m_\mathrm{p}c^2$ denotes the difference of the
rest-mass energies of the neutron and the proton, and $F_{\mathrm{e}^\mp}$ 
are the Fermi distribution functions of electrons or positrons.
The constants $g_\mathrm{V}=1$ and $g_\mathrm{A}=1.254$ are the
nucleon form factors for the vector and axial vector currents, respectively.
The quantities $\eta_\mathrm{np}$ and  $\eta_\mathrm{pn}$  
take into account the effects of fermion blocking of the nucleons. 
Making the approximations mentioned above, in
particular ignoring the recoil of the nucleon, 
\citet[][ Eq.~C14; see also \citealp{redpra98}]{bru85} derived:
\begin{multline}\label{eq:etapn}
\eta_\mathrm{np}\equ 2\int\frac{\dvol{p}}{(2\pi\hbar
  c)^3}F_\mathrm{n}(\epsilon)[1-F_\mathrm{p}(\epsilon)] \\
=\frac{n_\mathrm{p}-n_\mathrm{n}}{\exp{\left[\psi_\mathrm{p}-\psi_\mathrm{n}\right]}-1}
\,,
\end{multline}
where $n_\mathrm{n}$, $n_\mathrm{p}$ are the number densities
of neutrons and protons, respectively.
The corresponding degeneracy parameters $\psi_\mathrm{n}$ and $\psi_\mathrm{p}$
are calculated by inverting the relation ($\mathrm{N}$ stands for $\mathrm{n}$ or $\mathrm{p}$)
\begin{equation}
n_\mathrm{N}=\frac{1}{2\pi^2}\left(\frac{2\,m_\mathrm{N} c^2\,T}{(\hbar
    c)^2}\right)^{3/2} F_{1/2}(\psi_\mathrm{N})
\,,
\end{equation}
where $F_{1/2}$ is the Fermi integral of index $1/2$ \cite[see,
e.g.,][ Sect.~2.2]{latswe91}. 
The quantity $\eta_\mathrm{pn}$ in Eq.~(\ref{eq:kappaa_eanu}) is defined
accordingly by exchanging the subscripts $n$ and $p$ in
Eq.~(\ref{eq:etapn}).  
In the nondegenerate
regime, where blocking is unimportant, one verifies the familiar limits
$\eta_\mathrm{pn}=n_\mathrm{p}$ and $\eta_\mathrm{np}=n_\mathrm{n}$.

\medskip

Scattering of neutrinos off free nucleons ($\mathrm{n}$ or $\mathrm{p}$) is mediated by neutral
currents only, which makes the distinction between neutrinos and
antineutrinos unneccessary.
Neglecting nucleon recoil and nucleon thermal motions, the
isoenergetic kernel obeys  
${R}_{\rm IS}(\epsilon,\omega)\propto (1+\omega)$ 
\cite[e.g.][]{bru85,mezbru93:coll}, which implies
$\phi_{l>1}(\epsilon)\equiv 0$.  
The non-vanishing Legendre coefficients read
\begin{alignat}{3}
\phi_0(\epsilon)&=
 \frac{{\cal G}}{8\pi}\cdot&
    \begin{cases}
    \eta_\mathrm{nn}[1+3g_\mathrm{A}^2]
    &(\text{for n})\,, \\
    4\eta_\mathrm{pp}[(C_{\rm V}-1)^2+\frac{3}{4}g_\mathrm{A}^2]
    &(\text{for p})\,,
    \end{cases}     \label{eq:NNS_phi0}\\
\phi_1(\epsilon)&=
 \frac{{\cal G}}{24\pi}\cdot&
    \begin{cases}
    \eta_\mathrm{nn}[1-g_\mathrm{A}^2]
    &(\text{for n})\,, \\
    4\eta_\mathrm{pp}[(C_{\rm V}-1)^2-\frac{1}{4}g_\mathrm{A}^2]
    &(\text{for p})\,.
    \end{cases}      \label{eq:NNS_phi1}
\end{alignat}
The factors 
\begin{equation}\label{eq:etapp}
\eta_\mathrm{nn/pp}\equ T\frac{\partial n_\mathrm{n/p}}{\partial \mu_\mathrm{n/p}}
\end{equation}
approximately account for nucleon final state
blocking \cite[]{bru85}, and 
$C_{\rm V}=1/2+2\sin^2\theta_{\rm W}$ with $\sin^2\theta_{\rm W}=0.23$
for the Weinberg angle.
In order to accurately reproduce the limits
$\eta_\mathrm{nn/pp}=n_\mathrm{n/p}$ for nondegenerate (and nonrelativistic),
and $\eta_\mathrm{nn/pp}=n_\mathrm{n/p}\cdot 3T/(2 E^{\rm F}_\mathrm{n/p})$
for degenerate (and nonrelativistic)
nucleons ($E^{\rm F}_\mathrm{n/p}$ is the Fermi energy) we prefer using 
the analytic interpolation formula proposed by \cite{mezbru93:coll}  
\begin{equation}
\eta_\mathrm{nn/pp}=
 n_\mathrm{n/p}\frac{\xi_\mathrm{n/p}}{\sqrt{1+\xi_\mathrm{n/p}^2}},\quad 
\text{with}\quad \xi_\mathrm{n/p}\equ
\frac{3T}{2 E^{\rm F}_\mathrm{n/p}}
\,,
\end{equation}
instead of a direct numerical evaluation of Eq.~(\ref{eq:etapp}).

\subsubsection{Neutrino absorption and emission by heavy nuclei}\label{appx:opa.specifics.abs_nuclei}

As in the case of charged-current reactions with free nucleons,
neutrino absorption and emission by heavy nuclei 
($\nue + A' \rightleftharpoons A + \mathrm{e}^-$) is implemented 
by using Eq.~(\ref{eq:B_AE}).
For calculating the opacity of this process 
we adopt the description employed by \cite{bru85} and
\cite{mezbru93:coll}:  
\begin{multline}
\kappa_{\rm a}^*(\epsilon)=
{\cal G}\cdot g_\mathrm{A}^2\, \frac{2}{7}N_{\rm p}(Z)N_{\rm h}(N)\, 
    \frac{F_{\mathrm{e}^-}(\epsilon+Q')}{f^{\rm eq}_\nue(\epsilon)}
    \, n_{A}  \\
\cdot(\epsilon+Q')
    \sqrt{(\epsilon+Q')^2-m_\mathrm{e}^2c^4}\,,
\end{multline}
if $\epsilon\ge m_\mathrm{e}c^2-Q'$, and $\kappa_{\rm a}^*(\epsilon)=0$ else. 
Here
$Q'\equ m_{A'}c^2-m_{A}c^2\approx \mu_\mathrm{n}c^2-\mu_\mathrm{p}c^2
+\Delta$ denotes the
energy difference between the excited state $A'=(N+1,Z-1)$ and the
ground state $A=(N,Z)$. 
Following \cite{bru85}, we set $\Delta=3$~MeV for all nuclei.
The internal structure of the nucleus with charge $Z$ and 
neutron number $N=A-Z$ is represented by the term
$2/7\,N_{\rm p}(Z)N_{\rm h}(N)$, with the functions $N_{\rm p}(Z)$,
$N_{\rm h}(N)$ as given by \citet[ Eqs.~31, 32]{mezbru93:coll}.

\subsubsection{Coherent scattering of neutrinos off nuclei}\label{appx:opa.specifics.scat_nuclei}

We use reaction rates for coherent neutrino scatterings off nuclei
which include corrections due to the ``nuclear form factor''
\cite[following an approximation by][]{mezbru93:coll,brumez97}, and 
ion-ion correlations \cite[as described in][]{hor97}. 
For a detailed discussion of the numerical handling of \emph{both} 
corrections, see the appendix of
\cite{brumez97}. The result of the latter reference can readily be
used to calculate $B^{(1)}_{\rm IS,ions}$, the contribution of coherent
scattering to the rhs.~of the first order moment equation.
For the Boltzmann equation, 
we have to make a minor additional approximation:
The correction factor 
$\left\langle\mathcal{S}(\epsilon)\right\rangle_{\rm ion}$ as provided by \cite{hor97}
applies for the transport opacity and thus for the combination 
$\phi_0-\phi_1$ (which is given by \citealp{brumez97}) but not
necessarily for the 
Legendre coefficients $\phi_0$ and $\phi_1$ individually, which are
needed for calculating the collision integral $B_{\rm IS,ions}$ for
the Boltzmann equation. 
We nevertheless assume here that
$\phi_{0/1}(\epsilon)\propto \left\langle{\cal
    S}(\epsilon)\right\rangle_{\rm ion}$. This implies 
\begin{equation}
\phi_{0/1}(\epsilon)\approx
 \frac{1}{16\pi}\,{\cal G}\cdot A^2\, n_A \mathcal{Z}(A,Z)\cdot
{\cal F}_{0/1}(y)\left\langle\mathcal{S}(\epsilon)\right\rangle_{\rm ion} 
\,,
\end{equation}
with 
\begin{eqnarray}
{\cal F}_0(y)&\equ&
 \frac{2y-1+{\rm e}^{-2y}}{y^2}   \,,       \nonumber \\
{\cal F}_1(y)&\equ&
 \frac{2-3y+2y^2-(2+y){\rm e}^{-2y}}{y^3}   \,,       \\
y&\equ&
\frac{2}{5}\frac{(1.07A)^{2/3}\epsilon^2(10^{-13}{\rm cm})^2}{(\hbar
  c)^2} 
\,,\nonumber 
\end{eqnarray}
containing the effects of the nuclear form factor, and
\begin{equation}
\mathcal{Z}(A,Z)\equ
 \left[(C_{\rm A}-C_{\rm V})+(2-C_{\rm A}-C_{\rm V})\frac{2Z-A}{A}\right]^2 
\,.
\end{equation}
$C_{\rm V}$ is defined as before (see Eqs.~\ref{eq:NNS_phi0},
\ref{eq:NNS_phi1}) and $C_{\rm A}=1/2$. 
In effect, the angular dependence of the ion-ion correlation correction
is not exactly accounted for in the Boltzmann equation.
The approximation, however, only influences the \emph{closure relation} 
for the moment equations at a higher level of accuracy. 
It neither affects the Legendre coefficients at larger neutrino energies 
(where $\left\langle\mathcal{S}(\epsilon)\right\rangle_{\rm ion}$ is
close to unity) nor 
does it destroy the consistency of the Boltzmann equation and its
moment equations.  
Moreover, it has been shown that the overall effect of 
ion-ion correlations  in stellar core collapse is only
marginally relevant \cite[]{brumez97,ram00}.

\subsubsection{Inelastic scattering of neutrinos off charged leptons}\label{appx:opa.specifics.nes}

For the scattering of neutrinos off charged leptons,
we approximate the dependence of the scattering kernel on the 
scattering angle by truncating the Legendre expansions
(Eqs.~\ref{eq:B_scatt}, \ref{eq:B0_scatt}, \ref{eq:B1_scatt}) at
$l_\mathrm{max}=1$. 
For our purposes, this has been shown to provide a sufficiently
accurate approximation \cite[see][]{smicer96,mezbru93:nes}, in
particular also because the total scattering rate is correctly
represented.  
Expressions for the Legendre coefficients are taken from the
works of \cite{yuebuc77,mezbru93:nes,cer94,smicer96}:
\begin{equation}\label{eq:NES_phi}
\phi_l^{\rm out}(\epsilon,\epsilon')=
  \alpha_{\rm I} A_l^{\rm I}(\epsilon,\epsilon')
 +\alpha_{\rm II} A_l^{\rm II}(\epsilon,\epsilon')
\,,
\end{equation}
with $k\in \{{\rm I},{\rm II}\}$ and
\begin{multline}\label{eq:NES_A}
A_l^k(\epsilon,\epsilon')=
\frac{{\cal G}}{\epsilon^2\,\epsilon^{\prime 2}}\cdot \frac{1}{(2\pi\hbar c)^3}\,
\int_{\max(0,\epsilon'-\epsilon)}^\infty\dlin{E_\mathrm{e}}
F_\mathrm{e}(E_\mathrm{e})\cdot\\
\big(1-F_\mathrm{e}(E_\mathrm{e}+\epsilon-\epsilon')\big)H^k_l(\epsilon,\epsilon',E_\mathrm{e})
\,.
\end{multline}
The constants $\alpha_{\rm I/II}$, which depend on the neutrino
species and also the type of charged lepton (i.e.~electron or
positron) involved in the scattering process, are combinations of the
weak coupling constants \cite[see, e.g.,][]{cer94}. 
For the $\mu$ and $\tau$ neutrinos and antineutrinos, which are
treated as a single kind of neutrino in our code, we take the
arithmetic mean of the corresponding coupling constants.
For $l=0,1$ the functions $H^{\rm I/II}_l(\epsilon,\epsilon',E_\mathrm{e})$
(in units of $\mev^5$) are given by \citet[][ with the
corrections noted by \citealp{bru85}]{yuebuc77}. 

\medskip

Calculating the Legendre coefficients
$\phi^\mathrm{in/out}_l(\epsilon_j,\epsilon_{j'})$ is a computationally  
expensive task, since for all combinations $(\epsilon_j,\epsilon_{j'})$
and all radial grid points numerical integrals
over the energy of the charged leptons (Eq.~\ref{eq:NES_A}) have to be
carried out.
It is, however, sufficient to explicitly compute
$\phi^\mathrm{in/out}_l(\epsilon_j,\epsilon_{j'})$ 
for $\epsilon_j \le \epsilon_{j'}$.
The missing coefficients can be obtained by exploiting a number of
symmetry properties \cite[see][]{cer94}.
In doing so, not only the 
computational work is reduced by almost a factor of two, but even more 
importantly, detailed balance 
($B_\mathrm{NES}\equiv 0$, if $f\equiv f^\mathrm{eq}$) can be verified
for the employed approximation and discretization  of the
collision integral and its 
angular moments to within the roundoff error of the machine, provided
that all integrals over neutrino energies are approximated by simple
zeroth-order quadrature formulae (cf.~Eq.~\ref{eq:freqdis}).    
Similarly, one can also verify the conservation of
particle number (Eq.~\ref{eq:numbconsv}) in the corresponding finite
difference representation.

\medskip

In our neutrino transport code the dependence of the source terms on
the neutrino distribution 
and its angular moments is treated fully implicitly in time, i.e.~the
time index of the corresponding quantities in
Eqs.~(\ref{eq:B_scatt}--\ref{eq:B1_scatt}) reads $f^{n+1}$ and
$L^{n+1}_{l}$.
The thermodynamic quantities and the composition of the stellar matter
which are needed to calculate the Legendre coefficients for
inelastic scattering of neutrinos, however, are computed from 
$\varepsilon^*$ and $\ye^*$, representing the 
specific internal energy and electron fraction at the time level after
the hydrodynamic substeps (cf.~Sect.~\ref{sect:transp.rhd.schedule}). 
This allows one to save computer time since the Legendre coefficients
must be computed only once at the beginning of each transport time step.
Comparing to results obtained with a completely time-implicit
implementation of the source terms (i.e.~using $\varepsilon^{n+1}$ and
$\ye^{n+1}$ for determining the stellar conditions that enter the
computation of the Legendre coefficients) we have found no
differences in the solutions during the core-collapse phase, where
neutrino-electron scattering plays an important role in redistributing
neutrinos in energy space and thus couples the neutrino transport to the
evolution of the stellar fluid. 

\subsubsection{Neutrino pair production/annihilation by thermal $\mathrm{e}^+\,\mathrm{e}^-$
  pairs}\label{appx:opa.specifics.pairs} 

Following the suggestion of \cite{ponmir98} we
truncate the Legendre series~(\ref{eq:B_TP}--\ref{eq:B1_TP}) for the
pair-production kernels  at $l_\mathrm{max}=2$.
The Legendre coefficients are given by \cite[]{bru85,ponmir98}
\begin{equation}\label{eq:Pairs_phi}
\begin{split}
\phi^\mathrm{p}_l(\epsilon_j,\epsilon_{j'})&={\cal G}\cdot
\frac{T^2}{1-\mathrm{e}^x} \\
&\cdot \frac{1}{(2\pi\hbar c)^3}\,
\left( \alpha_1^2\Psi_l(\epsilon_j,\epsilon_{j'})
      +\alpha_2^2\Psi_l(\epsilon_{j'},\epsilon_j)
\right)
\,,
\end{split}
\end{equation}
with $x\equ (\epsilon+\epsilon')/T$ and ${\cal G}$ defined by Eq.~(\ref{eq:G_const}).
Explicit expressions for the (dimensionless) functions
$\Psi_l(\epsilon_j,\epsilon_{j'})$ and details about their efficient
calculation are given by \cite{ponmir98}. 
Also the coefficients $\alpha_1,\alpha_2$ for the different neutrino
flavours can be found there \cite[ Table 1]{ponmir98}.

\subsubsection{Nucleon-nucleon bremsstrahlung}\label{appx:opa.specifics.brems}

Legendre coefficients for the nucleon-nucleon bremsstrahlung process,
assuming nonrelativistic nucleons, are
calculated by using the rates given in \cite{hanraf98}. 
The non-vanishing coefficients read: 
\begin{eqnarray}
\phi^\mathrm{p}_0(\epsilon_j,\epsilon_{j'})&=&
{\cal G}\cdot\frac{3\,C_{\rm A}^2}{(2\pi)^2}\,
\big(\mathrm{e}^{-x}-1\big)\cdot \label{eq:Brems_phi0}\\
&&\sum_{\{\mathrm{N}\}}
n_\mathrm{N}\cdot\frac{\Gamma_\sigma}{(\epsilon+\epsilon')^2 +
  0.25\,(\Gamma_\sigma\,g)^2}\cdot s(x)  \,,\nonumber\\
\phi^\mathrm{p}_1(\epsilon_j,\epsilon_{j'})&=&
-\frac{1}{9}\phi^\mathrm{p}_0(\epsilon_j,\epsilon_{j'})\label{eq:Brems_phi1} 
\,,
\end{eqnarray}
with
\begin{equation}\label{eq:Brems_gamma}
\Gamma_\sigma\equ\frac{8\sqrt{2\pi}\alpha_\pi^2}{3\pi^2}
\eta_*^{3/2}\frac{T^2}{m_\mathrm{N} c^2}(\hbar c)^3, \quad
\eta_*\equ\frac{(p_\mathrm{F}\,c)^2}{2 m_\mathrm{N} c^2 \,T}
\,.
\end{equation}
In Eq.~(\ref{eq:Brems_gamma}), 
$p_\mathrm{F}=\hbar\cdot(3\pi^2\,n_\mathrm{N})^{1/3}$
denotes the Fermi momentum for the nucleons, $\alpha_\pi=15$ is the
pion-nucleon ``fine-structure constant'' \cite[see][]{hanraf98}, and 
$m_\mathrm{N}$ is the nucleon mass.
The analytic form of the dimensionless fit functions $g$ and $s(x)$,
which both depend on the nucleon density $n_\mathrm{N}$ and temperature $T$ of the
stellar medium can be found in \cite{hanraf98}.
The sum in Eq.~(\ref{eq:Brems_phi0}) runs over the individual processes 
$\nu\bar\nu\,\mathrm{nn}\rightleftharpoons \mathrm{nn}; (\mathrm{N}=\mathrm{n})$, 
$\nu\bar\nu\,\mathrm{pp}\rightleftharpoons \mathrm{pp}; (\mathrm{N}=\mathrm{p})$, and
$\nu\bar\nu\,\mathrm{np}\rightleftharpoons \mathrm{np}; (\mathrm{N}=\mathrm{np})$.
The latter process is approximately taken into account by using the
particle density 
$n_\mathrm{np}\equ \sqrt{n_\mathrm{n} n_\mathrm{p}}$ and multiplying the
corresponding contribution to the sum in Eq.~(\ref{eq:Brems_phi0}) by
a factor of $28/3$ \cite[cf.][]{thobur00}.  

\bigskip

The production processes of neutrino-antineutrino pairs both by
the annihilation of  $\mathrm{e}^+\,\mathrm{e}^-$ pairs and by the 
nucleon-nucleon bremsstrahlung are implemented into the discretized
equations~(Eqs.~\ref{eq:MEj_fd}, \ref{eq:MEh_fd}, \ref{eq:STe_fd},
\ref{eq:STy_fd}) by applying the techniques  described for the
inelastic scattering reactions (see Sect.~\ref{appx:opa.specifics.nes}).


\section{Equation of state and nuclear burning}\label{appx:eos}

In this section we give basic information about the equation of state
that is currently used in our simulations of stellar core collapse and
supernovae. 

\subsection{Numerical handling of the equation of state:}

We employ the equation
of state (EoS) of \cite{latswe91} with a value of $K_{\rm
  s}=180~\mev$ for the incompressibility modulus of bulk nuclear
matter. The other parameters can be found in the paper of \cite{latswe91}. 
Since this EoS assumes
matter to be in nuclear statistical equilibrium (NSE) it is
applicable only for sufficiently high temperatures and densities.
A more general equation of state, which consistently extends 
to lower temperatures and densities not being available, we have
supplemented the EoS of \cite{latswe91} 
with an EoS that describes a mixture of ideal gases of nucleons and
nuclei with given abundances, plus ideal gases of electrons and
positrons of arbitrary degeneracy and relativity, plus photons
\cite[]{jan99}. 
The latter EoS includes Coulomb lattice corrections for the pressure, 
energy density, entropy, and adiabatic index. 

In general, the regime of NSE is bounded from below by a complicated 
hyperplane in $(\rho,T,\ye)$-space. 
For our purposes, however,  it is sufficient to assume that
the transition between the two regimes (and thus the two equations
of state) takes place at a fixed value of the density $\rho_0$: 
If the density of a fluid element fulfils the condition 
$\rho > \rho_0 = 6\times 10^7$~\gcm, the temperature in a supernova
core is usually large enough ($T\gtrsim 0.5$~MeV) to ensure that NSE
holds. 
Therefore the EoS of \cite{latswe91} can be used at such conditions. 
If the density drops below the value $\rho_0$, the non-NSE equation of
state is invoked.
Since both EoSs employ a similar description of the physical
properties of stellar matter at densities in the range between 
$10^7$~\gcm and $10^8$~\gcm, the merging of them across the
$(T,\ye)$-plane at $\rho_0$ 
is sufficiently smooth as far as, e.g., the pressure, internal energy
density and chemical potentials as functions of density are concerned.

A particular complication, however, arises with the chemical
composition of the stellar plasma.
\cite{latswe91} describe the baryonic part of the EoS as a mixture of
nucleons, $\alpha$-particles and a  heavy nucleus with in general
non-integer mass and charge numbers $(A,Z)$. 
The latter is considered
to be representative of a mixture of heavy nuclei that coexist at
given $(\rho,T,\ye)$.
Assuming NSE, the corresponding number densities of the
nuclear constituents and the mass and charge numbers of the  
representative heavy nucleus are given as a function of the local values
of $\rho$, $T$, and $\ye$.
At the transition from NSE to non-NSE, the nuclear composition 
freezes out and one would ideally want to retain the baryonic abundances
as given by the EoS of \cite{latswe91} at $\rho_0$. While this is no
problem in case of a Lagrangian hydrodynamics code, where the grid
cells follow the evolution of individual fluid elements with specific
information about the chemical composition attributed, the use of an
Eulerian or moving grid requires a more complicated numerical
procedure. Here, additional advection equations (similar to
Eq.~\ref{eq:hydro.ye}) for the different nuclear components must be
integrated to trace the temporal evolution of the composition on the
whole numerical grid. With a finite, fixed number of such conservation
equations to be solved in the code, one cannot allow for an arbitrarily
large ensemble of different nuclear species with non-integer mass and
charge numbers, as provided at freeze-out from NSE at $\rho_0$ by the  
EoS of \cite{latswe91}. Instead, we predefine a discrete set of nuclei 
$\{(A_k,Z_k)\}_{k=1,\dots,N_\mathrm{nuc}}$ at the beginning of the
simulation, with $k=1$ for neutrons, $k=2$ for protons, $k=3$ for
$\alpha$-particles and $k=4,\dots,N_\mathrm{nuc}$ for a number of 
suitably chosen heavy nuclei. When the conditions in a computational
cell change from NSE to non-NSE, we must map  the NSE
composition 
$n_\mathrm{n}(\rho_0), n_\mathrm{p}(\rho_0), n_{\alpha}(\rho_0), 
n_{(A(\rho_0),Z(\rho_0))}(\rho_0)$ as given by the EoS of
\cite{latswe91} to our discrete sample of species 
$\{(A_k,Z_k)\}_{k=1,\dots,N_\mathrm{nuc}}$ with the constraints of charge
neutrality and baryon number conservation.

The following procedure turned out to yield very satisfactory
results. It is applied at each time step and in cells of the
computational grid, which are close to a possible breakdown of NSE
(i.e. which are close to $\rho_0$). Applying this procedure also in
grid cells where NSE still holds and the EoS of \cite{latswe91} is
used, makes sure that the compositional information is available for a
transition to the non-NSE regime, if the freeze-out condition
($\rho<\rho_0$) is reached during a hydrodynamic time step.
First we identify the densities of neutrons and $\alpha$-particles
with their NSE values: $n_\mathrm{n}=n_\mathrm{n}(\rho)$, 
$n_{\alpha}=n_{\alpha}(\rho)$. For given density $\rho$ and electron
fraction $\ye$, the number densities of protons and nuclei, $n_\mathrm{p}$
and $n_{(A_j,Z_j)}$, respectively, can then be determined from the 
requirements of charge neutrality and baryon number conservation. 
The index $(A_j,Z_j)$ points to a particular nucleus 
$(A_j,Z_j)\in \{(A_k,Z_k)\}_{k=4...N_\mathrm{nuc}}$,
which is chosen from the set of non-NSE nuclei. 
It is associated with the representative heavy nucleus 
$(A(\rho),Z(\rho))$ according to the conditions:
\begin{align}\label{eq:nucmap}
|A_j-A| &\le |A_k-A| \quad \forall k\in \{4,\dots,N_\mathrm{nuc}\}\,,\nonumber\\
A_j-Z_j &\ge A-Z \,, \\
Z_j     &\le Z    \,.\nonumber
\end{align}
The first condition ensures that $(A_j,Z_j)$ is selected as the
nucleus of the ensemble $(A_j,Z_j)\in \{(A_k,Z_k)\}_{k=4...N_\mathrm{nuc}}$
which is closest to $(A(\rho),Z(\rho))$ concerning its mass
number. A unique solution requires that no isobars exist in the
ensemble. The second and third conditions guarantee that 
$n_\mathrm{p}\ge 0$ and $n_{(A_j,Z_j)}\ge 0$.

\bigskip

Both the EoS of \cite{latswe91} and the non-baryonic part of our
low-density equation of state are stored in tabular form for our
calculations. Our table of the EoS of \cite{latswe91} 
has 180 and 120  
logarithmically spaced entries within the density range
$5\times 10^7~\gcm\le\rho\le 3\times 10^{15}~\gcm$ and the temperature
range $0.1~\mev\le T \le 80~\mev$, respectively,
and 50 equally spaced entries for $0.001\le \ye\le 0.6$. 
Similarly, the table for the 
lepton plus photon EoS \cite[]{jan99} has 441 entries for
$10^{-10}~\gcm \le \rho \ye \le 10^{12}~\gcm$ and 141 entries
for $8.6\times 10^{-6}~\mev \le T \le 86~\mev$.
Intermediate values are obtained by trilinear interpolation 
in $\log \rho$, $\log T$, $\ye$ and bilinear
interpolation in $\log (\rho \ye)$, $\log T$, respectively. 

Using EoSs in a discretized form we have performed the test
calculation suggested by \citet[][ \S 4]{swe96} for exploring the
accuracy of the EoS evaluation.
From our tests we can exclude any serious ``unphysical
entropy production or loss in otherwise adiabatic flows''
\cite[]{swe96}.
For several combinations of initial values for $\ye$ and $s$ that are
typical of conditions encountered in core-collapse simulations, we
found the
deviations from adiabaticity to be negligibly small. The relative
change of the entropy per baryon was less than $10^{-3}$ for a
density increase by a factor of $10^4$.

\subsection{Nuclear burning, dissociation and recombination}
Nuclear burning, dissociation of nuclei and the recombination of free
nucleons and
$\alpha$-particles to heavy nuclei can change the baryonic composition
in the regime below the transition density ($\rho_0 = 6\times
10^7$~\gcm) from the EoS of \cite{latswe91} to the low-density EoS.
We take into account complete burning of carbon, nitrogen, oxygen,
magnesium and silicon as well as nuclear
dissociation and recombination in approximate ways. 

\paragraph{Burning:}
If the temperature in a grid cell
exceeds a threshold value characteristic of a certain nuclear reaction,
the burning element is assumed to be instantaneously converted to the
end product of the reaction.
This is a reasonably good approximation because the burning time
scale, which is an extremely steep function of the temperature,
is small compared with the time scale for hydrodynamic
changes.
Moreover, in dynamical supernova models one is mainly interested in
the energy release associated with nuclear reactions, but not in
detailed information about the nucleosynthetic yields.

When the condition for silicon burning applies, which is taken to be
$T \ge 4.5\times 10^9$~K \cite[]{hixthi99,mezlie01}, we simply add
up the local mass fractions of $\element[][28]{Si}$ and 
$\element[][56]{Ni}$:
$X^{n+1}_{\element[][56]{Ni}}=X^{n}_{\element[][56]{Ni}}+
                                 X^{n}_{\element[][28]{Si}}$, and set
$X^{n+1}_{\element[][28]{Si}}=0$. 
Analogously, we treat the burning of $\element[][12]{C}$ to
$\element[][24]{Mg}$ with a threshold temperature of $2.5\times 10^9$~K
and of $\element[][16]{O}$, $\element[][20]{Ne}$,
$\element[][24]{Mg}$ to $\element[][28]{Si}$ at $T=3.5\times 10^9$~K
\cite[]{wooheg02}.
\begin{figure}[!h]
\epsfxsize=8.8cm \epsfclipon \epsffile{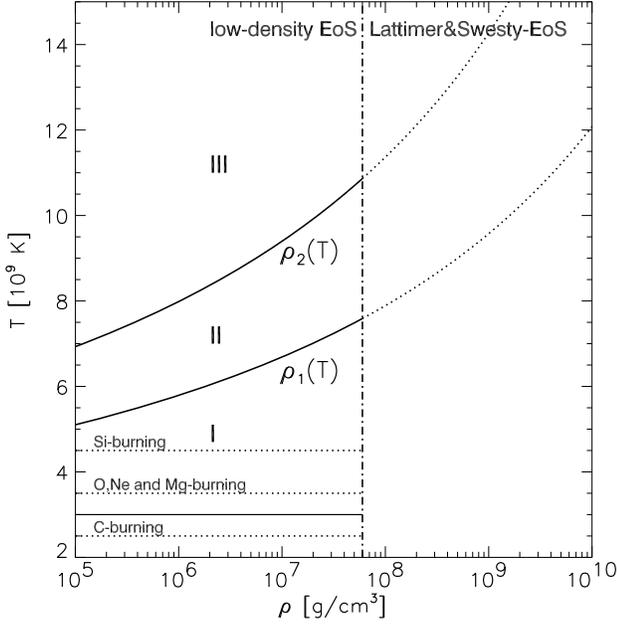} 
\caption[]{Area in the $\rho$-$T$-plane in the vicinity of the transition
  between the EoS of \cite{latswe91} and the low-density EoS.
  The transition density 
  $\rho_0$ is indicated by the vertical dash-dotted line. 
  The curves $\rho_{1}(T)$ and $\rho_{2}(T)$  bound three separate
  regions (I--III) in the  
  $\rho$-$T$-plane, where different prescriptions are used for
  determining the nuclear composition in the low-density regime.
  Region I contains heavy nuclei and possibly free nucleons, in region
  II the composition is determined by $\alpha$-particles and free
  nucleons, and in region III all nuclei and $\alpha$-particles are
  dissolved. 
  Below the temperature marked by the solid horizontal line 
  the recombination of $\alpha$-particles and free nucleons is
  suppressed. 
  The threshold temperatures of the different burning reactions are
  given by the horizontal dotted lines. 
} 
\label{fig:EoS}
\end{figure}

\paragraph{Nuclear dissociation and recombination:}
When the density drops into the range $\rho < \rho_0$ and the
temperature fulfills $T>3\times 10^9$~K at the same time the baryonic
composition is adjusted to account for the photo-disintegration of
nuclei and the recombination of free nucleons and $\alpha$-particles
to $\element[][56]{Ni}$. We distinguish between three different
regimes (I--III), separated by curves $\rho_{1}(T), \rho_{2}(T)$ in the
$\rho$-$T$-plane (see Fig.~\ref{fig:EoS}):
\begin{eqnarray}
\log(\rho_1(T))&\equ& 11.62 + 1.5\,\log(T_9)-39.17/T_9\,,\label{eq:EoS_rho_1} \\
\log(\rho_2(T))&\equ& 10.60 + 1.5\,\log(T_9)-47.54/T_9\,,\label{eq:EoS_rho_2}
\end{eqnarray} 
where the logarithms are to base 10 and $T_9\equ T/10^9\mathrm K$.
The function $\rho_{1}(T)$ is derived from the equations of
Saha equilibrium assuming a mixture of half (by mass)
$\element[][56]{Fe}$ and half $\alpha$-particles and 
free neutrons.
Analogously, $\rho_{2}(T)$ is defined by half the mass being in
$\alpha$-particles and half in free nucleons \cite[see][
Chap.~18]{shateu83}.
In detail the nuclear composition in the three regions is obtained
from the following prescriptions:

\begin{itemize}
\item[I)]
If $\rho > \rho_1(T)$, all free nucleons and a fraction $0 \le
f_\mathrm{I} \le 1$ of possibly existing $\alpha$-particles in a
corresponding grid cell are assumed to form $\element[][56]{Ni}$.
Pairing a suitable number of free neutrons and protons we therefore
set for the new composition:
\begin{align}
X^{n+1}_\mathrm{n}&=\max\{X^{n}_\mathrm{n}-X^{n}_\mathrm{p},0\} \,,\nonumber \\
X^{n+1}_\mathrm{p}&=\max\{X^{n}_\mathrm{p}-X^{n}_\mathrm{n},0\} \,,\nonumber \\
X^{n+1}_{\element[][4]{He}}&=(1-f_\mathrm{I})\cdot
X^{n}_{\element[][4]{He}} \,,\nonumber \\  
X^{n+1}_{\element[][56]{Ni}}&=X^{n}_{\element[][56]{Ni}}
                   +f_\mathrm{I}\cdot X^{n}_{\element[][4]{He}}
                   +2\cdot\min\{X^{n}_\mathrm{n},X^{n}_\mathrm{p}\} \,.
\label{eq:pseudonse_I}
\end{align}

\item[II)]
If $\rho_2(T) \le \rho \le \rho_1(T)$,
$\alpha$-particles are formed by the disintegration of heavy
nuclei and the recombination of free nucleons. The new composition is
given by: 
\begin{align}
X^{n+1}_\mathrm{n}     &=\max\{X^{*}_\mathrm{n}-X^{*}_\mathrm{p},0\} \,,\nonumber \\ 
X^{n+1}_\mathrm{p}     &=\max\{X^{*}_\mathrm{p}-X^{*}_\mathrm{n},0\} \,,\nonumber \\ 
X^{n+1}_{\element[][4]{He}}&=X^{*}_{\element[][4]{He}}+2\cdot\min\{X^{*}_\mathrm{n},X^{*}_\mathrm{p}\}\,,\nonumber \\ 
X^{n+1}_{k}&=(1-f_\mathrm{II})\cdot X^{n}_{k}\quad 
\forall k\in \{4,\dots,N_\mathrm{nuc}\}
\,,\nonumber  
\end{align}
with
\begin{align}
X^{*}_\mathrm{n}&=X^{n}_\mathrm{n}+f_\mathrm{II}\cdot\sum_{k=4}^{N_\mathrm{nuc}}
\frac{\max\{A_k-2 Z_k,0\}}{A_k}\cdot X^{n}_{k} \,,\nonumber \\ 
X^{*}_\mathrm{p}&=X^{n}_\mathrm{p}+f_\mathrm{II}\cdot\sum_{k=4}^{N_\mathrm{nuc}}
\frac{\max\{2 Z_k-A_k,0\}}{A_k}\cdot X^{n}_{k} \,,\nonumber \\ 
X^{*}_{\element[][4]{He}}&=X^{n}_{\element[][4]{He}}+f_\mathrm{II}\cdot
2\sum_{k=4}^{N_\mathrm{nuc}}\frac{\min\{A_k-Z_k,Z_k\}}{A_k}\cdot X^{n}_{k} \,.
\label{eq:pseudonse_II}
\end{align}
The factor $0 \le f_\mathrm{II} \le 1$ is the degree of dissociation
of heavy nuclei which is determined as outlined below. 

\item[III)]
If $\rho < \rho_2(T)$, all heavy nuclei and a fraction $0 \le
f_\mathrm{III} \le 1$ of the $\alpha$-particles is disintegrated into
free nucleons: 
\begin{align}
X^{n+1}_\mathrm{n}&=X^{n}_\mathrm{n}
           +f_\mathrm{III}\cdot\frac{1}{2}X^{n}_{\element[][4]{He}}
           +\sum_{k=4}^{N_\mathrm{nuc}}
                      \frac{A_k-Z_k}{A_k}\cdot X^{n}_{k} \,,\nonumber \\ 
X^{n+1}_\mathrm{p}&=X^{n}_\mathrm{p}
           +f_\mathrm{III}\cdot\frac{1}{2}X^{n}_{\element[][4]{He}}
           +\sum_{k=4}^{N_\mathrm{nuc}}
    \frac{Z_k}{A_k}\cdot X^{n}_{k} \,,\nonumber \\ 
X^{n+1}_{\element[][4]{He}}&=(1-f_\mathrm{III})\cdot X^{n}_{\element[][4]{He}}\,,\nonumber \\ 
X^{n+1}_{k}&=0
\quad \forall k\in \{4,\dots,N_\mathrm{nuc}\} \,.
\label{eq:pseudonse_III}
\end{align}

\end{itemize}

\bigskip
\noindent
Nuclear transmutations release (or consume) nuclear
binding energy. Consequences of this effect should be taken into
account in the hydrodynamical simulation. Since our EoS
defines the internal energy density such that contributions from
nuclear rest masses are included, the conversion between nuclear
binding energy and thermal energy will automatically lead to
corresponding changes of the temperature, pressure, entropy etc. of
the stellar gas.

The factors $f_\mathrm{I}$, $f_\mathrm{II}$ and $f_\mathrm{III}$ are
equal to unity in the corresponding regions I, II or III, respectively
of the $\rho$-$T$-plane but vary between 0 and 1 at the boundary
curves (Eqs.~\ref{eq:EoS_rho_1}, \ref{eq:EoS_rho_2}). This means that
the composition changes take place only on the curves 
$\rho_1(T)$ and $\rho_2(T)$. The degree of dissociation or
recombination (expressed by the values 
$0 \le f_\mathrm{I}, f_\mathrm{II},f_\mathrm{III}\le 1$) is limited by
the available internal energy.
Starting out with given internal energy density (at a fixed value of
$\rho$) and taking into account temperature variations due to changes
of the nuclear composition, the numerical algorithm maximizes
$f_\mathrm{I}$, $f_\mathrm{II}$ or $f_\mathrm{III}$, while keeping the
temperature on the boundary curves until the dissociation or
recombination is complete.

\end{appendix}

\bibliographystyle{aa}
\bibliography{lit}

\end{document}